# ANOMALIES OF PHASE DIAGRAMS AND PHYSICAL PROPERTIES OF ANTIFERRODISTORTIVE PEROVSKIDE OXIDES

**Maya D. Glinchuk[1] and Anna N. Morozovska[2*]**

[1] Institute for Problems of Materials Science, NAS of Ukraine,
Krjijanovskogo 3, 03142 Kyiv, Ukraine

[2] Institute of Physics, NAS of Ukraine, 46, pr. Nauky, 03028 Kyiv, Ukraine

The influence of rotomagnetic (RM), rotoelectric (RE) and magnetoelectric (ME) coupling on phase diagram and properties of antiferrodistortive (AFD) perovskite oxides was reviewed. The main examples we consider in the review are typical AFD perovkites, such as incipient ferroelectrics $EuTiO_3$, $SrTiO_3$, $Eu_xSr_{1-x}TiO_3$, multiferroic $BiFeO_3$ and $Bi_{1-x}R_xFeO_3$ (x = La, Nd). The strong influence of RM, RE and ME couplings on the physical properties and phase diagrams including antiferromagnetic (AFM), ferroelectric (FE) and structural AFD phases has been revealed in the framework of Landau-Ginzburg-Devonshire (LGD) theory, as well as the prediction of novel (double and triple) multiferroic phases has been demonstrated.

In the review we are especially focused on

(a) the possibility to induce FM (FE) phase in $EuTiO_3$ (as well as in other paraelectric AFM oxides) by the application of an electric (magnetic) field due to the ME coupling;

(b) the analysis of the size effects and novel phases in $Eu_xSr_{1-x}TiO_3$ nanosystems, where the LGD predicts the presence of the triple AFD-FE-FM(AFM) phase at low temperatures;

(c) the appearance of improper spontaneous polarization and pyroelectricity in the vicinity of antiphase domain boundaries, structural twin walls, surfaces and interphases in the AFD phase of non-ferroelectric $SrTiO_3$ induced by the flexoelectricity and rotostriction;

(d) the occurrence of low symmetry monoclinic phase with in-plane FE polarization in thin strained $Eu_xSr_{1-x}TiO_3$ films and its stabilization over wide temperature range by AFD oxygen octahedron tilts due to flexoelectric and rotostriction coupling;

(e) discussion of a surprisingly strong size-induced increase of AFM transition temperature caused by the joint action of RM coupling with elastic stress accumulated in the intergrain spaces of $BiFeO_3$ dense ceramics.

Noteworthy, the results obtained within LGD approach show the possibility of controlling multiferroicity, including FE, FM and AFM phases in bulk and nanosized AFD ferroics, with help of size effects and/or electric/magnetic field application. Since theoretical results are in qualitative agreement with experimental results, we conclude that LGD theory can be successfully applied to many AFD perovskite oxides.

---

[*] Corresponding author: anna.n.morozovska@gmail.com

# CONTENT

# LIST OF ABBREVIATIONS

AFB - antiphase phase boundaries  
AFD – antiferrodistortive  
AFM – antiferromagnetic  
BFO - bismuth ferrite, $BiFeO_3$  
ETO – europium titanate $EuTiO_3$  
ES - electrostriction  
FE – ferroelectric  
FM – ferromagnetic  
FiM - ferrimagnetic  
LGD - Landau-Ginzburg-Devonshire  
ME – magnetoelectric  
MS – magnetostrictive (or magnetostriction)  
PE - paraelectric  
PM - paramagnetic  
RE – rotoelectric  

RM - rotomagnetic
RS - rotostriction
STO – strontium titanate SrTiO3
TB - twin boundaries

# I. INTRODUCTION

Antiferrodistorted perovskite oxides can possess octahedra oxygen rotations characterized by spontaneous octahedra tilt angles, which in turn can be described by an axial vector $\Phi_i$ (see a typical schematics in the **Figure 1.1**). Following Gopalan and Litvin [1] the AFD symmetry is in fact a "rotosymmetry" that includes 69 rotation groups. Typical AFD perovskites with octahedrally tilted phases are incipient ferroelectrics SrTiO3, CaTiO3, antiferromagnet incipient ferroelectric EuTiO3, antiferromagnetic ferroelectric BiFeO3, ferroelectric Pb(Zr,Ti)O3 and antiferroelectric PbZrO3.

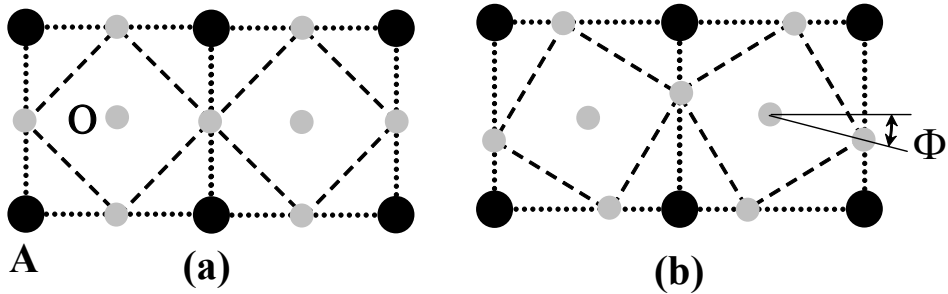

**Figure 1.1. (a)** Atomic ordering in ABO3 perovskite structure in the high temperature parent phase. **(b)** AFD ordering in the low temperature phase determined by the oxygen octahedron projection tilt Φ. The tilt value is typically opposite for the neighbouring oxygen octahedrons ABO3. (Reproduced from [E. A. Eliseev et al, Journal of Applied Physics 118, 144101 (2015)], with the permission of AIP Publishing).

The antiferrodistortive (**AFD**), ferroelectric (**FE**), ferromagnetic (**FM**) or antiferromagnetic (**AFM**) orderings in incipient ferroelectrics are related via the different types of biquadratic coupling. Firstly Balashova and Tagantsev [2] considered a multiferroic with two abstract scalar order parameters coupled biquadratically and reported about its versatile phase diagrams. Then Haun et al [3] and Tagantsev [4] had shown that a spontaneous polarization vector $P_i$ can appear inside antiphase domain boundaries and elastic twins of SrTiO3 and CaTiO3 due to "rotoelectric" (**RE**) biquadratic coupling term, $\xi_{ijkl} P_i P_j \Phi_k \Phi_l$. The RE coupling term was later regarded as Houchmandazeh-Laizerowicz-Salje (HLS) coupling [5]. Daraktchiev et al [6] considered biquadratic magnetoelectric (**ME**) coupling described by the term $\eta_{ijkl} P_i P_j M_k M_l$ and considered the coupling as the reason of magnetization appearance inside the FM domain wall in a non-ferromagnetic media. The "direct" rotomagnetic (**RM**) coupling described by the terms $\left(\xi_{ijkl}^{M} M_i M_j + \xi_{ijkl}^{L} L_i L_j\right)\Phi_k \Phi_l$ where **M** is magnetization, **L** is antiferromagentic order parameter, was introduced by Eliseev et al [7]. The microscopic background of RM coupling between the tilt **Φ** and AFM order **L** is related with the dependence of exchange interaction on the bond angle between magnetic and nonmagnetic ions in $ABO_3$ compounds [8, 9].

The ME, RE and RM biquadratic couplings are universal for all AFD materials with non-polar and non-magnetic parent symmetry. ME and RE coupling influence on their properties are relatively well studied, but not the RM coupling. For the first time Bussmann-Holder et al [10, 11] revealed experimentally the magnetic field impact on the AFD phase transition temperature and magnetic susceptibility of $EuTiO_3$. Recently Morozovska et al studied theoretically and experimentally RM coupling in fine-grained mutiferroic $BiFeO_3$ [[12]].

Detailed description of the obtained results is given in the following sections of the review text. Namely **Section II** analyzes the calculations of the phase diagrams and physical properties of AFD perovskite oxides, which are incipient ferroelectrics. In **Section III** the interfacial polarization and pyroelectricity induced by flexoelectric effect and rotostriction (**RS**) in AFD structures are analyzed. Possible new multiferroics based on $Eu_xSr_{1-x}TiO_3$ nanowires and nanotubes are proposed in **Section IV**. Low-symmetry monoclinic FE phase stabilized by oxygen octahedra rotation in strained $Eu_xSr_{1-x}TiO_3$ thin films is considered in **Section V**. **Section VI** represents interesting phenomena related with electric field induced FM phase in paraelectric antiferromagnet $EuTiO_3$. Multiferroics properties of $BiFeO_3$ and its solid solutions with La or Nd are considered with respect to linear or nonlinear AFD-AFM effect in **Section VII**. **Section VIII** is a conclusion.

## II. LANDAU-GINZBURG-DEVONSHIRE THEORY FOR CALCULATIONS OF PHASE DIAGRAMS AND PHYSICAL PROPERTIES OF ANTIFERRODISTORTED PEROVSKITE INCIPIENT FERROELECTRICS

As example, we will consider bulk quantum paraelectric $EuTiO_3$ (**ETO**) with relatively well-known parameters as prototype of the possible group of paraelectric antiferromagnets. ETO is a low temperature antiferromagnet [13, 14, 15] with Neel temperature about 5.5 K and AFD material below 285 K [16, 17, 18, 19, 20]. Further theoretical consideration utilizes the Landau-Ginzburg-Devonshire (**LGD**) theory that is based on the phase stability analysis of thermodynamic potential.

LGD potential is a series expansion on powers of the order parameters, which are polarization vector **P**, sum and difference of sublattices magnetizations vectors, **M** and **L**, and oxygen octahedra rotation angle pseudo-vector $\Phi$ for ETO with cubic high temperature parent phase [21, 22]:

$$g = g_{FE} + g_{AFD} + g_M + g_{Coupling} \qquad (2.1)$$

FE contribution to the energy (2.1) is:

$$g_{FE} = \frac{\alpha_P(T)}{2}P^2 + \frac{\beta_P}{4}P^4 - E_i P_i \qquad (2.2)$$

Here $P^2 = P_1^2 + P_2^2 + P_3^2$, $E_i$ is an external electric field component. For incipient ferroelectric expansion coefficient $\alpha_P$ depends on the absolute temperature $T$ in accordance with Barrett law, $\alpha_P(T) = \alpha_T^{(P)}\left(\left(T_q^{(P)}/2\right)\coth\left(T_q^{(P)}/2T\right) - T_c^{(P)}\right)$, where $\alpha_T^{(P)}$ is constant, temperature $T_q^{(P)}$ is the so-called quantum vibration temperature, $T_c^{(P)}$ is the "effective" Curie temperature corresponding to the polar soft modes in bulk quantum paraelectrics. Coefficient $\beta_P$ is regarded temperature independent.

AFD energy is

$$g_{AFD} = \frac{\alpha_\Phi(T)}{2}\Phi^2 + \frac{\beta_\Phi}{4}\Phi^4 \qquad (2.3)$$

Here $\Phi^2 = \Phi_1^2 + \Phi_2^2 + \Phi_3^2$. The tilt vector expansion coefficient $\alpha_\Phi$ depends on the absolute temperature $T$ in the form $\alpha_\Phi(T) = \alpha_T^{(\Phi)}\left(T_q^{(\Phi)}/2\right)\left(\coth\left(T_q^{(\Phi)}/2T\right) - \coth\left(T_q^{(\Phi)}/2T_S\right)\right)$, where $\alpha_T^{(\Phi)}$ is constant, temperature $T_q^{(\Phi)}$ is the characteristic temperature, $T_S$ is the AFD transition temperature (see e.g. ref.[23]).

AFM energy is [24]:

$$g_M = \begin{pmatrix} \frac{\alpha_T^{(M)}(T-T_C)}{2}M^2 + \frac{\alpha_T^{(L)}(T-T_N)}{2}L^2 - \mu_0 M H + \\ \frac{\beta_L}{4}L^4 + \frac{\beta_M}{4}M^4 + \frac{\xi_{LM}}{2}L^2 M^2 + \frac{\gamma_L}{6}L^6 + \frac{\gamma_M}{6}M^6 \end{pmatrix} \qquad (2.4)$$

Here two order parameters are introduced, namely FM, $M = (M_a + M_b)/2$, and AFM, $L = (M_a - M_b)/2$, ones. $M_a$ and $M_b$ are the components of magnetizations of two equivalent sub-lattices. $H$ is the external magnetic field. Using molecular field approximation, one could show that the equality $\alpha_T^{(L)} = \alpha_T^{(M)}$ is valid for the sub-lattices antiferromagnets. $T_C$ is the seeding FM Curie temperature and $T_N$ is the seeding Neel temperature for bulk material without AFD ordering. In two sub-lattices antiferromagnets the negative $T_C$ value can be determined experimentally from the temperature dependence of inverse magnetic susceptibility in paramagnetic phase [24].

Biquadratic coupling energy consists of ME, RE, RM and higher order coupling terms:

$$g_{Coupling} = \frac{P^2}{2}\left(\eta_{FM}M^2 + \eta_{AFM}L^2\right) + \frac{\xi_{RE}}{2}\Phi^2 P^2 + \frac{\Phi^2}{2}\left(\xi_{RM}^M M^2 + \xi_{RM}^L L^2\right) + \frac{\gamma_{PM}}{2}\left(M^2 - L^2\right)^2 P^2 \quad (2.5)$$

Following Lee *et al.* [25] one can assume that the coefficients of FM and AFM order parameters contributions to ME coupling have equal absolute values and opposite signs, i.e. $\eta_{AFM} = -\eta_{FM}$. The biquadratic RE coupling coefficient $\xi_{RE}$ and RM coupling coefficients, $\xi_{RM}^M$ and $\xi_{RM}^L$, and higher order coupling coefficient $\gamma_{PM}$ are regarded temperature-independent [3, 4, 5]. Hereinafter we also assume that $\xi_{RM}^M = -\xi_{RM}^L$ as a consequence of two magnetic sub-lattices equivalence. The higher order biquadratic ME coupling term, $\frac{\gamma_{PM}}{2}(M^2 - L^2)^2 P^2$, is included in Eq.(2.5) in order to reach better fitting of the experimental results. In the case of incipient ferroelectric perovskite, which we consider, namely SrTiO$_3$, CaTiO$_3$, EuTiO$_3$, and the latter is a cubic PM-PE phase at temperatures higher than AFD transition. Hence we could not see any symmetry grounds to include the bilinear or trilinear couplings [26, 27, 28] into the parent phase energy.

The equations of state for the tilt $\Phi$, magnetization $M$, AFM order parameter $L$ and polarization $P$ were obtained from the minimization of the free energy (2.1)-(2.5). Dielectric, magnetic and ME susceptibility tensors can be calculated from expressions $\chi_{ij}^E = \partial P_i/\partial E_j$, $\chi_{ij}^M = \partial M_i/\partial H_j$ and $\chi_{ij}^{ME} = \partial P_i/\partial H_j$ correspondingly. These properties are tilt-dependent due to the RE and RM coupling.
.

### 2.1. RM effect in the vicinity of the AFD phase transition

In accordance with both Bussmann-Holder et al [10, 11] experiments and our calculations [7], magnetic susceptibility should have the change of the slope for the second order transition at the point of AFD transition (see **Figure 2.1**). Symbols in the **Figure 2.1** represent experimental results taken from Ref. [11]; fitting from Ref.[7] is shown by solid lines. The main result is the determination of $\xi_{RM}^M = -1.75 \times 10^{16}$ N/(m$^2$A$^2$).

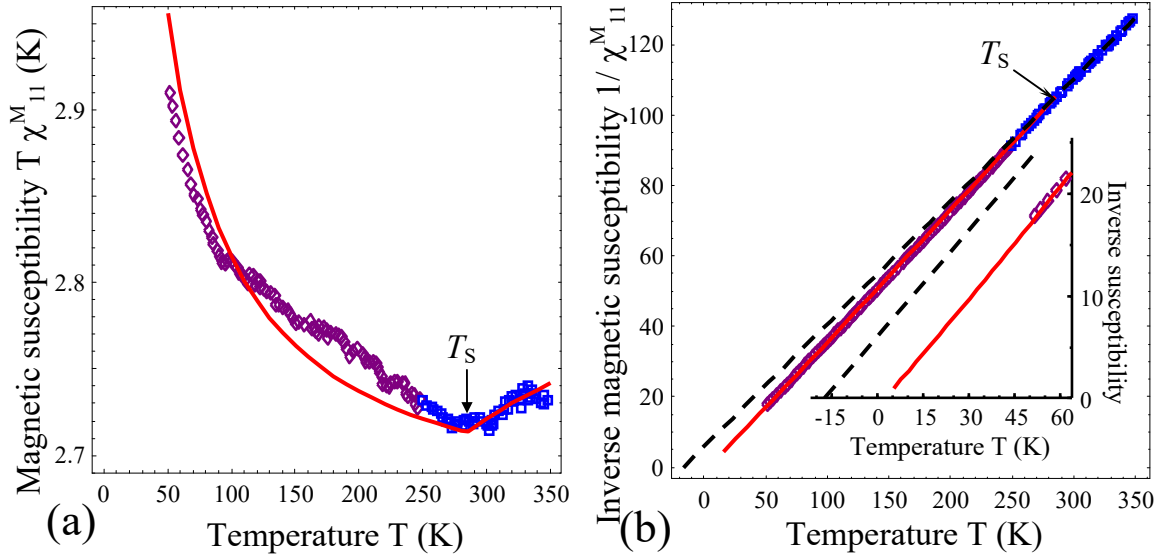

**Figure 2.1**. **Temperature dependence of ETO magnetic susceptibility.** Symbols represent experimental results of the susceptibility component $\chi_{11}^M$ taken from Ref. [11] at very small measuring magnetic field *H*. Best fitting (solid curves) [7] for dimensionless component of magnetic susceptibility $\chi_{11}^M$ multiplied by T **(a)** and the inverse magnetic susceptibility **(b)** correspond to the following values of parameters: $T_S$= 285 K, $T_C = -17$ K and $\xi_{RM}^M = -1.75 \times 10^{16}$ N/(m$^2$A$^2$). Dashed curve is the extrapolation of the dependence just above the transition temperature $T_S$ to lower temperatures. (Reproduced from [E. A. Eliseev et al, Journal of Applied Physics 118, 144101 (2015)], with the permission of AIP Publishing).

Effects shown in the **Figures 2.1** appeared in the close vicinity of AFD transition temperature (285 K) are small enough, since the AFD order parameter is small here. More pronounced and intriguing peculiarities of the dielectric susceptibility, magnetization and AFD order parameter can be caused by RM coupling at lower temperatures. Corresponding examples are demonstrated in the next section**.**

### 2.2. External field control of the phase diagram

**Figures 2.2a** and **2.2c** illustrates the influence of external magnetic field on AFD antiferromagnets (prototype is ETO) magnetic properties at zero electric field (compare with results of Ref.[29]). Solid and dashed curves are calculated in [7]; symbols are obtained from experimental data [15] by indirect way. Let us underline adequate agreement between the theoretical modelling and experimental results. At zero electric field RM coupling shifts the AFM phase boundary to the lower temperatures region from 25.2 K to 5.5 K, as follows from the comparison of **Figure 2.2a** with **2.2c.** The seeding Neel temperature calculated without RM coupling (i.e. without inclusion of the tilt influence on magnetization) is about 25 K for ETO. With RM coupling $\xi_{RM}^L = 1.75 \times 10^{16}$ N/(m$^2$A$^2$) the Neel temperature appeared equal to experimental value 5.5 K. Our modelling describes the experimental data for AFM-PM phases boundary at temperatures more than 4 K relatively good only with RM coupling included. Thus we can regard this fact as the direct evidence of the coupling existence.

**Figures 2.2b** and **2.2d** show the influence of external electric field on the ferroic magnetic properties at zero magnetic field. All panels of the **Figure 2.2** correspond to the temperature region $T<T_S$, so that AFD phase is always present here. Phase diagrams contain AFD paramagnetic (**PM**), ferromagnetic (**FM**), ferrimagnetic (**FiM**) and antiferromagnetic (**AFM**) phases separated by solid or dashed curves, which in fact should be interpreted as the dependences of the critical magnetic (**Figure 2.2a, 2.2c**) and electric (**Figure 2.2b, 2.2d**) fields on temperature. At zero magnetic field phase diagrams contain a tetra-critical point between FM, FiM, AFM and PM phases inside AFD phase (**Figure 2.2b**) and three-critical point between FM, AFM and PM phases inside AFD phase (**Figure 2.2d**). All the panels of **Figure 2.2** illustrates the possibility to govern the phase diagram with AFM, FM, FiM and PM phases inside AFD phase by external magnetic and electric fields, the latter phenomena originating from RM, RE and ME couplings.

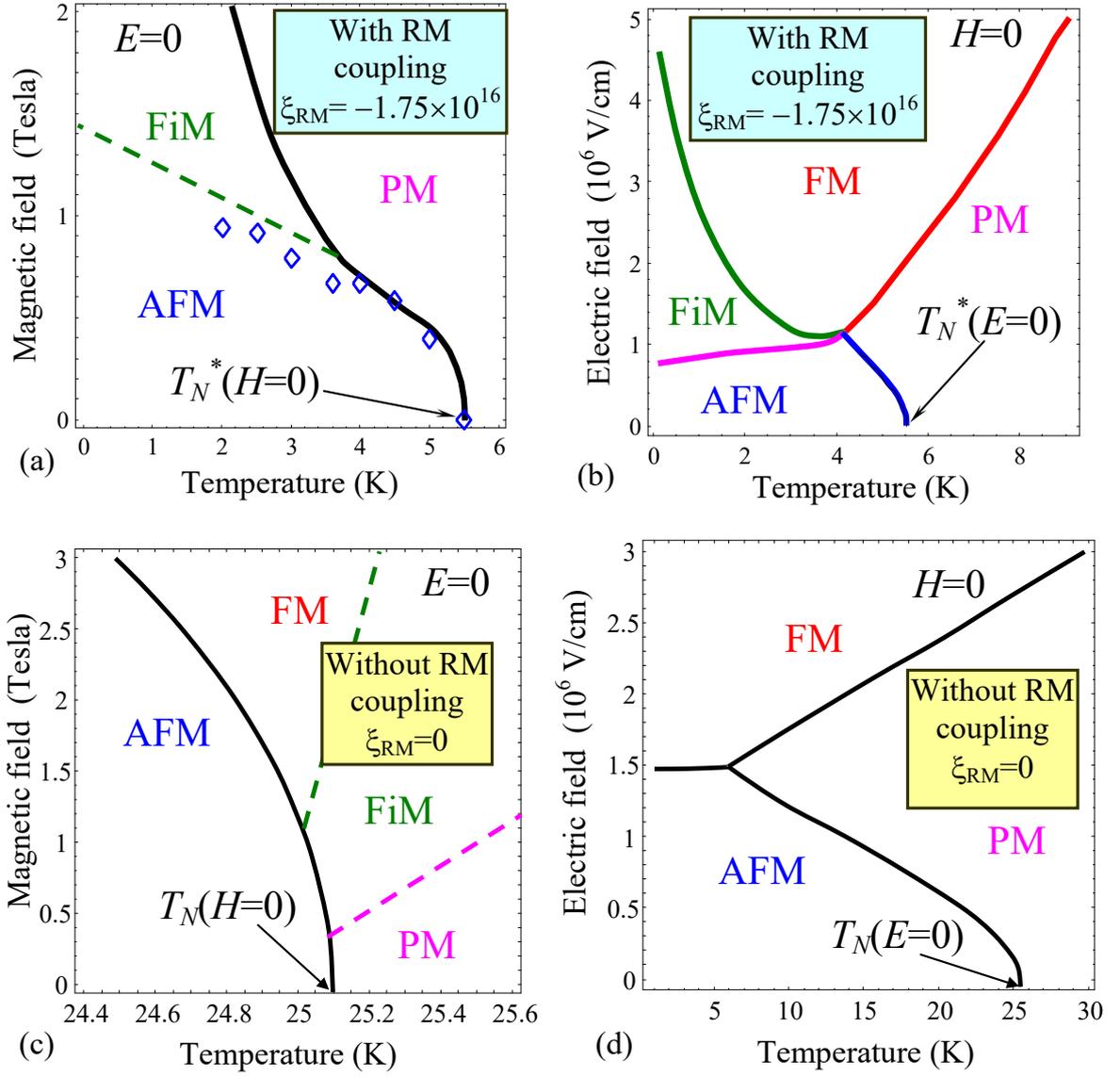

**Figure 2.2. External field control of multiferroic properties.** Phase diagram in the coordinates of temperature versus external electric **(a,c)** and magnetic **(b,d)** fields. The AFD ordered phase is always present in the PM, FM, FiM and AFM phases. Solid and dashed curves are calculated by us, symbols in plot (a) are experimental data from the ref.[15]. The position of the dashed lines corresponding to the AFM-FiM and PM-FM boundaries are approximate due to the absence of strict criteria for determination of FiM-PM and PM-FM transition in the presence of magnetic field. Plots **(a)** and **(b)** are calculated with RM-coupling ($\xi_{RM}^{M} = -\xi_{RM}^{L}$, $\xi_{RM}^{L} = -1.75 \times 10^{16}$ N/(m$^2$A$^2$)). Plots **(a)** and **(c)** are calculated without RM coupling ($\xi_{RM}^{M} = \xi_{RM}^{L} = 0$) (Reproduced from [E. A. Eliseev et al. Journal of Applied Physics 118, 144101 (2015)], with the permission of AIP Publishing).

### 2.3. Phase diagram in coordinates temperature – RM coupling

Theoretical results depicted in the **Figures 2.2** were obtained on the basis of quantitative calculations with the help of Eqs.(2.1)-(2.5). To make more clear the physical reasons the shift of Neel temperature and other results it appeared useful to look for analytical formulas by minimization of Eq.(2.1) with respect to Eqs.(2.2)-(2.5). The transitions from PM phase to AFM and FM phases are of the second order under the conditions $\left(\xi_{RM}^L\right)^2 < \beta_\Phi \beta_L$ and $\left(\xi_{RM}^M\right)^2 < \beta_\Phi \beta_M$, respectively. Renormalized Neel and Curie temperatures were obtained in Ref.[7] in the form:

$$T_N^* = \frac{T_N - \alpha_T^{(L)} \xi_{RM}^L \left(\alpha_T^{(\Phi)}/\beta_\Phi\right) T_S}{1 - \alpha_T^{(L)} \xi_{RM}^L \left(\alpha_T^{(\Phi)}/\beta_\Phi\right)}, \qquad T_C^* = \frac{T_C - \alpha_T^{(M)} \xi_{RM}^M \left(\alpha_T^{(\Phi)}/\beta_\Phi\right) T_S}{1 - \alpha_T^{(M)} \xi_{RM}^M \left(\alpha_T^{(\Phi)}/\beta_\Phi\right)} \qquad (2.6)$$

Keeping in mind that $\xi_{RM}^M = -\xi_{RM}^L$, dependences of $T_N^*$ and $T_C^*$ on RM coupling coefficient $\xi_{RM}^L$ could be seen from the corresponding phase diagram (**Figure 2.3**). Allowing for the condition $\alpha_T^{(L)} \xi_{RM}^L \left(\alpha_T^{(\Phi)}/\beta_\Phi\right) \ll 1$ it is seen that the dependences of $T_N^*$ and $T_C^*$ on the coupling coefficient $\xi_{RM}^L$ are close to linear function. The decrease of $T_N^*$ and increase of $T_C^*$ originate from negative sign of $\xi_{RM}^M$ and of positive sign of $\xi_{RM}^L$. Since the opposite signs are characteristic feature of any two sublattices antiferromagnets, the decrease $T_N^*$ and increase of $T_C^*$ and their intersection at some $\xi_{RM}^L$ has to be valid for others antiferromagnets with two sublattices.

As one could see from the **Figure 2.3** the boundary between AFM and FM phases (dotted curve) starts from the crossing point $T_N^* = T_C^*$ corresponding to the critical value $\xi_{RM}^L = 1.87 \times 10^{16}$ N/(m²A²). The latter is absent in ETO, because its RM coefficient, $\xi_{RM}^L = 1.75 \times 10^{16}$ N/(m²A²), is smaller than the critical value.

Note, that **Figure 2.3** shows that the ground state is very sensitive to the $\xi_{RM}^L$ critical value, but we are quite sure that the situation is material specific. In fact the ETO ground state, that is AFM, has very close energy to the FM one. The fact is supported by recent DFT calculations [30]. Noteworthy the proximity of the apparent values Curie (virtual FM) and Neel (actual AFM transition) temperatures, namely $T_C$=3.8 K and $T_N$=5.5 K respectively. Therefore high sensitivity of ETO ground state to the perturbations is related to the close values of its Curie and Neel temperatures. Since the condition $\left(\xi_{RM}^M\right)^2 < \beta_\Phi \beta_M$ is already broken above the critical value of the coupling coefficient, the transition between PM and FM phases is of the first order in this region of parameters. The condition $\left(\xi_{RM}^M\right)^2 < \beta_\Phi \beta_M$ is already broken for ETO, but the FM phase is not realized in the material.

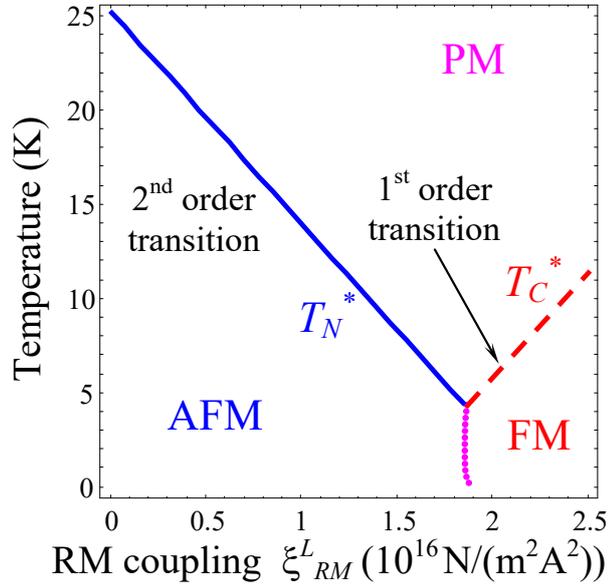

**Figure 2.3.** Phase diagram in the coordinates temperature - RM coupling coefficient $\xi_{RM}^L$ calculated at $\xi_{RM}^M = -\xi_{RM}^L$. Solid line corresponds to the boundary between AFM and PM phases, dashed line represent the boundary between FM and PM phases and the dotted line in boundary between AFM and FM phases. (Reproduced from [E. A. Eliseev et al, Journal of Applied Physics 118, 144101 (2015)], with the permission of AIP Publishing).

To resume the section, we have to underline the following issues. The main driving force of RM effects was shown to be biquadratic coupling between AFD and AFM order parameters. RM effect influence on the properties appeared much stronger in the region with coexistence of two abovementioned long-range orders, i.e. in the multiferroic state. In particular the value of observed Neel temperature was shown to be defined by RM coupling. Without the coupling Neel temperature $T_N$=25.2 K, and it decreases linearly to observed in ETO value 5.5 K with the coupling increase, while magnetic Curie temperature increases also linearly from $T_C= -17$ K to the value larger than 5.5. K at RM coupling coefficient $\xi_{RM}^L > 1.87 \times 10^{16}$ N/(m²A²). Therefore the possibility of transformation from AFM to FM phase transition appears in the material with RM coupling larger than that in ETO.

The RM coupling shifts and deforms the equilibrium lines on the phase diagrams in the coordinates temperature, external magnetic and electric fields. It is worth to underline that the value of critical electric field, required to induce the transition from AFM phase to FM phase, calculated with RM coupling, appeared to be essentially smaller than the one calculated without RM coupling. Note the appearance of the three-critical point in AFD phase region under electric field application, namely FM, AFM and PM phases coexist in the point. Multiferroic properties essentially depend on RM coupling values without external fields. In particular for small or high enough RM coupling values the phase transition becomes of the second or the first order respectively. Among other interesting anomalies we would like to underline the anomalies of the temperature dependence of magnetic and dielectric permittivities arising with RM coefficient increase [7].

# III. IMPACT OF FLEXOELECTRIC EFFECT, ROTOSTRICTION AND FREE CARRIERS IN FERROELECTRICS AND FERROELASTICS ON INTERFACIAL POLARIZATION AND PYROELECTRICITY

**3.1. Interfacial polarization and pyroelectricity in antiferrodistortive structures induced by the flexo-roto field**

Unique multifunctional properties of oxide interfaces, such as the superconductivity of 2-dimensional electron gas [31, 32, 33], charged domain walls [34], magnetism [35, 36] and multiferroicity at oxide interfaces [37] and thin strained films [38]. Gradients of various order parameters such as strain, octahedral rotations, polarization, and magnetization are inherent to surfaces and interfaces, which can couple to induce new phenomena not present in the relevant bulk materials [39]. The influence of strain [38, 40] and strain gradients [41, 42, 43] in inducing FE polarization is well known. Improper ferroelectricity induced by coupling to octahedral rotations has been predicted in $YMnO_3$ [44], $Ca_3Mn_2O_7$ [45] and $CaTiO_3$ [46] oxides and their multilayers [47]. Interfaces and surfaces of antiferrodistorted perovskite oxides possess gradients of oxygen octahedral rotations, characterized by spontaneous octahedral tilt angles, $\Phi_i$ [1]. Immediately omnipresent flexoelectric effect [41, 43, 48, 49, 50, 51] (that is the appearance of a FE polarization due to a strain gradient) and rotostriction (that is a quadratic coupling between octahedral rotations and strain) couples at such interfaces. The coupling can lead to a FE polarization at an interface or surface [52, 53, 54] across which the octahedral rotation varies, which is the subject of this section.

It has been predicted that a spontaneous polarization vector $P_i$ can appear inside structural walls due to biquadratic Houchmandazeh-Laizerowicz-Salje (HLS) coupling term $\eta_{ijkl}P_i P_j \Phi_k \Phi_l$ [3, 4, 5], but it is absent in the bulk. HLS coupling leads to a polarization appearance inside antiphase boundaries in $SrTiO_3$ below 50 K [4]. Zubko et al [55] experimentally observed strong changes of the apparent flexoelectric coefficient in $SrTiO_3$ at much higher temperatures, namely below the AFD phase transition temperature (105 K), and supposed that one of its possible reasons is the polarization appearance at the domain walls between twins. Recently Salje et al [56] directly observed ferrielectric polarization at ferroelastic domain boundaries in $CaTiO_3$ by aberration-corrected TEM at room temperature.

Above argumentation backgrounds calculations based on the LGD free energy, to study the impact of flexoelectric coupling on the spontaneous polarization in the vicinity of structural domain walls in non-ferroelectric tilted perovskites such as incipient ferroelectrics $SrTiO_3$, $CaTiO_3$, and EuTiO$_3$. Below we discuss results for 90-degree twin boundaries (**TB**), 180-degree antiphase phase boundaries (**APB**) and surfaces in SrTiO$_3$ based on our papers [52 - 54].

The free energy density has the form [52]:

$$F_b = a_i(T)P_i^2 + a_{ij}^u P_i^2 P_j^2 + \ldots + \frac{g_{ijkl}}{2}\left(\frac{\partial P_i}{\partial x_j}\frac{\partial P_k}{\partial x_l}\right) - P_i\left(\frac{E_i^d}{2} + E_i^{ext}\right) - q_{ijkl}u_{ij}P_k P_l + \frac{c_{ijkl}}{2}u_{ij}u_{kl}$$

$$+ b_i(T)\Phi_i^2 + b_{ij}^u \Phi_i^2 \Phi_j^2 - \eta_{ijkl}^u P_i P_j \Phi_k \Phi_l + \frac{v_{ijkl}}{2}\left(\frac{\partial \Phi_i}{\partial x_j}\frac{\partial \Phi_k}{\partial x_l}\right) \quad (3.1)$$

$$- r_{ijkl}^{(\Phi)} u_{ij} \Phi_k \Phi_l + \frac{f_{ijkl}}{2}\left(\frac{\partial P_k}{\partial x_l}u_{ij} - P_k \frac{\partial u_{ij}}{\partial x_l}\right) - \Phi_i \tau_i^d$$

$\Phi_i$ is the components ($i=1, 2, 3$) of an axial tilt vector corresponding to the octahedral rotation angles, $\tau_i^d$ is de-elastification torque [1]; $u_{ij}(\mathbf{x})$ is the strain tensor. The summation is performed over all repeated indices. Coefficients $a_i(T)$ and $b_i(T)$ temperature dependence can be fitted with Barrett law for quantum paraelectrics [57]: $a_1(T) = \alpha_T T_q^{(E)}\left(\coth(T_q^{(E)}/T) - \coth(T_q^{(E)}/T_0^{(E)})\right)$ and $b_1(T) = \beta_T T_q^{(\Phi)}\left(\coth(T_q^{(\Phi)}/T) - \coth(T_q^{(\Phi)}/T_S)\right)$. Gradients coefficients $g_{ijkl}$ and $v_{ijkl}$ are regarded positive for commensurate ferroics. $f_{ijkl}$ is the forth-rank tensor of flexoelectric coupling, $q_{ijkl}$ is the forth-rank electrostriction tensor, $r_{ijkl}^{(\Phi)}$ is the rotostriction tensor. The biquadratic coupling between $\Phi_i$ and polarization components $P_i$ are defined by the constants $\eta_{ijkl}$. The flexoelectric effect tensor $f_{ijkl}$ and rotostriction tensor $r_{ijkl}^{(\Phi)}$ have nonzero components in all phases and for any symmetry of the system. Tensors form for cubic $m3m$ symmetry is well-known; in particular in Voight notations $f_{12}$, $f_{11}$ and $f_{44}$ are nonzero [55] similarly to elastic constants and electrostriction tensors [58]. Note, that the inclusion of the flexoelectric Lifshitz term in the free energy is critical for all effects discussed below.

External field is $E_i^{ext}$. In general case polarization distribution $P_i(x_i)$ can induce the depolarization field $E_i^d$ inside the wall. In the dielectric limit $E_i^d$ obeys electrostatic equation:

$$\varepsilon_0 \varepsilon_b \frac{\partial E_i^d}{\partial x_i} = -\frac{\partial P_i}{\partial x_i}, \quad (i=1, 2, 3) \quad (3.2)$$

where $\varepsilon_0 = 8.85 \times 10^{-12}$ F/m is the universal dielectric constant, $\varepsilon_b$ is the "base" isotropic lattice permittivity, different from the FE soft mode permittivity [59, 60, 61, 62]. Semiconductor case is considered elsewhere [63] (which is an extended electronic version of published paper [52]).

Euler-Lagrange equations of state are obtained from the minimization of the free energy (3.1) as

$$\frac{\partial F_b}{\partial \Phi_i} - \frac{\partial}{\partial x_j}\left(\frac{\partial F_b}{\partial(\partial \Phi_i/\partial x_j)}\right) = 0, \qquad \frac{\partial F_b}{\partial P_i} - \frac{\partial}{\partial x_j}\left(\frac{\partial F_b}{\partial(\partial P_i/\partial x_j)}\right) = 0, \qquad (3.3a)$$

$$\frac{\partial F_b}{\partial u_{ij}} - \frac{\partial}{\partial x_k}\left(\frac{\partial F_b}{\partial(\partial u_{ij}/\partial x_k)}\right) = \sigma_{ij}. \qquad (3.3b)$$

Where $\sigma_{ij}(\mathbf{x})$ is the stress tensor that satisfies mechanical equilibrium equation $\partial \sigma_{ij}(\mathbf{x})/\partial x_j = 0$. Note, that the stress tensor, polarization and tilt gradients vanish far from the domain walls.

Elastic equation of state could be rewritten via the strains $u_{ij}(\mathbf{x})$ as follows:

$$u_{mn} = s_{mnij}\sigma_{ij} + R^{(\Phi)}_{mnkl}\Phi_k\Phi_l + Q_{mnkl}P_k P_l - F_{mnkl}\frac{\partial P_k}{\partial x_l}. \qquad (3.4)$$

Where $s_{mnij}$ is the elastic compliances tensor; $R^{(\Phi)}_{ijkl} = s_{ijmn}r^{(\Phi)}_{mnkl}$ is the rotostriction strain tensor; $Q_{ijkl} = s_{ijmn}q_{mnkl}$ is the electrostriction strain tensor; $F_{ijkl} = s_{ijmn}f_{mnkl}$ is the flexoelectric strain tensor [52]. The latter term corresponds to converse flexoelectric effect.

The inhomogeneous strain $u_{ij}(\mathbf{x})$ given by Eq.(3.4) induces the polarization variation $\delta P_i(\mathbf{x})$ across the structural APB and TB, domain walls, defects and interfaces due to the direct flexoelectric effect:

$$\delta P_i(\mathbf{x}) \sim a^{-1}_{iv} f_{mnvl}\frac{\partial u_{mn}}{\partial x_l} \sim -a^{-1}_{iv} f_{mnvl} R^{(\Phi)}_{mnpq}\frac{\partial(\Phi_p\Phi_q)}{\partial x_l}. \qquad (3.5)$$

The term $f_{mnvl}\dfrac{\partial u_{mn}}{\partial x_l}$ denotes direct flexoelectric effect. Note, that Eq.(3.5) is valid only for zero electric field, including both external and depolarization fields. The proportionality in Eq.(3.5) suggests that the product of the flexoelectric $f_{mnvl}$ and rotostriction $R^{(\Phi)}_{mnpq}$ coefficients leads to the appearance of spontaneous polarization, which will be abbreviated as flexo-roto-effect.

Below we consider several one-dimensional problems, which follow from general results of the previous section, namely a typical 180-degree APB and 90-degree TB.

### 3.1.1. Flexo-roto-effect manifestation at the APB

In the octahedral tilted phase at $T < T_S$, the one-component spontaneous tilt, $\Phi^S_3$, appears in a bulk SrTiO3, other components, $\Phi_1$ and $\Phi_2$, can be nonzero in the vicinity of APB. "Easy" APB with $\Phi_3(x_3) \neq 0$, $\Phi_2 \equiv 0$, $\Phi_1 \equiv 0$ (see **Fig. 3.1a**) induces nonzero odd or even distribution of polarization $P_3(x_3)$, while $P_1 \equiv 0$, $P_2 \equiv 0$. "Hard" APB with $\Phi_1(x_1) \neq 0$, $\Phi_3(x_1) \neq 0$, $\Phi_2 \equiv 0$ (see **Fig. 3.1b**) induces nonzero odd or even distributions of polarization $P_1(x_1)$ and even distribution of $P_3(x_1)$, while $P_2 \equiv 0$. Note, that classification "easy" and "hard" APB comes from Ref.[4].

Flexo-roto field, which induces the polarization component parallel to the wall, is shown at the bottom plots **Fig. 3.1** at two different temperatures $T_1 < T_2$. Note, that Vasudevarao et al [38] observed and calculated by phase-field various orientations of the ferroelastic APB in SrTiO$_3$.

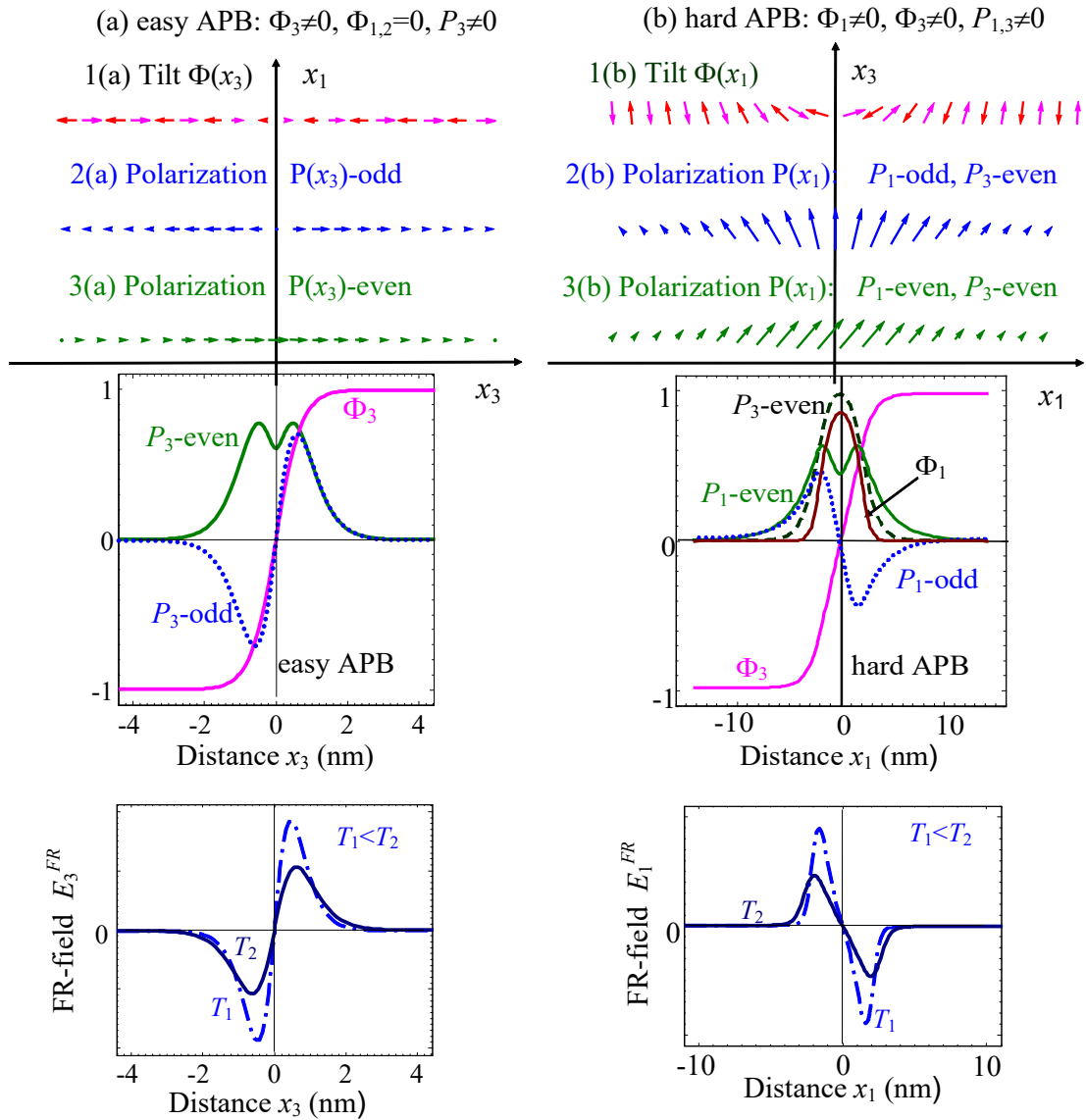

**Figure 3.1.** Schematics of the polarization appearance inside easy (a) and hard (b) APB. $x_1$=[100], $x_3$=[001] and $x_2$=[010] (not shown), are crystallographic axes directions of SrTiO$_3$. (Reproduced from [A. N. Morozovska et al, Phys.Rev.B. **85**, 094107 (2012)], with the permission of APS Publishing)

Under the absence of the flexoelectric field the spontaneous polarization is zero at temperatures higher than the effective Curie temperature $T_C^{APB}$ (see curves 1,2 calculated at $F_{ij} \equiv 0$ and $\eta_{ij} \neq 0$ in **Figs. 3.2**). The flexoelectric field rather weakly influences on the polarization component $P_3$ (compare curves 1, 2 with curves 3, 4 in **Fig.3.2a**). However the flexoelectric field $E_1^{FR}$ strongly increases the component $P_1$ below $T_S$, since $P_1 \sim E_1^{FR}$ (compare curves 1, 2 with curves 3, 4 in **Fig.3.2b**). Actually, for the case $F_{ij} \neq 0$ the component $P_1$ appears below $T_S$, firstly quasi-linearly increases with temperature decrease, then has a pronounced jump at $T_C^{APB}$ and then saturates at temperatures $T \ll T_q$. The break at $T_C^{APB}$ originates from the appearance of reversible polarization component $P_3$ below $T_C^{APB}$. The component $P_1 \sim E_1^{FR} \sim \dfrac{\partial \Phi^2}{\partial x_1}$ in the vicinity of $T_S$. Note, that Tagantsev et al [4] analytically predicted spontaneous polarization about 4 μC/cm² at hard APB below (35-40) K without considering flexo-roto-effect contribution. Allowing for the flexo-roto-effect we obtained $P_3 \sim 8$ μC/cm² and $P_1 \sim 0.1$ μC/cm² at hard APB below $T_C^{APB} \approx 50$ K. At temperatures $T < T_C^{APB}$ the amplitude of $P_1$ is much smaller than the amplitude of $P_3$ due to the strong depolarization field $E_1^d(x_1)$.

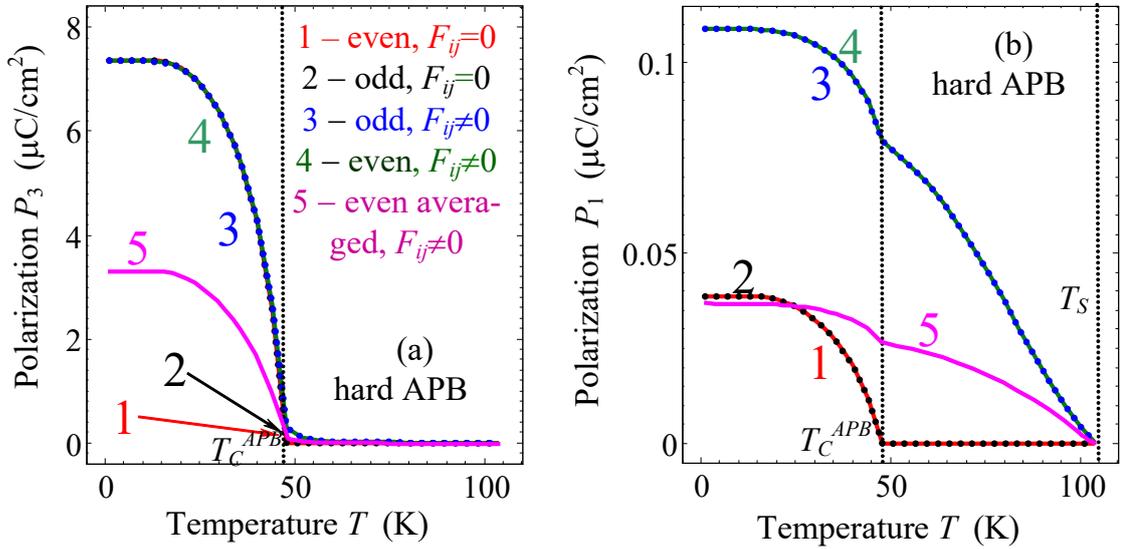

**Figure 3.2.** Temperature dependences of spontaneous polarization components $P_3$ and $P_1$ (a,b) calculated for hard APB in SrTiO₃ without free screening charges. Temperature dependences are calculated for nonzero flexoelectric effect $F_{ij} \neq 0$ and biquadratic coupling $\eta_{ij} \neq 0$ (curves 3, 4, 5) and for the case of nonzero biquadratic coupling $\eta_{ij} \neq 0$ and zero flexoelectric effect $F_{ij} \equiv 0$ (curves 1, 2). Curves 1-4 are maximal values, curves 5 – are the even-type distributions averaged across APB width. (Reproduced from [A. N. Morozovska et al, Phys.Rev.B. **85**, 094107 (2012)], with the permission of APS Publishing)

### 3.1.2. Flexo-roto-effect manifestation at 90-degree TB

90-degree twins can have structure with rotation vector parallel (**Fig.3.3a**) or perpendicular (**Fig.3.3b**) to the domain wall plane in the immediate vicinity of the plane. Far from the wall the tilt vectors of twins are perpendicular. We will regard parallel twins as "**hard**" TB, since they have higher energy and perpendicular twins as "**easy**" TB, since they have much lower energy as demonstrated in Ref.[52]. The flexo-roto field $\widetilde{E}_1^{FR}(\widetilde{x}_1)$ exist for 90-degree twins; it is an odd function with more complex structure than the one for APBs, (see bottom **Figs.3.3a** and **3.3b** and compare them with the bottom **Figs.3.1a** and **3.1b**).

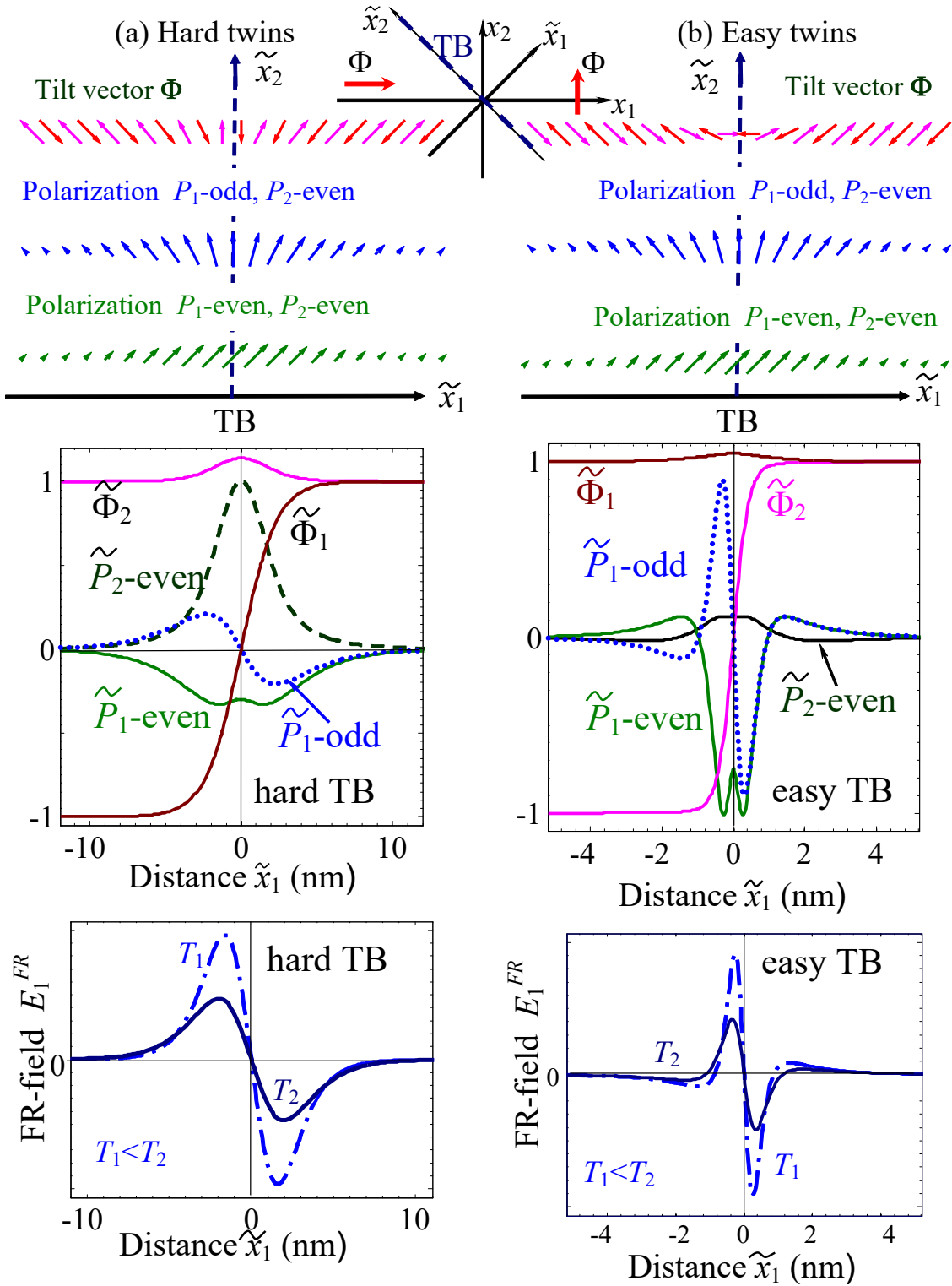

**Figure 3.3.** Schematics of 90-degree TB: rotation vector $\Phi$ is parallel (a) or perpendicular (b) to the domain wall plane in the immediate vicinity of the wall plane. Polarization appears inside the twins. TB plane $\tilde{x} = 0$ (denoted as TB-plane) is in the centre. Flexo-roto fields are shown at the bottom plots. (Reproduced from [A. N. Morozovska et al, Phys.Rev.B. **85**, 094107 (2012)], with the permission of APS Publishing)

Polarization spatial distribution across hard TB and its temperature behavior are qualitatively similar to the ones calculated for hard APB. However, numerical values of polarization and pyroelectric coefficient for hard TBs are typically smaller than for hard APBs (compare **Figs. 3.3** with **Figs. 3.2**). The difference originated from the smaller effective flexoelectric field, which in turn originate from smaller stress gradients. Differences in the stress gradients originate from the different orientation of the tilt vector $\Phi$ inside hard TB and APB.

Temperature dependences of the maximal and average spontaneous polarization values calculated inside hard TB are shown in **Figs. 3.4(a-b)**. Under the absence of the flexoelectric field spontaneous polarization and pyroelectric coefficient are zero at temperatures higher than the effective Curie temperature $T_C^{TB}$ (see curves 1, 2 calculated at $F_{ij} \equiv 0$ and $\eta_{ij} \neq 0$). The flexo-roto effect rather weakly influences on the polarization component $\tilde{P}_2$. For the case $F_{ij} \neq 0$ the component $\tilde{P}_1 \sim E_1^{FR}$ appears below $T_S$, firstly quasi-linearly increases with temperature decrease, then non-linearly increases, then has a pronounced jump at $T_C^{TB}$ and then saturates at low temperatures $T \ll T_q$. The jump at $T_C^{TB}$ originates from the appearance of reversible FE polarization component $\tilde{P}_2$ below $T_C^{TB}$ The maximal values of polarization are very close for odd and even types of solutions in the dielectric limit. Allowing for the flexo-roto-effect contribution we obtained $\tilde{P}_2 \sim 2$ μC/cm² and $\tilde{P}_1 \sim 0.02$ μC/cm² below $T_C^{TB}$. Without the flexo-roto-effect $\tilde{P}_2$ is still ~2 μC/cm² at low temperatures, but $\tilde{P}_1 < 0.005$ μC/cm².

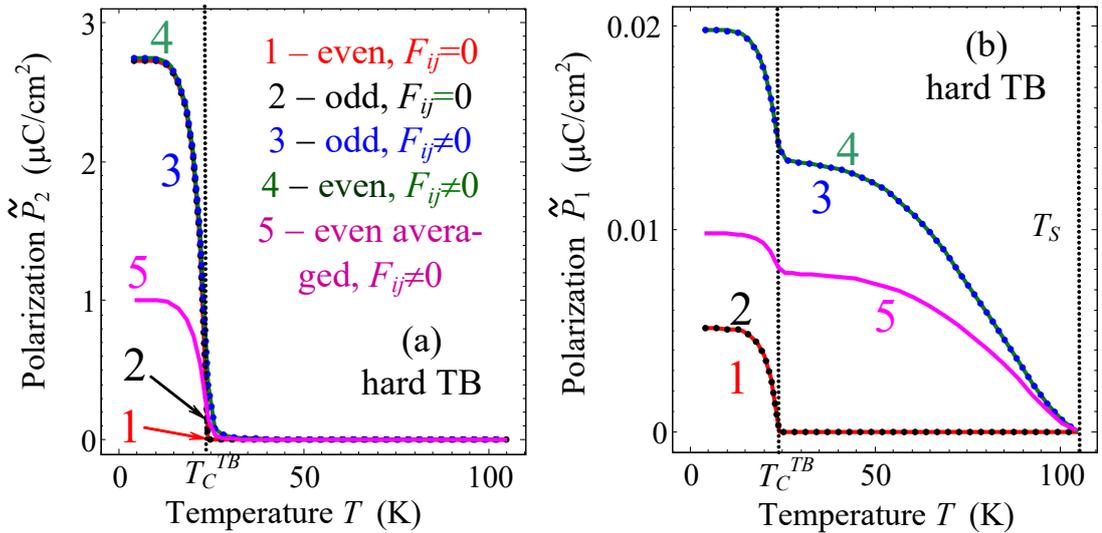

**Figure 3.4.** Temperature dependences of spontaneous polarization components $\tilde{P}_2$ and $\tilde{P}_1$ maximal values (a,b) calculated for hard TB in SrTiO₃ without free carriers. Temperature dependences are calculated for nonzero flexoelectric effect $F_{ij} \neq 0$ and biquadratic coupling $\eta_{ij} \neq 0$ (curves 3, 4, 5) and for the case of nonzero biquadratic coupling $\eta_{ij} \neq 0$ and zero flexoelectric effect $F_{ij} \equiv 0$ (curves 1, 2). Curves 1-4 are maximal values, curves 5 – are the even even-type distributions averaged across TB width. (Reproduced from [A. N. Morozovska et al, Phys.Rev.B. **85**, 094107 (2012)], with the permission of APS Publishing)**.**

To summarize the section 3.1, let us underline that polarization across APB and TB in SrTiO$_3$ originate from the flexo-roto effect in the temperature range $T_C^{APB,TB} < T < T_S$ and should exist in other ferroelastics incipient-ferroelectrics like in CaTiO$_3$ and EuTiO$_3$. In summary, the coupling of flexoelectric and rotostriction effects can give rise to the appearance of a significant improper spontaneous polarization and pyroelectricity across a structural antiphase boundary and twins, and by extension across interfaces in otherwise non-ferroelectric perovskites such as CaTiO$_3$, SrTiO$_3$, and EuTiO$_3$.

### 3.2. Surface polar states and pyroelectricity in ferroelastics induced by flexo-roto field

Theoretical analysis based on the LGD theory is used to show that the joint action of flexoelectric effect and rotostriction leads to a large spontaneous in-plane polarization (~ 1-5 μC/cm$^2$) and pyroelectric coefficient (~10$^{-3}$ C/m$^2$K) in the vicinity of surfaces of otherwise non-ferroelectric ferroelastics, such as SrTiO$_3$, with static octahedral rotations [53]. The origin of the improper polarization and pyroelectricity is "flexo-roto" field, its strength is proportional to the convolution of the flexoelectric and rotostriction tensors with octahedral tilts and their gradients. Flexo-roto field should exist at surfaces and interfaces in all structures with static octahedral rotations, and thus it can induce surface polar states and pyroelectricity in a large class of otherwise nonpolar materials.

Using LGD approach one can analyze the behaviour on the polar ($P_i$) and structural ($\Phi_i$) order parameter components in the presence of ferroelastic surface via equations of state, Eqs.(3.3). Allowing for the surface energy, $F_S = \int_S \left(a_i^S P_i^2 + b_i^S \Phi_i^2\right) d^2 r$, equations Eqs.(3.3) should be supplemented with the boundary conditions at $x_3 = 0$ for the tilt and polarization vectors:

$$\left(2b_i^S \Phi_i - v_{i3kl}\frac{\partial \Phi_k}{\partial x_l}\right)\bigg|_{x_3=0} = 0, \qquad \left(2a_i^S P_i - g_{i3kl}\frac{\partial P_k}{\partial x_l} + \frac{f_{jki3}}{2}u_{jk}\right)\bigg|_{x_3=0} = 0 \qquad (3.6)$$

Third kind boundary conditions (3.6) reflect the surface energy contribution into the tilt and polarization vector components slope near the surface that can be characterized by so-called extrapolation lengths $\sim v_{i3kl}/2b_i^S$ and $g_{i3kl}/2a_i^S$. The additional source of polarization in Eq.(3.6), $f_{jki3}u_{jk}/2$, originated from the flexoelectric effect. Surface energy coefficients $a_i^S$ and $b_i^S$ ($i$=1 − 3) are regarded positive and weakly temperature dependent. Note that the values of $b_i^S$ could essentially influence near surface behaviour of the structural order. For instance, the most likely case $b_3^S \ll b_{1,2}^S$ favours the octahedral rotations around the axis normal to the surface.

Allowing for the flexoelectric effect boundary condition for elastic stress at mechanically free flat surface acquires the form [64]:

$$\left( \sigma_{3i} - \frac{f_{j3i3}}{2}\frac{\partial P_j}{\partial x_3} + \frac{f_{j3il}}{2}\frac{\partial P_j}{\partial x_l} \right)\bigg|_{x_3=0} = 0. \tag{3.7}$$

Hereafter we chose tetragonal SrTiO$_3$ (with AFD transition temperature $T_S$= 105 K, space group I4/mcm) for numerical simulations, since all necessary parameters including gradient coefficients and flexoelectric tensor are known for the material. Unfortunately exact values of gradient coefficients and flexoelectric tensor are unknown for other ferroelastics like CaTiO$_3$ or EuTiO$_3$, but the extension of the obtained results will be valid qualitatively for them, making the flexo-roto field induced polar states at surfaces and interfaces a general phenomenon in nature.

Now let us calculate the depth of the induced polarization penetration from the free surface $x_3 \equiv z = 0$. For the case when 4-fold axis is parallel to the mono-domain SrTiO$_3$ surface, the most thermodynamically preferable situation is: two z-dependent components of the tilt vector, in-plane $\Phi_\parallel(z)$ and out-of plane $\Phi_\perp(z)$, and z-dependent in-plane polarization $P_\parallel(z)$ that does not cause any depolarization field ($E_\parallel^d = 0$, see the sketch of the problem geometry in **Fig. 3.5a**). Also one may consider out-of-plane polarization $P_\perp(z)$, but without enough concentration of free carriers its value is strongly affected by the depolarization field $E_\perp^d = -P_\perp(z)/\varepsilon_0\varepsilon_b$. We calculated numerically that $P_\parallel(z)$ values are at least $10^3$ times higher than $P_\perp(z)$ values without screening by free carriers.

From **Fig. 3.6** we can conclude that the flexo-roto fields do induce polar state under the surface at distances $z \leq 2L_\Phi$ in ferroelastics. Note, that $L_\Phi \sim 3$ nm for SrTiO$_3$ at T<90 K [4, 52] determines the nanometer scale of the surface polar state. So, the typical thickness of polar state is about 7 lattice constants, making continuum theory results at least semi-quantitatively valid. Surface-induced polarization appears at temperatures lower than $T_S$ and it increases as the temperature decreases (compare different curves in **Fig. 3.6a**). Surface polarization and maximal values increase as the extrapolation length $\lambda_P$ increases (compare different curves in **Fig. 3.6b**). It is seen that spontaneous polarization can reach noticeable values ~ 1 – 10 μC/cm$^2$ in the gradient region $z \leq 2L_\Phi$ at temperatures lower than 60 K.

To resume the gradient nature of flexo-roto field induces a significant improper spontaneous polarization and pyroelectricity in the vicinity of surfaces and interfaces of otherwise non-ferroelectric ferroelastics such as SrTiO3, and by extension in CaTiO3 and EuTiO3. In SrTiO3 the flexo-roto effect leads to a large spontaneous polarization (~1 – 5 μC/cm²) and pyroelectric coefficient (~$10^{-3}$ C/m²K). The strength of the flexo-roto field is proportional to the convolution of the flexoelectric and rotostriction tensors with the gradients of octahedral rotations, which are structural order parameters. The strength of the surface flexo-roto polarization is proportional to the convolution of the flexoelectric and rotostriction tensors with octahedral rotations on the surface. Flexo-roto effects should exist at surfaces in all structures with static rotations, which are abundant in nature, it allows for contribution into polar interfaces in a large class of nonpolar materials.

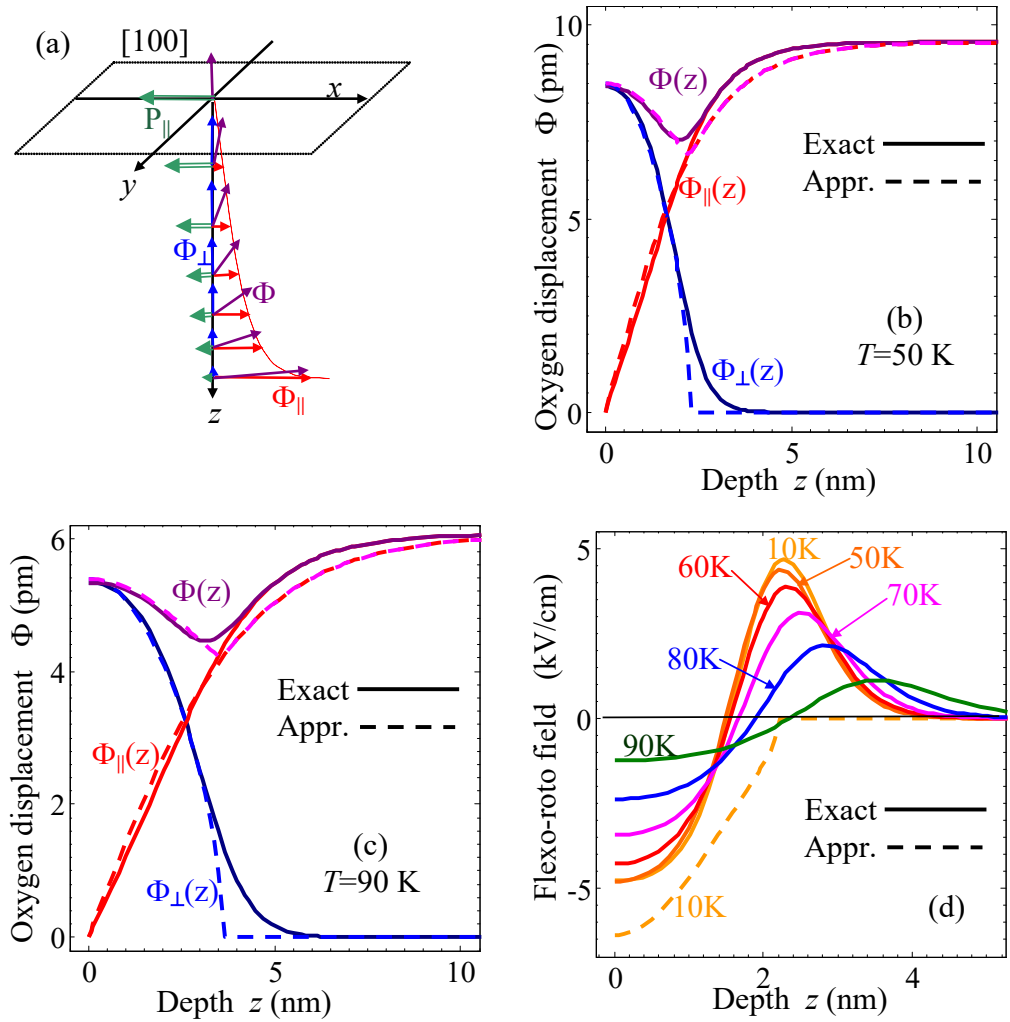

**Figure 3.5**. (a) Sketch of the problem geometry in the vicinity of SrTiO3 [100] cut. 4-fold axis is parallel to the surface. (b, c) Depth z-profile of the structural tilt components $\Phi_\perp(z)$, $\Phi_\parallel(z)$ and absolute value $\Phi(z) = \sqrt{\Phi_\perp^2(z) + \Phi_\parallel^2(z)}$ (labels near the curves) calculated numerically from coupled (solid curves) and analytically from decoupled equations (dashed curves) at temperature T= 50 K (b) and 90 K (c), extrapolation length $\lambda_\parallel = 0$. (d) Flexo-roto field $E_{FR}^B$ calculated at different temperatures 10, 50, 60, 70, 80

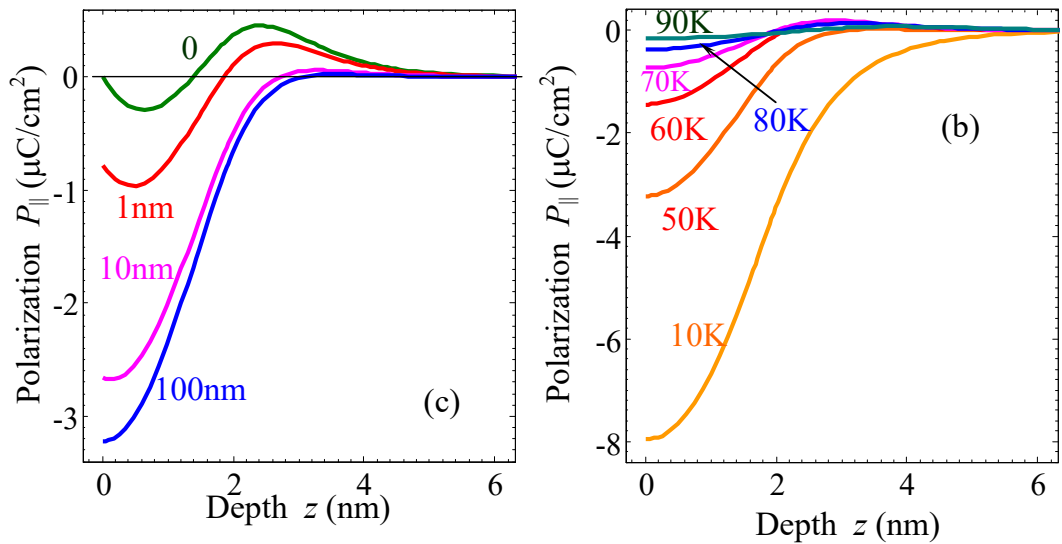

**Figure 3.6.** **(a)** Surface-induced polarization $P_{\parallel}(z)$ vs. the depth z from the surface calculated at different temperatures 10, 50, 60, 70, 80 and 90 K (numbers near the curves) for polarization extrapolation length $\lambda_P = 0$. **(b)** Polarization $P_{\parallel}(z)$ vs. the depth z calculated for different length $\lambda_P = 0$, 1 nm, 10 nm, 100 nm (figures near the curves) and temperature 50 K. (Reproduced from [A. N. Morozovska et al, Appl. Phys. Lett. **100**, 142902 (2012)], with the permission of AIP Publishing)

### 3.3. Impact of free charges on interfacial pyroelectricity in ferroelastics

Interfacial polarization spatial distribution and its temperature behaviour calculated for semiconductor $SrTiO_3$ with $n_0 = (10^{24} - 10^{26})$ m$^{-3}$ appeared semi-qualitatively similar to the ones calculated in dielectric $SrTiO_3$ (see Ref.[52]). Naturally, free electrons and mobile oxygen vacancies effectively screen the depolarization field across polar interfaces and strongly enhance the components of interfacial polarization conjugated with depolarization field. However several other important differences exist.

Numerical values of polarization and pyroelectric response across easy APB and easy TB appeared much higher (up to 100 times for $n_0 = 10^{26}$ m$^{-3}$!) for semiconductor $SrTiO_3$ than for the dielectric one. The polarization component, that is perpendicular to the wall plane, increases with the carrier concentration increase due to much smaller depolarization field, which decrease comes from the screening carriers.

The values of polarization component parallel to the hard APB or hard TB plane also increase (up to 10 times at higher temperatures) with the carrier concentration increase. Since the component is not directly affected by the depolarization field the increase originated from the coupling with parallel component [52].

The wall width of easy APB and TB (as well as the width of the polarization component perpendicular to the hard APB or TB plane) significantly increases in the semiconductor SrTiO$_3$, up to 5 times for the polarization component perpendicular to the wall plane. The reason for the width increase with free carriers concentration increase is the depolarization field decrease. The parallel component depends on the perpendicular one via the biquadratic coupling terms, therefore the wall width of the polarization component parallel to the hard APB and hard TB also increases with the carriers concentration increase, but the effect is weaker.

Temperature dependence of the maximal polarization at easy APB or TB (as well as the polarization component perpendicular to the hard APB or TB plane) has no saturation at low temperatures in semiconductor SrTiO$_3$, in contrast to the dielectric SrTiO$_3$ (see e.g. Ref.[52]). Moreover, maximal polarization of easy antiphase boundaries super-linearly increases with the temperature decrease. The increase originated from the decrease of the depolarization field with the temperature increase. Really, in accordance with Debye equation $\frac{\partial^2 \varphi}{\partial x_i^2} \approx \frac{\varphi}{R_d^2} + \frac{\partial P_i}{\varepsilon_0 \varepsilon^b \partial x_i}$ for depolarization field $E_i^d = -\partial \varphi / \partial x_i$, the field decreases with the temperature increase, since the Debye screening radius $R_d = \sqrt{\varepsilon^b \varepsilon_0 k_B T / (2 e^2 n_0)}$ decreases.

It is seen from **Figs. 3.7a** and **3.7b** that APB energy is rather weakly dependent on the polarization distribution and its screening conditions, while TB energy looks almost independent on these factors, since the polarization component perpendicular to the TB plane is very small. The explanation is the weak dependence of the wall energy on the electric contribution. Similarly to the case of dielectric SrTiO$_3$, $P_3$-even polarization distributions have the highest energy, but the electric energy difference between all types of polarization solutions are very small and super-linearly decreases with temperature increase; for TB (not shown) the difference is even smaller than for APB.

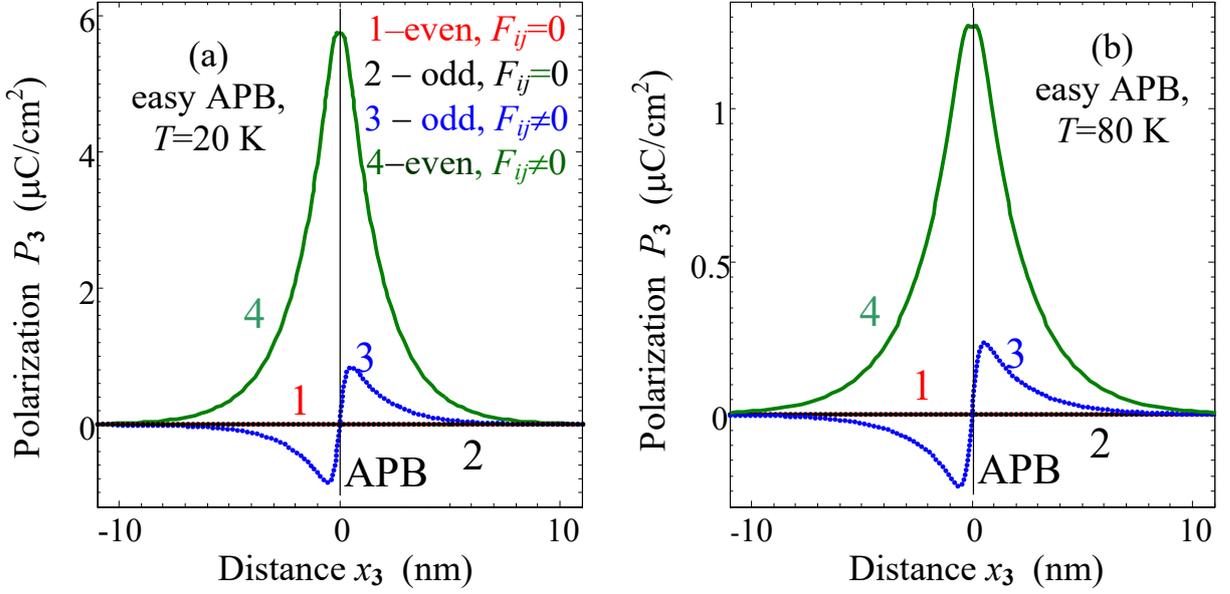

**Figure 3.7.** Spontaneous polarization distribution across easy APB calculated at temperatures $T$=20 K (a) and 80 K (b) for SrTiO$_3$ parameters; concentration of free carriers $n_0$=10$^{26}$ m$^{-3}$. Perpendicular to APB component $P_3$-odd (dotted curves) and $P_3$-even (solid curves) are calculated for nonzero flexoelectric effect $F_{ij} \neq 0$ and biquadratic coupling $\eta_{ij} \neq 0$ (curves 3, 4) and for the case of nonzero biquadratic coupling $\eta_{ij} \neq 0$ and zero flexoelectric effect $F_{ij} \equiv 0$ (curves 1, 2). (Reproduced from [A. N. Morozovska et al, Ferroelectrics, **438:1**, 32-44 (2012)], with the permission of T&F Publishing).

Note, that there are other possible reasons for polar surface states in nonpolar materials such as SrTiO$_3$: space charge due to defect chemistry and band gap differences between surfaces and bulk [65], surface reconstruction and atom clustering [66], surface piezoelectricity [67, 68] and strained polar regions that extends into the bulk at a distance much larger than a few nanometers [69]. In accordance with these and other studies, combined rotostriction and flexoelectricity cannot not be the sole contribution to the polar surface stats in ferroelastics. However the conclusion is that the surfaces of all ferroelastics with octahedral tilts should be intrinsically polar in the low temperature octahedrally tilted phase. Notably, it was theoretically shown that flexoelectric coupling combined with a rotostriction effect can lead to a spontaneous polarization within ferroelastic twin walls [52] and the wall – surface junctions [70]. The predicted interfacial FE phase was recently validated by experimental measurements [71] of domain wall damping and elastic softening of twin walls in SrTiO$_3$.

## IV. NEW MULTIFERROICS BASED ON Eu$_x$Sr$_{1-x}$TiO$_3$ NANOWIRES AND NANOTUBES

The search for new multiferroic materials with large ME coupling are very interesting for fundamental studies and important for applications based on the magnetic field control of the material dielectric permittivity, information recording by electric field, and non-destructive readout by magnetic field [13, 14]. Solid solutions of different quantum paraelectrics (such as $Eu_xSr_{1-x}TiO_3$ or $Eu_xCa_{1-x}TiO_3$) subjected to elastic strains can be promising for multiferroic applications. Multiferroic properties of $Eu_xSr_{1-x}TiO_3$ nanotubes and nanowires [21] have been predicted using LGD theory. In this section we discuss the possibility of inducing ferroelectricity and ferromagnetism in $Eu_xSr_{1-x}TiO_3$ nanosystems within LGD theory with account of AFD ordering in the system.

The intrinsic surface stress can induce ferroelectricity, ferromagnetism and increase corresponding phase transition temperatures in conventional ferroelectrics and quantum paraelectric nanorods, nanowires [72, 73, 74, 75, 76] and binary oxides [77]. The surface stress is inversely proportional to the surface curvature radius and directly proportional to the surface stress tensor. The surface stress depends both on the growth conditions, surface termination morphology [78, 79] and surface reconstruction [80, 81], which affect the surface tension value or even causes of surface stresses.

Using LGD theory Morozovska et al. [22] predicted the FE-FM multiferroic properties of $EuTiO_3$ nanowires originated from the intrinsic surface stress without consideration of AFD order. Since the AFD order parameter strongly influences the phase diagrams, polar and pyroelectric properties of quantum paraelectric $SrTiO_3$ [4, 52, 53, 82], similar influence is expected for $EuTiO_3$ [10] and $Eu_xSr_{1-x}TiO_3$. Therefore, a fundamental study of the possible appearance of the polar, magnetic and multiferroic phases in $Eu_xSr_{1-x}TiO_3$ solid solution considering AFD order seems necessary. Recently the transition from PE cubic phase to AFD phase in solid solution $Eu_xSr_{1-x}TiO_3$ has been studied by means of Electron Paramagnetic Resonance [83].

**Figure 4.1** illustrates the nanosystems considered in this study, namely nanotubes clamped to the rigid core, where the outer sidewall of the tube is mechanically free and electrically open, i.e., non-electroded (**Fig. 4.1a**) and freestanding nanowires (**Fig. 4.1b**). In the nanotube cases, technologically convenient materials for a rigid core can be ZnO, Si, SiC ultra-thin nanowires. Perovskite-type cores like $LaAlO_3$, $LaAlSrO_3$, $DyScO_3$, $KTaO_3$ or $NdScO_3$ are more sophisticated to design. The misfit strains appeared at the $Eu_xSr_{1-x}TiO_3$ tube-core interface are approximately -6% for ZnO, -1.7% for Si, 10% for SiC, -4% for $LaAlSrO_3$, -3% for $LaAlO_3$, +0.9% for $DyScO_3$, +2.1% for $KTaO_3$ and +2.6% for $NdScO_3$ core. In this study we considered the axial polarization $P_3$ directed along the tube axis $z$, while the radial polarization $P_\rho$ perpendicular to the surface of the tube is neglected due to the strong depolarization field $E_\rho^d \propto -P_\rho/\varepsilon_0\varepsilon_b$ that appears for the component in the case of non-electroded tube sidewalls.

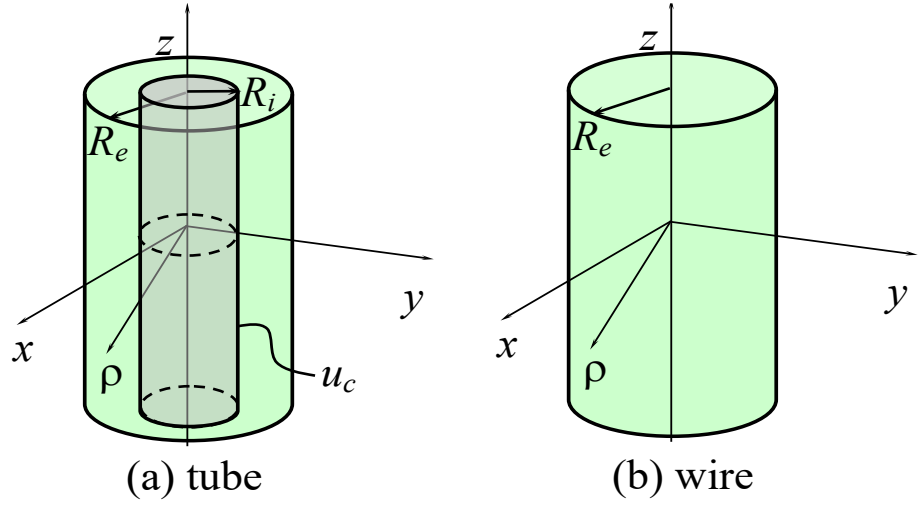

**Figure 4.1. (a)** Schematics of a nanotube clamped on a rigid core. Mismatch strain $u_c$ can exist at the tube-core interface. The tube outer radius is $R_e$, the inner radius is $R_i$, $\rho$ is the polar radius. **(b)** Schematics of a nanowire. (Reproduced from [E. A. Eliseev et al. J. Appl. Phys. **113**, 024107 (2013)], with the permission of APS Publishing)

### 4.1. Landau-Ginzburg-Devonshire theory for Eu$_x$Sr$_{1-x}$TiO$_3$

The LGD free energy density $G$ of Eu$_x$Sr$_{1-x}$TiO$_3$ solid solution depends on the polarization vector **P**, oxygen octahedra tilt vector **Φ**, magnetization vector **M** and antimagnetization vector **L** as:

$$G = G_{grad} + G_S + G_{elastic} + G_{ME} + G_M + G_{P\Phi} \qquad (4.1)$$

where $G_{grad}$ is the gradient energy, $G_S$ is the surface energy, $G_{elastic}$ is elastic energy, $G_{ME}$ is the biquadratic ME energy, $G_M$ is magnetization-dependent energy, and $G_{P\Phi}$ is polarization-and-tilt-dependent energy.

The form of $G_{grad} + G_S$ is the same as listed in Ref.[22]. The elastic energy is given as $G_{elastic} = -s_{ijkl}\sigma_{ij}\sigma_{kl}/2$, where elastic compliances $s_{ijkl}(x) = x s_{ijkl}^{EuTiO_3} + (1-x) s_{ijkl}^{SrTiO_3}$; $\sigma_{ij}$ is the elastic stress tensor. The biquadratic ME coupling energy density ($G_{ME}$) is given as

$$G_{ME} = \int_V d^3 r \frac{P_3^2}{2}\left(\gamma_{FM} M^2 + \gamma_{AFM} L^2\right) \qquad (4.2)$$

Here $P_3$ is FE polarization, $M^2 = M_1^2 + M_2^2 + M_3^2$ is the square of FM magnetization, and $L^2 = L_1^2 + L_2^2 + L_3^2$ is the square of AFM order parameter vector.

Magnetic properties are observed in EuTiO$_3$ and are absent in SrTiO$_3$. Therefore, composition dependence of the biquadratic ME coupling coefficients $\gamma_{FM}(x)$ and $\gamma_{AFM}(x)$ should be included into Eq.(4.2). Here we assume a linear dependence on Eu content ($x$) above percolation threshold ($x_{cr}$) (see e.g. Ref. [84]), namely $\gamma_{AFM}(x) = \gamma_{AFM}^{EuTiO_3}(x - x_{cr}^A)/(1 - x_{cr}^A)$ and $\gamma_{FM}(x) = \gamma_{AFM}^{SrTiO_3}(x - x_{cr}^F)/(1 - x_{cr}^F)$ at content $x_{cr}^{F,A} \leq x \leq 1$; while $\gamma_{AFM}(x) = 0$ and $\gamma_{FM}(x) = 0$ at $x < x_{cr}^{A,F}$. The percolation threshold concentration $x_{cr}$ can be estimated from the percolation theory [84]. For the simple cubic sub-lattice of magnetic ions (Eu) $x_{cr}^{F} \approx 0.24$ for FM order [84], while the percolation threshold is supposed to be higher for AFM ordering, $x_{cr}^{A} \approx 0.48$ (see e.g. [85]). Following Lee *et al.* [25], $\eta_{AFM}^{EuTiO_3} \approx -\eta_{FM}^{EuTiO_3} > 0$ as anticipated for equivalent magnetic Eu ions with antiparallel spin ordering in a bulk EuTiO3 (see e.g. section II).

The magnetization-dependent part of the free energy is [22, 77]:

$$G_M = \int_V d^3r \left( \begin{array}{c} \frac{\alpha_M}{2} M^2 + \frac{\alpha_L}{2} L^2 + \frac{\beta_M}{4} M^4 + \frac{\beta_L}{4} L^4 + \frac{\lambda}{2} L^2 M^2 \\ -\sigma_{mn} \left( Z_{mnkl} M_k M_l + \tilde{Z}_{mnkl} L_k L_l \right) \end{array} \right) \qquad (4.3)$$

where, coefficient $\alpha_M(T,x) = \alpha_C(T - T_C(x))$, $T$ is absolute temperature, $T_C(x) = T_C^0 (x - x_{cr}^F)/(1 - x_{cr}^F)$ is the solid solution FM Curie temperature defined at $x_{cr}^F \leq x \leq 1$. $T_C^0$ is the Curie temperature for bulk EuTiO3. Coefficient $\alpha_L(T,x) = \alpha_N(T - T_N(x))$, where Néel temperature $T_N(x) = T_N^0 (x - x_{cr}^A)/(1 - x_{cr}^A)$ is defined at $x_{cr}^A \leq x \leq 1$. $T_N^0$ is the Néel temperature for bulk EuTiO3. The magnetic Curie and Néel temperatures are zero at $x < x_{cr}^{F,A}$. For equivalent amount of magnetic Eu ions with antiparallel spin ordering it can be assumed that $\alpha_C \sim \alpha_N$. The positive coupling term $\frac{\lambda}{2} L^2 M^2$ prevents the appearance of FM as well as ferrimagnetic (FiM) phases at low temperatures ($T < T_C$) under the condition $\sqrt{\beta_M \beta_L} < \lambda$. Coefficients $\beta_M$, $\beta_L$, $\lambda$ are regarded $x$-independent. $Z_{mnkl}$ and $\tilde{Z}_{mnkl}$ represents magnetostriction and antimagnetostriction tensors respectively.

The polarization and AFD parts of the free energy bulk density is

$$G_{P\Phi} = \int_V d^3r \left( \frac{\alpha_P}{2} P_3^2 + \frac{\beta_P}{4} P_3^4 - Q_{ij33} \sigma_{ij} P_3^2 + \frac{\alpha_\Phi}{2} \Phi_3^2 + \frac{\beta_\Phi}{4} \Phi_3^4 - R_{ijkl} \sigma_{ij} \Phi_k \Phi_l + \frac{\eta_{i3}}{2} \Phi_i^2 P_3^2 \right) \qquad (4.4)$$

Here $P_i$ is the polarization vector, and $\Phi_i$ is the structural order parameter (rotation angle of oxygen octahedron measured as displacement of oxygen ion). The biquadratic coupling between the structural order parameter $\Phi_i$ and polarization components $P_i$ are defined by the tensor $\eta_{ij}$. [4, [86]]. Biquadratic coupling tensor and higher order expansion coefficients are regarded composition dependent: $\beta_{P,\Phi}(x) = x\beta_{P,\Phi}^{EuTiO_3} + (1-x)\beta_{P,\Phi}^{SrTiO_3}$, $\eta_{ij}(x) = x\eta_{ij}^{EuTiO_3} + (1-x)\eta_{ij}^{SrTiO_3}$. $Q_{ijkl}(x) = xQ_{ijkl}^{EuTiO_3} + (1-x)Q_{ijkl}^{SrTiO_3}$ and $R_{ijkl}(x) = xR_{ijkl}^{EuTiO_3} + (1-x)R_{ijkl}^{SrTiO_3}$ are the electrostriction and rotostriction tensors components respectively, which also depend linearly on the composition $x$. Coefficients $\alpha_P(T,x)$ and $\alpha_\Phi(T,x)$ depend on temperature in accordance with Barrett law [87] and composition $x$ of Eu$_x$Sr$_{1-x}$TiO3 solid solution as $\alpha_P(T,x) = x\alpha_P^{EuTiO_3}(T) + (1-x)\alpha_P^{SrTiO_3}(T)$ and $\alpha_\Phi(T,x) \approx \alpha_{\Phi T}(T - T_S(x))$, where $T_S(x) \approx 113.33 + 390.84x - 621.21x^2 + 398.87x^3$ in accordance with [83]. Coefficient $\alpha_m(T) = (T_q^m/2)(\coth(T_q^m/2T) - \coth(T_q^m/2T_c^m))$, where sub- and superscript $m = P, \Phi$. Temperatures $T_q^m$ are so called quantum vibration temperatures for SrTiO3 and EuTiO3 respectively, related with either polar (P) or oxygen octahedron rotations ($\Phi$) modes, $T_c^m$ are the "effective" Curie temperatures corresponding to polar soft modes in bulk EuTiO3 and SrTiO3.

For tetragonal FE, FM, AFM and cubic elastic symmetry groups, coefficients $\alpha$ are renormalized by the surface tension [72-74], misfit strains [88] and biquadratic coupling with a structural order parameter [52, 53, 82]. For considered geometry, the renormalization is:

$$\alpha_{RP}(T,x) = \alpha_P(T,x) + \frac{4Q_{12}(x)\mu}{R_e} - \frac{Q_{11}(x) + Q_{12}(x)}{s_{11}(x) + s_{12}(x)} \frac{R_i^2}{R_e^2} u_c - \eta_{11}(x) \frac{\alpha_\Phi(T,x)}{\beta_\Phi}, \quad (4.5a)$$

$$\alpha_{MR}(T,R,x) \approx \alpha_C \left( T - \left( T_C^0 - \frac{W}{\alpha_C} \frac{4\mu}{R_e} + \frac{Z_{11} + Z_{12}}{\alpha_C(s_{11}(x) + s_{12}(x))} \frac{R_i^2}{R_e^2} u_c \right) \frac{x - x_{cr}^F}{1 - x_{cr}^F} \theta(x - x_{cr}^F) \right), \quad (4.5b)$$

$$\alpha_{LR}(T,R,x) \approx \alpha_N \left( T - \left( T_N^0 - \frac{\widetilde{W}}{\alpha_C} \frac{4\mu}{R_e} + \frac{\widetilde{Z}_{11} + \widetilde{Z}_{12}}{\alpha_C(s_{11}(x) + s_{12}(x))} \frac{R_i^2}{R_e^2} u_c \right) \frac{x - x_{cr}^A}{1 - x_{cr}^A} \theta(x - x_{cr}^A) \right). \quad (4.5c)$$

In Eqs. (4.5), $R_e$ is the tube outer radius, $R_i$ is the tube inner radius; $\mu$ is the surface tension coefficient, that is regarded as positive; and $u_c$ is misfit strain at the tube-core interface. For the practically important case of the ferroelectric tube deposited on a rigid dielectric core, the tube and core lattices mismatch or the difference of their thermal expansion coefficients determines $u_c$ value allowing for the possible strain relaxation for thick tubes. If the spontaneous (anti)magnetization is directed along z-axes, the parameters in Eqs.(4.5b-c) are $\widetilde{W} = +\widetilde{Z}_{12}$, $W = +Z_{12}$, where $Z_{ij}$ and $\widetilde{Z}_{ij}$ are the magnetostriction and anti-magnetostriction coefficients. When the spontaneous (anti)magnetization is along the {x,y} plane, the parameters are $\widetilde{W} = -(\widetilde{Z}_{12} + \widetilde{Z}_{11})/2$, $W = -(Z_{12} + Z_{11})/2$ [22]. Function $\theta(x - x_{cr})$ is the Heaviside step-function [89], i.e. $\theta(x \geq 0) = 1$ and $\theta(x < 0) = 0$.

The terms in Eq.(4.5) proportional to $\mu/R_e$ originated from the intrinsic surface stress, while the terms proportional to $u_c R_i^2/R_e^2$ are the strains induced by the rigid core. The size, misfit strain and composition dependence of the ordered phases stability can be obtained from the condition $\alpha_R(T,x) < 0$. In particular, the term $4Q_{12}(x)\mu/R_e$ in Eq.(4.5a) is negative because $Q_{12}(x) < 0$; so it leads to a reduction in $\alpha_{RP}(T,x)$ and thus favors FE phase appearance for small $R_e$. Since $Q_{11}(x) + Q_{12}(x) > 0$, the term $\sim (Q_{11}(x) + Q_{12}(x))(R_i^2/R_e^2)u_c$ in Eq.(4.5a) leads to a reduction in $\alpha_{RP}(T,x)$ and thus favors FE phase appearance for positive $u_c$. The numerical values of material parameters used in the LGD model are listed in **Tables 1** and **2** in Ref.[21]. Following

Ref.[25], misfit strain $u_c$ varies in the range – 5% to + 5%. The surface tension coefficient μ for $Eu_xSr_{1-x}TiO_3$ can be estimated as high as 30 N/m based on experimental data for ferroelectric $ABO_3$ perovskites (36.6 N/m [90] or even ~50 N/m [91] for $PbTiO_3$, 2.6-10 N/m for $PbTiO_3$ and $BaTiO_3$ nanowires [92], 9.4 N/m for $Pb(Zr,Ti)O_3$ [93]).

### 4.2. Phase diagrams of $Eu_xSr_{1-x}TiO_3$ nanosystems

**Figure 4.2a** shows the predicted phases for $Eu_xSr_{1-x}TiO_3$ bulk solid solution. The phase diagram shows the presence of five different phases: PM, AFD, AFD-FM, AFD-AFM, and AFD-FiM. The magnetic phases AFD-FM, AFD-AFM and AFD-FiM exist at temperatures lower than 10 K. Calculations [21] for the bulk solid solution do not differentiate between in-plane, out-of-plane or mixed FE phases.

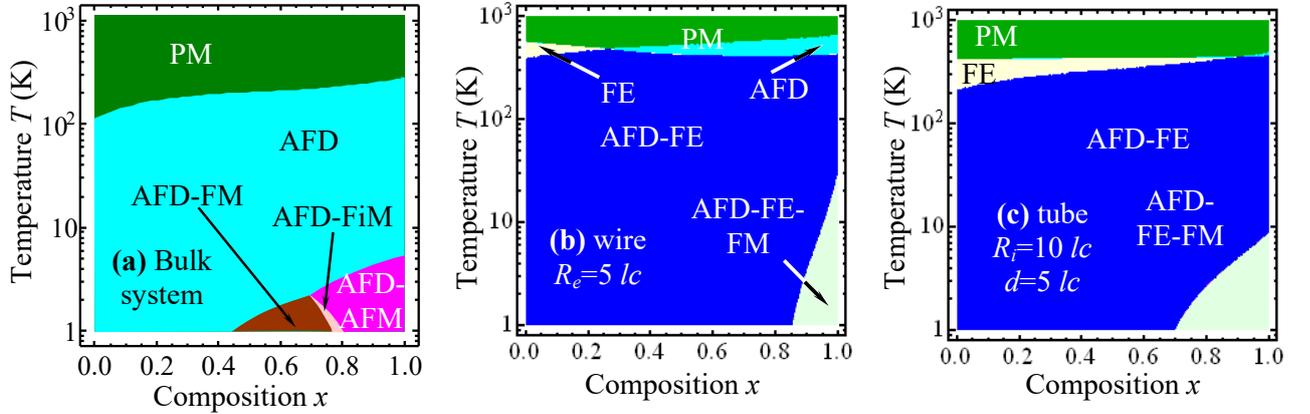

**Figure 4.2.** Predicted temperature-composition phase diagrams of **(a)** bulk $Eu_xSr_{1-x}TiO_3$ system, where Para, AFD, AFD-FM, AFD-AFM, and AFD-FiM phases are present, **(b)** $Eu_xSr_{1-x}TiO_3$ wire of radius 5 *lc*, **(c)** $Eu_xSr_{1-x}TiO_3$ nanotube of radius 10 *lc*. The tubes of thickness for (c) is 5 *lc* with a tensile misfit strain $u_c$ = +3%. The surface tension coefficient μ = 30 N/m for nanowire and nanotubes. The existing phases in the nanosystems are Para, FE, AFD, AFD-FE, and AFD-FE-FM. (Reproduced from [E. A. Eliseev et al. J. Appl. Phys. **113**, 024107 (2013)], with the permission of APS Publishing)

Analyzing the phase diagrams predicted for the bulk systems (**Fig. 4.2a**), we observed the unexpected appearance of the FM ordering (AFD-FM and AFD-FiM phases) for Eu content $x > 0.4$. Therefore, we predict Sr-diluted ferromagnetism for Eu composition $x$ from 0.45 to 0.75 or ferrimagnetism for $x \approx 0.8$. The FM ordering may originate from spin canting [94, 95], especially if the energies of different magnetic orderings (A-, C-, F-, and G-types) are very close. Therefore, bulk solid solution $Eu_xSr_{1-x}TiO_3$ should be included to the multiferroic family. The results presented in **Figures 4.2b-4.2d** clearly show that the triple AFD-FE-FM phase is present at high Eu content and absent at Sr-rich $Eu_xSr_{1-x}TiO_3$. Therefore, the increase in Sr content in $Eu_xSr_{1-x}TiO_3$ dilutes the spins and reduces the overall magnetization. In addition, the crossover of the AFD magnetic phases AFD-AFM → AFD-FiM → AFD-FM originated from the magnetic percolation model, namely due to the different percolation thresholds for ferromagnetism $x_{cr}^F \approx 0.24$ and antiferromagnetism $x_{cr}^A \approx 0.48$, which is in agreement with classical percolation theory [84]. We hope that this prediction will be verified either experimentally or from the first-principles calculations.

For $Eu_xSr_{1-x}TiO_3$ nanowires and nanotubes, our calculations demonstrated that several ordered phases can be thermodynamically stable under tensile strain (see **Figs. 4.2b-4.2c** respectively), namely PE, FE, AFD, AFD-FE, and AFD-FE-FM. Note, that FE, AFD-FE, and AFD-FE-FM phases are absent in the bulk $Eu_xSr_{1-x}TiO_3$, since ferroelectric ordering appearance in incipient ferroelectrics is possible for small sizes only.

Therefore, it is important to emphasize that the misfit strain existing between the nanotube-core interface allows the possibility of controlling the phase diagram of the $Eu_xSr_{1-x}TiO_3$ nanotubes. Such possibility is absent for nanowires. The misfit strain - composition phase diagrams of $Eu_xSr_{1-x}TiO_3$ nanotubes with internal radius 10 *lc*, outer radius 15 *lc*, and thickness *d*=5 *lc* are shown in **Fig. 4.3** at two different temperatures, at 4 K (low temperature), and at 300 K (room temperature). From **Fig. 4.3a** it is clear that the FM properties of $Eu_xSr_{1-x}TiO_3$ nanotubes can appear at about $x_c >$ 0.8, which is much higher than the percolation threshold of $x_{cr}^F \approx 0.24$ at low temperatures $T < 4$ K and positive tensile strains ($u_c > 0$). The region of AFD-FE-AFM stability becomes narrower with the increase in temperature and it disappears at higher temperatures. At low temperatures ($\leq 10$ K) there are two stable multiferroic phases, namely AFD-FE-AFM and AFD-FE-FM. Additional calculations (data not shown) proved that only pure $EuTiO_3$ can be AFM for given sizes and strains at $T > 10$ K. At room temperature (**Fig. 4.3b**) the disordered Para phase appears at $x < 0.8$ and $u_c < +1\%$. Such enlarged region of the PE phase occurs at room temperature because of the absence of the axial ferroelectric polarization $P_3$ in the compressed nanotubes (similar effect is reported for compressed ferroelectric films [88]). Due to the strong depolarization field $E_\rho^d \sim -P_\rho/\varepsilon_0\varepsilon_b$, the ferroelectric phase with radial polarization $P_\rho$ perpendicular to the surface of the tube/wire may appear at very high compressive strains $u_c < -5$ % [96].

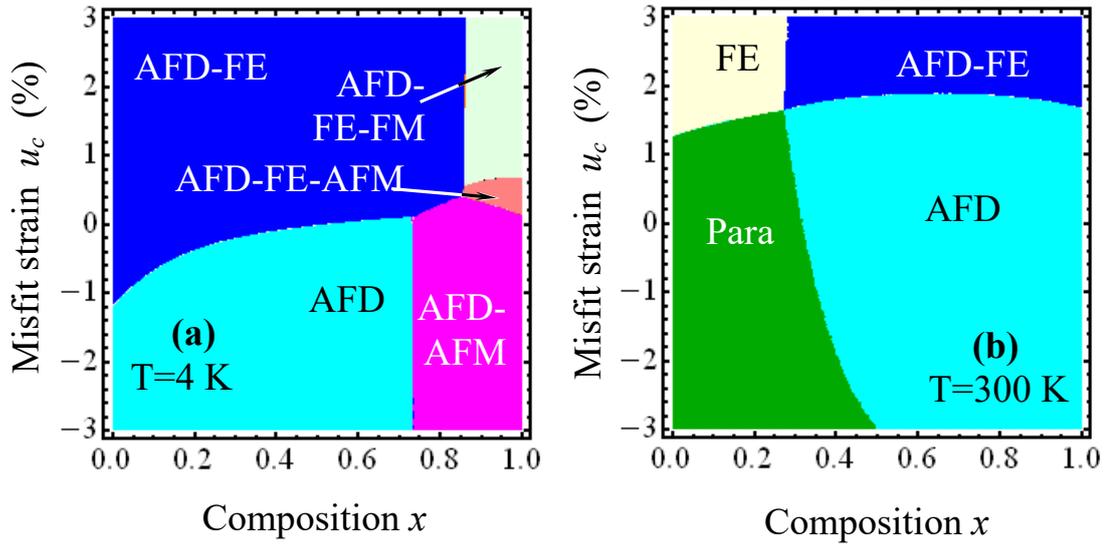

**Figure 4.3.** The misfit strain–composition phase diagrams of $Eu_xSr_{1-x}TiO_3$ nanotube with internal radius 10 $lc$, outer radius 15 $lc$, and thickness $d$=5 $lc$ at **(a)** temperature $T = 4$ K, and **(b)** at $T = 300$ K. (Reproduced from [E. A. Eliseev et al. J. Appl. Phys. **113**, 024107 (2013)], with the permission of APS Publishing)

### 4.3. Spontaneous polarization and magnetization

The spontaneous polarization and magnetization vs. composition $x$ of Eu in $Eu_xSr_{1-x}TiO_3$ nanowires and nanotubes are shown in **Fig. 4.4** for fixed radii, tensile misfit strain and different temperatures (specified near the individual curves). Note that spontaneous magnetization is absent at compressive strains and thus the case with tensile misfit strain of +3% is considered. It is observed that spontaneous polarization increases with the reduction in temperature. The magnitude of spontaneous polarization increases with the increase in Eu content for most of the temperatures. However, the trend is not followed above 280 K, which is the temperature of the structural phase transition in bulk $EuTiO_3$ (~280 K). Spontaneous magnetization abruptly appears with Eu content more than the threshold value $x_c$ and at temperatures less than 10 K. Such abrupt composition-induced FM phase transition is of the first order. It is clear from **Fig. 4.4** that FE phase exists at all $x$ and temperatures less than 300 K. The jumps of spontaneous polarization values at low temperatures (4 K data in **Figs. 4.4a-4.4b**) matches with the simultaneous appearance of spontaneous magnetization phases, i.e. they indicate magnetoelectric FE-FM phase transition.

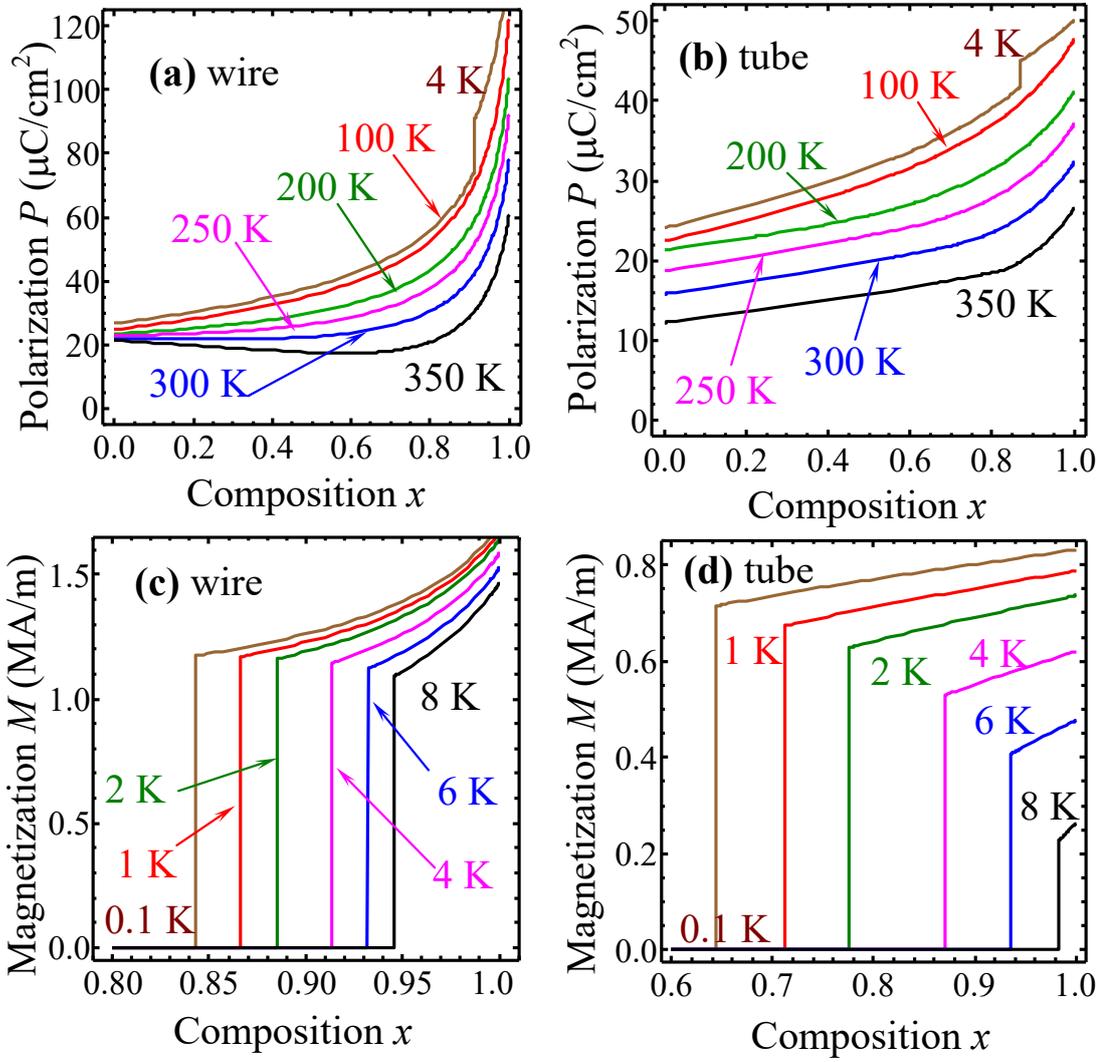

**Figure 4.4.** Change in spontaneous polarization vs. composition $x$ of $Eu_xSr_{1-x}TiO_3$ **(a)** nanowire, and **(b)** nanotube at different temperatures. Change in spontaneous magnetization vs. composition $x$ of $Eu_xSr_{1-x}TiO_3$ **(c)** nanowire, and **(d)** nanotube at different temperatures. The nanowire is of radius 5 $lc$ while the nanotube is with internal radius 10 $lc$ and thickness 5 $lc$. The results are shown for tensile misfit strain of +3%. The temperature values are specified near the curves. (Reproduced from [E. A. Eliseev et al. J. Appl. Phys. **113**, 024107 (2013)], with the permission of APS Publishing)

In this study we have calculated the phase diagrams of bulk $Eu_xSr_{1-x}TiO_3$, and $Eu_xSr_{1-x}TiO_3$ nanosystems (nanotubes, and nanowires) using phenomenological LGD theory. For bulk $Eu_xSr_{1-x}TiO_3$ solid solution, the FM phase is predicted to be stable at low temperatures for the concentration $0.4 < x < 0.8$, while AFM phase is stable at $0.8 < x < 1$. Within the *mesoscopic* phenomenological approach, the AFM to FM transition is guided by the difference in the critical percolation concentrations $x_{cr}^F \approx 0.24$ for FM state and $x_{cr}^A \approx 0.48$ for AFM state. The AFM to FM critical percolation concentration ratio ($x_{cr}^A/x_{cr}^F$) $\approx 2$ is in agreement with general percolation theory as well as with the intuitively clear fact that the percolation threshold for FM ordering should be lower than the AFM ordering [84, 85]. Even though our results only indicate to the possible microscopic origin of the diluted magnetism in Eu$_x$Sr$_{1-x}$TiO$_3$, we hope that our results will stimulate further research to investigate the associated mechanisms using low temperature experiments and *ab initio* simulations.

We hope that our predictions will stimulate experimental and computational studies of Eu$_x$Sr$_{1-x}$TiO$_3$ nanosystems, where the coupling between structural distortions, polarization and magnetization can lead to the versatility and tenability of the ME properties.

## V. LOW-SYMMETRY MONOCLINIC FERROELECTRIC PHASE STABILIZED BY OXYGEN OCTAHEDRA ROTATION IN STRAINED EU$_X$SR$_{1-X}$TIO$_3$ THIN FILMS

Epitaxial strains imposed on commensurate complex oxide thin films by substrates can lead to the emergence of a broad range of new properties [97] such as ferroelectricity [98, 99], magnetism [35], octahedral tilts [100], and multiferroicity [101] as well as of new phases with strong polar or magnetic long-range order which are absent in the corresponding bulk ferroelastics and quantum paraelectrics [97-100, 102].

The main focus of this section is on strained films of quantum paraelectric Eu$_x$Sr$_{1-x}$TiO$_3$ with structural AFD order in the bulk [83]. The vector nature of the AFD order parameter can strongly influence the phase stability, polar and pyroelectric properties of quantum paraelectrics [4, 52] at interfaces [53] or in thin film bulk [82, 102]. Since Eu$_x$Sr$_{1-x}$TiO$_3$ films are solid solutions of quantum paraelectrics EuTiO$_3$ and SrTiO$_3$, they may exhibit not only all the interesting structural and polar mode interactions of individual EuTiO$_3$ and SrTiO$_3$ films, but also new phenomena and properties.

Phase diagrams of strained films are usually complicated by new phases, which are absent in their bulk counterparts. Among these emergent new phases, low symmetry monoclinic phases are of particular interest due to the relative large number of possible ferroelectric and ferroelastic twin variants and wall orientations compared to higher symmetry phases, which give rise to possible dramatic enhancements in piezoelectric coefficients. Monoclinic phases with in-plane and out-of-plane polarization components of different amplitudes have been predicted theoretically in epitaxial BaTiO$_3$ films [88, 103, 104]. In the strained incipient ferroelectric SrTiO$_3$ films only tetragonal and orthorhombic phases were shown to be stable [102]. However, the addition of Eu to SrTiO$_3$ thin films results in the stabilization of monoclinic phases [23], and flexoelectric coupling with rotostriction effect enrich the behaviour in the Eu$_x$Sr$_{1-x}$TiO$_3$ solid solution systems.

### 5.1. Landau-Ginzburg-Devonshire potential for Eu$_x$Sr$_{1-x}$TiO$_3$

Let us consider a short-circuited Eu$_x$Sr$_{1-x}$TiO$_3$ film of thickness *h* that is clamped on to a rigid substrate (**Fig.5.1**). The lattice mismatch between the film and substrate leads to an in-plane strains $u_m$ at the interface. AFD structural order is characterized by the spontaneous displacement of oxygen atoms, that can also be viewed as oxygen octahedron rotation (measured as displacement of oxygen ion or "tilts"), described by an axial vector $\Phi_i$ ($i$=1, 2, 3) [1]. Polarization is described by vector $P_i$.

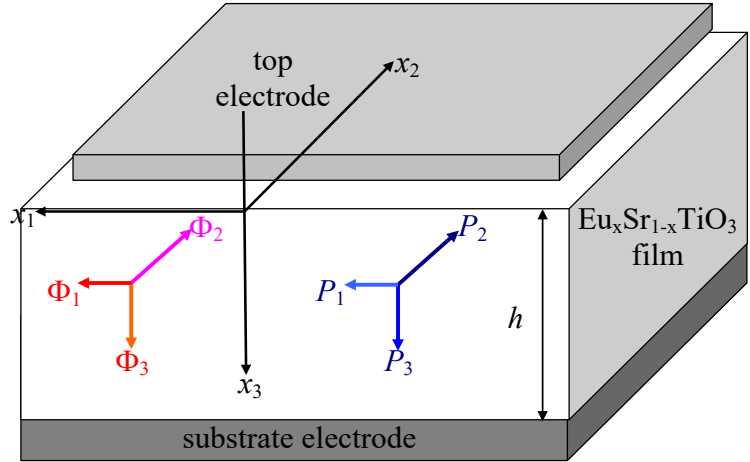

**Figure 5.1.** Schematics of a short-circuited $Eu_xSr_{1-x}TiO_3$ film clamped on a rigid substrate. (Reproduced from [A. N. Morozovska et al, Phys.Rev. **B 87**, 134102 (2013)], with the permission of APS Publishing)

Gibbs potential density of $Eu_xSr_{1-x}TiO_3$ solid solution as a function of polarization and oxygen octahedra tilt vectors is written as [21]:

$$G = G_S + \int_0^h \left(G_{grad} + G_{flexo} + G_{elastic} + G_{P\Phi}\right) dx_3 \quad (5.1)$$

where $G_S = a_i^S\left(P_i^2(0) + P_i^2(h)\right) + b_i^S\left(\Phi_i^2(0) + \Phi_i^2(h)\right)$ is the surface contribution; $G_{grad} = \frac{g_{ijkl}}{2}\left(\frac{\partial P_i}{\partial x_j}\frac{\partial P_k}{\partial x_l}\right) + \frac{v_{ijkl}}{2}\left(\frac{\partial \Phi_i}{\partial x_j}\frac{\partial \Phi_k}{\partial x_l}\right)$ is the gradient term, and $G_{flexo} = \frac{F_{ijkl}}{2}\left(\sigma_{ij}\frac{\partial P_k}{\partial x_l} - P_k\frac{\partial \sigma_{ij}}{\partial x_l}\right)$ is the flexoelectric term. $F_{ijkl}$ is the forth-rank tensor of flexoelectric coupling that was determined experimentally for $SrTiO_3$ in a wide temperature range by Zubko et al [55]. $G_{elastic}$ is elastic contribution, and $G_{P\Phi}$ is polarization-and-tilt-dependent term. The form of $G_{grad} + G_S$ is the same as listed in Ref. [22]. The elastic contribution is $G_{elastic} = -s_{ijkl}\sigma_{ij}\sigma_{kl}/2$, where $s_{ijkl}(x) = xs_{ijkl}^{EuTiO_3} + (1-x)s_{ijkl}^{SrTiO_3}$ are elastic compliances; $\sigma_{ij}$ is the elastic stress tensor. The polarization and structural parts of the 2-4-power Landau-potential density for cubic m3m parent phase is [21]:

$$G_{P\Phi} = \begin{pmatrix} \alpha_P P_i^2 + \beta_{Pij} P_i^2 P_j^2 - Q_{ijkl}\sigma_{ij}P_k P_l + \alpha_\Phi \Phi_i^2 \\ + \beta_{\Phi ij}\Phi_i^2\Phi_j^2 - R_{ijkl}\sigma_{ij}\Phi_k\Phi_l + \frac{\xi_{ik}}{2}\Phi_i^2 P_k^2 \end{pmatrix} \quad (5.2)$$

The biquadratic coupling between the structural order parameter $\Phi_i$ and polarization components $P_i$ is regarded as Houchmandazeh-Laizerowicz-Salje coupling which is defined by the tensor $\xi_{ik}$ [3, 4, 5]. And this coupling was considered as the reason for the appearance of magnetization inside a FM domain wall in a non-ferromagnetic media [6]. Both biquadratic coupling tensor and higher order expansion coefficients are regarded composition dependent, i.e. $\beta_{P,\Phi}(x) = x\beta_{P,\Phi}^{EuTiO_3} + (1-x)\beta_{P,\Phi}^{SrTiO_3}$ and $\xi_{ij}(x) = x\xi_{ij}^{EuTiO_3} + (1-x)\xi_{ij}^{SrTiO_3}$. $Q_{ijkl}(x) = xQ_{ijkl}^{EuTiO_3} + (1-x)Q_{ijkl}^{SrTiO_3}$ and $R_{ijkl}(x) = xR_{ijkl}^{EuTiO_3} + (1-x)R_{ijkl}^{SrTiO_3}$ are the electrostriction and rotostriction tensors components respectively, which are also assumed to depend linearly on the composition $x$. Coefficient $\alpha_P(T,x)$ depends on the temperature $T$ in accordance with Barrett law [87] and composition $x$ of Eu$_x$Sr$_{1-x}$TiO$_3$ solid solution as $\alpha_P(T,x) = x\alpha_P^{EuTiO_3}(T) + (1-x)\alpha_P^{SrTiO_3}(T)$ and $\alpha_P(T) = \alpha_T^{(P)}(T_q^{(P)}/2)(\coth(T_q^{(P)}/2T) - \coth(T_q^{(P)}/2T_c^{(P)}))$. Temperature $T_q^{(P)}$'s are so-called quantum vibration temperatures for SrTiO$_3$ and EuTiO$_3$ respectively, which are related to polar modes. Temperature $T_c^{(P)}$'s are the "effective" Curie temperatures corresponding to polar soft modes in bulk EuTiO$_3$ and SrTiO$_3$. To account for the experiment and Barrett law, the dependence of coefficient $\alpha_\Phi(T,x)$ on temperature and composition $x$ of Eu$_x$Sr$_{1-x}$TiO$_3$ solid solution is written as $\alpha_\Phi(T,x) = \alpha_T^{(\Phi)}(x)(T_q^{(\Phi)}(x)/2)(\coth(T_q^{(\Phi)}(x)/2T) - \coth(T_q^{(\Phi)}(x)/2T_S(x)))$. Also linear extrapolations, e.g., $\alpha_T^{(\Phi)}(x) = x \cdot \alpha_{T\Phi}^{EuTiO_3} + (1-x)\alpha_{T\Phi}^{SrTiO_3}$ and $T_q^{(\Phi)}(x) = x \cdot T_{q\Phi}^{EuTiO_3} + (1-x)T_{q\Phi}^{SrTiO_3}$ can be used.

To neglect surface gradient effects in the numerical calculations, we assume that extrapolation lengths are much greater than the film thickness $h$. To account for possible dislocations, effective misfit strain [105] can be introduced as $u_m^*(h) = u_m$ at $h_d < h$ and $u_m^*(h) = u_m h_d/h$ at $h_d \geq h$, where $h_d$ is the critical thickness for dislocation formation.

### 5.2. Phase diagrams of Eu$_x$Sr$_{1-x}$TiO$_3$ thin films

Numerical calculations of the Eu$_x$Sr$_{1-x}$TiO$_3$ thin films polar, structural properties and phase diagrams were performed in Ref.[23] as a function of temperature $T$, composition $x$ and misfit strain $u_m^*$. Some of the most representative are presented in **Figures 5.2-5.3.** The gradient effects, which may appear in the vicinity of surfaces and domain boundaries, are ignored here for the calculation of homogeneous Eu$_x$Sr$_{1-x}$TiO$_3$ films. Designation $P_i\Phi_j$ in **Figures 5.2-5.3** represents the nonzero components of order parameters in a given phase.

The temperature-composition phase diagrams of Eu$_x$Sr$_{1-x}$TiO$_3$ bulk, compressed ($u_m = -2\%$) and tensiled ($u_m = 2\%$) thin films are shown in **Figures 5.2a – 5.2c**, respectively. Two features were observed in these phase diagrams, namely a morphotropic-like boundary between AFD in-plane and out-of-plane phases and a thermodynamically stable ferroelectric monoclinic phase. The boundary between AFD phases $\Phi_1$ and $\Phi_3$ in the weakly strained films only, i.e. at $|u_m| \leq 0.01\%$, is morphotropic-like, and the film becomes spontaneously twinned. Note that the phases $\Phi_1$ and $\Phi_3$ are undistinguishable in the bulk since they are essentially the two variants of the tetragonal phase which are energetically equivalent. However, biaxial stresses exist in the thin epitaxial films clamped to a rigid substrate and the symmetry between the in-plane and out-of-plane directions is broken. Thus, the AFD phases with the order parameter pointed along these two directions become thermodynamically non-equivalent.

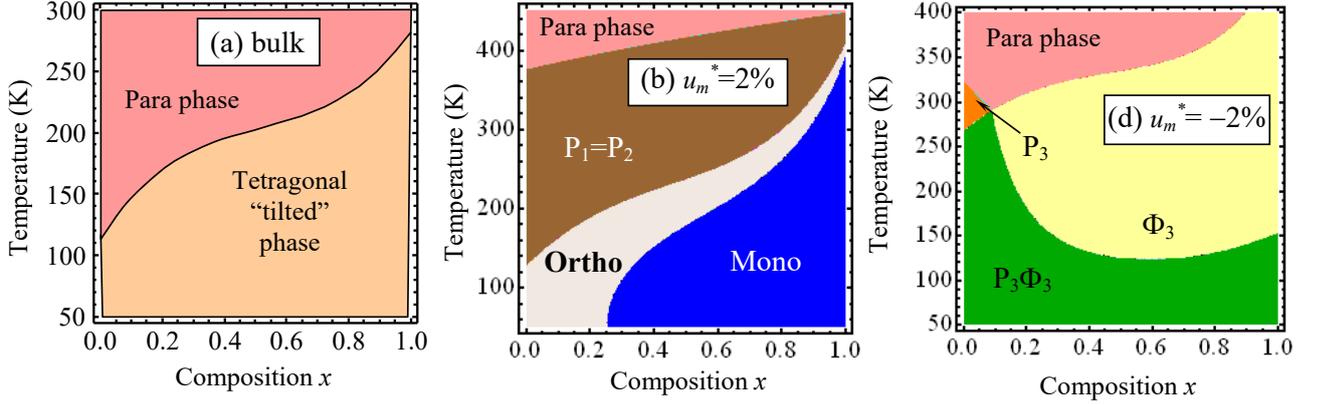

**Figure 5.2.** Temperature - composition phase diagrams of $Eu_xSr_{1-x}TiO_3$ bulk **(a)** and thin films **(b-c)** with misfits $u_m^* = +2\%$ **(b)**, $u_m^* = -2\%$ **(c)**. (Reproduced from [A. N. Morozovska et al, Phys.Rev. **B 87**, 134102 (2013)], with the permission of APS Publishing)

$Eu_xSr_{1-x}TiO_3$ films phase diagrams are mainly in-line with the earlier theoretical calculations for $SrTiO_3$ [102] and experiment [69]. predicted that compressive strains can induce out-of-plane tetragonal ferroelectric phase in $SrTiO_3$. Jang et al. confirmed the ferroelectricity in $SrTiO_3$ films on a (110) $NdGaO_3$ substrate with an average biaxial compressive strain of −1.18% under a fully commensurate condition. The absence of ferroelectricity in $SrTiO_3$ films grown on compressive $(La,Sr)(Al,Ta)O_3$ (LSAT) substrates, the lattice constant of which is close to that of $NdGaO_3$, may be related to the increase of AFD transition temperature [106]. Different polar properties of $SrTiO_3$/LSAT and $SrTiO_3$/$NdGaO_3$ may originate from the strong structural anisotropy of orthorhombic $NdGaO_3$ in comparison with cubic LSAT substrates.

The analytical expressions for the order parameters in the monoclinic phase with polarization components $P_1 \neq P_2 \neq 0$ and tilts $\Phi_1 \neq \Phi_2 \neq 0$ was derived

$$P_1^2 - P_2^2 = a_m \sqrt{\frac{P_m^4 - \phi^2 \Phi_m^4}{a_m^2 - \phi^2}}, \quad \Phi_1^2 - \Phi_2^2 = \sqrt{\frac{P_m^4 - \phi^2 \Phi_m^4}{a_m^2 - \phi^2}}, \quad (5.3)$$

where $P_m \equiv \sqrt{P_1^2 + P_2^2}$ and $\Phi_m \equiv \sqrt{\Phi_1^2 + \Phi_2^2}$. Evident expressions for $a_m$, $\phi^2$, $\Phi_m$ and $P_m$ are given in **Appendix B** of Ref.[23]. The stability of monoclinic phase (i.e. its minimal energy) was examined by the minimization of the Eu$_x$Sr$_{1-x}$TiO$_3$ free energy with respect to $P_1, P_2$ and $\Phi_1, \Phi_2$ without any additional assumptions. Our numerical calculations indeed showed that the monoclinic phase with $|P_1| \neq |P_2| \neq 0$ and $|\Phi_1| \neq |\Phi_2| \neq 0$ is thermodynamically stable in the region.

Notably, the monoclinic phase region is strongly dependent on Eu content $x$ and temperature. **Figures 5.3a** and **5.3b** show phase diagrams of Eu$_x$Sr$_{1-x}$TiO$_3$ thin films in the coordinates of misfit strain–composition for $T$=50 K and 200 K. The para-phase region increases with temperature (compare **Figs. 5.3a** and **5.3b**). The boundary $\Phi_1/\Phi_3$ occurs at very small misfit strains $|u_m| \leq 0.01\%$ and is almost independent of composition until the transition from the AFD to para-phase takes place. Different orthorhombic phases ($P_1 = P_2 \neq 0$ and $\Phi_1 = \Phi_2 \neq 0$, and $P_1 = P_2 \neq 0$) dominate at small $x$. As $x$ increases, the monoclinic phase replaces the orthorhombic phase region. The monoclinic phase exists in tensile strained EuTiO$_3$ films ($u_m \approx 2\%$) up to temperatures 400 K and higher.

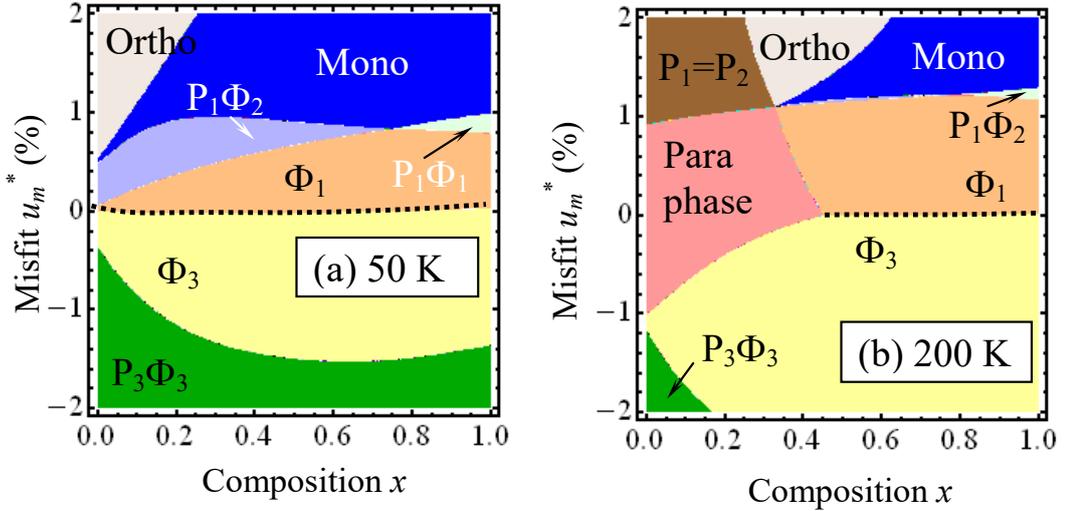

**Figure 5.3.** The misfit strain–composition phase diagrams of Eu$_x$Sr$_{1-x}$TiO$_3$ thin films at temperature 50 K **(a)** and 200 K **(b)**. (Reproduced from [A. N. Morozovska et al, Phys.Rev. **B 87**, 134102 (2013)], with the permission of APS Publishing)

Theoretical calculations [23] show that the favourable condition of the monoclinic phase appearance in Eu$_x$Sr$_{1-x}$TiO$_3$ is the *negative sign* of biquadratic coupling tensor coefficients $\xi_{ik}$. Also LGD-expansion coefficients $\alpha^*_{Pi}$ and $\alpha^*_{\Phi i}$ should be negative. These conditions could be readily reached in the strained films because the coefficients are essentially renormalized by misfit strains.

The conditions $\xi_{ij}^* < 0$ are valid if $\xi_{ij} < 0$ because the renormalization of $\xi_{ik}$ by misfit effect is usually small. The opposite signs of the coupling tensor $\xi_{ik}$ in SrTiO$_3$ and EuTiO$_3$ can explain the increase of the monoclinic phase region with the increase of Eu content, *x*. Thus we can conclude that simultaneous presence of both octahedra tilts and polarization in epitaxial Eu$_x$Sr$_{1-x}$TiO$_3$ films stabilize in-plane monoclinic phase at moderate and high tensile strains $u_m > 1\%$.

It should be noted that the monoclinic phase can also appear as the intermediate phase between the phases with higher order symmetry [107]. For example, the monoclinic phase was found in Pb(Zr,Ti)O$_3$ by Noheda et al. [108] at the morphotropic boundary between tetragonal and rhombohedral phases. It was demonstrated [109] that the monoclinic phase in Pb(Zr,Ti)O$_3$ is accompanied by the octahedral tilts, at least at lower temperatures. Local inhomogeneity can stabilize monoclinic phase as well. The monoclinic phase was predicted in the superlattices BaTiO$_3$/SrTiO$_3$ [110] as a consequence of complex electrostatic and elastic interactions within an inhomogeneous domain structure in the multilayered ferroelectric film.

## VI. ELECTRIC FIELD INDUCED FERROMAGNETIC PHASE IN PARAELECTRIC ANTIFERROMAGNETS

The search for new multiferroic materials with large ME coupling leads to rich new physics, in addition to exciting potential applications involving magnetic field control of the dielectric properties, as well as electric field control of magnetization [13, 14, 101, 111]. Electric field *E* control of ferromagnetism is a hot topic for the scientists around the world, because its multiple potential applications in magnetic memory storage, sensorics and spintronics [30, 112, 113]. Recently, Ryan et al. [30] considered the possibility of a reversible control of magnetic interactions in EuTiO$_3$ thin strained films by applying *E*-field. Because of Ti displacement from its central position under the *E*-field, changes of the spatial overlap between the electronic orbitals of the ions, and thus of the magnetic exchange coupling is expected. The density functional theory calculation shows that the competition between FM and AFM interactions is resolved in favour of FM for paraelectric EuTiO$_3$ film on compressive substrate, when applied *E*-field exceeds a critical value estimated as $E_{cr} = 0.5 \times 10^6$ V/cm. Notably, the mechanism proposed in Ref. [30] is based on the ME coupling.

Analytical calculations of the influence of the *E*-field on the EuTiO$_3$ phase diagram in the framework of LGD theory was performed in Ref.[29]. Below we consider the ME coupling characteristic for EuTiO$_3$ as the main mechanism of *E*-field influence on the phase diagram using LGD theory. The magnetization and polarization-dependent part of the LGD free energy is [21, 22]:

$$G_M = \int_V d^3r \left( \begin{array}{l} \dfrac{\alpha_P}{2}P_3^2 + \dfrac{\beta_P}{4}P_3^4 - E_3 P_3 + \dfrac{\alpha_M}{2}M^2 + \dfrac{\alpha_L}{2}L^2 + \dfrac{\beta_M}{4}M^4 + \dfrac{\beta_L}{4}L^4 \\ + \dfrac{\lambda}{2}L^2 M^2 + \dfrac{P_3^2}{2}\left(\eta_{FM}M^2 + \eta_{AFM}L^2\right) + \dfrac{\alpha_\Phi}{2}\Phi^2 + \dfrac{\beta_\Phi}{4}\Phi^4 + \dfrac{\xi}{2}\Phi^2 P_3^2 \end{array} \right) \quad (6.1)$$

Here $P_3$ is ferroelectric polarization component, $E_3$ is external electric field component, $M^2 = M_1^2 + M_2^2 + M_3^2$ is FM magnetization square and $L^2 = L_1^2 + L_2^2 + L_3^2$ is the square of the AFM order parameter, correspondingly. The terms $\dfrac{P_3^2}{2}\left(\eta_{FM}M^2 + \eta_{AFM}L^2\right)$ represent biquadratic ME coupling between order parameters. Expansion coefficient $\alpha_P$ depends on the absolute temperature $T$ in accordance with Barrett law, namely $\alpha_P(T) = \alpha_T^{(P)}\left(T_q^{(P)}/2\right)\left(\coth\left(T_q^{(P)}/2T\right) - \coth\left(T_q^{(P)}/2T_c^{(P)}\right)\right)$. Here $\alpha_T^{(P)}$ is constant, temperatures $T_q^{(P)}$ is the so-called quantum vibration temperature related with polar soft modes, $T_c^{(P)}$ is the "effective" Curie temperature corresponding to the polar modes in bulk EuTiO$_3$. Coefficient $\beta_P$ is regarded as temperature independent.

The expansion coefficient $\alpha_M$ depends on the temperature in accordance with the Curie law, namely $\alpha_M(T) = \alpha_C(T - T_C)$, where $T_C$ is the FM Curie temperature. Note that this dependence determines the experimentally observed inverse magnetic susceptibility in paramagnetic phase of EuTiO$_3$. The temperature dependence of the expansion coefficient $\alpha_L$ is $\alpha_L(T) = \alpha_N(T - T_N)$, where $T_N$ is the Neel temperature for bulk EuTiO$_3$. For equivalent permutated magnetic Eu ions with antiparallel spin ordering, it can be assumed that $\alpha_C \approx \alpha_N$. The LM-coupling coefficient $\lambda$ should be positive, because only the positive coupling term $\lambda L^2 M^2/2$ prevents the appearance of FM phases at low temperatures $T < T_C$ under the condition of $\sqrt{\beta_M \beta_L} < \lambda$ regarded valid hereafter [30]. Coefficients $\beta_L$ and $\beta_M$ are regarded as positive and temperature independent. Following Lee *et al.* we assume that ME coupling coefficients of FM and AFM are equal and positive, i.e. $\eta_{AFM} \approx -\eta_{FM} > 0$ for numerical calculations, as anticipated for equivalent magnetic Eu ions with antiparallel spin ordering in a bulk EuTiO$_3$.

$\Phi$ is the structural order parameter (AFD displacement). The corresponding expansion coefficient $\alpha_\Phi$ depends on the absolute temperature $T$ in accordance with Barrett law, $\alpha_\Phi(T) = \alpha_T^{(\Phi)}\left(T_q^{(\Phi)}/2\right)\left(\coth\left(T_q^{(\Phi)}/2T\right) - \coth\left(T_q^{(\Phi)}/2T_S\right)\right)$ [23]. The biquadratic coupling coefficient $\xi$ is regarded as temperature-independent [3, 4, 5].

Considering the case of incipient ferroelectric and in order to obtain analytical results, one could suppose a linear dependence of polarization $P_3$ on the applied electric field via linear dielectric susceptibility $\chi$

$$P_3 \approx \chi E_3, \qquad \chi = \frac{1}{\alpha_P + \xi\Phi^2 + \eta_{FM}M^2 + \eta_{AFM}L^2}. \tag{6.2}$$

Equations of state for the absolute value of the magnetization *M*, and the antimagentization *L* can be obtained from the minimization of the free energy (6.1). They are $(\alpha_M + \eta_{FM}P_3^2)M + \beta_M M^3 + \lambda L^2 M = 0$ and $(\alpha_L + \eta_{AFM}P_3^2)L + \beta_L L^3 + \lambda L M^2 = 0$. The formal solution of these equations contains the possible *E*-field induced phase transition, namely the appearance of the mixed magnetic phases with order parameters:

$$M = \sqrt{\frac{\alpha_L \lambda - \alpha_M \beta_L + (\lambda\eta_{AFM} - \beta_L \eta_{FM})P^2}{\beta_M \beta_L - \lambda^2}}, \tag{6.3}$$

$$L = \sqrt{\frac{\alpha_M \lambda - \alpha_L \beta_M + (\lambda\eta_{FM} - \beta_M \eta_{AFM})P^2}{\beta_M \beta_L - \lambda^2}}. \tag{6.4}$$

The critical values of polarization could be found by substituting into the equations either *M*=0 or *L*=0, i.e.:

$$P_{cr}\big|_{M=0} = \sqrt{\frac{\alpha_M \beta_L - \alpha_L \lambda}{\lambda\eta_{AFM} - \beta_L \eta_{FM}}}, \qquad P_{cr}\big|_{L=0} = \sqrt{\frac{\alpha_L \beta_M - \alpha_M \lambda}{\lambda\eta_{FM} - \beta_M \eta_{AFM}}}. \tag{6.5}$$

Expressions (6.5) correspond to the lower and upper critical fields respectively:

$$E_{cr}\big|_{M=0} = \frac{1}{\chi}\sqrt{\frac{\alpha_M \beta_L - \alpha_L \lambda}{\lambda\eta_{AFM} - \beta_L \eta_{FM}}}, \qquad E_{cr}\big|_{L=0} = \frac{1}{\chi}\sqrt{\frac{\alpha_L \beta_M - \alpha_M \lambda}{\lambda\eta_{FM} - \beta_M \eta_{AFM}}} \tag{6.6}$$

Note, that LM-coupling constant $\lambda$, $\beta_M$ and $\beta_L$ are positive, as required for the stability of free energy (6.1). The conditions $\eta_{FM} < 0$ and $\eta_{AFM} > 0$, $\alpha_M(T) > 0$ and $\alpha_L(T) < 0$ are sufficient for the absolute stability of the FM phase at applied electric fields higher than the critical field $E_{cr}\big|_{M=0}$.

The complex behaviour of *M* and *L* induced by $E_3$ can be explained by the phase diagram of bulk EuTiO$_3$ in the coordinates of temperature and external electric field, as shown in **Figure 6.1a**. One can see from the diagram that the FM phase stability region starts at electric fields greater than 0.5 MV/cm at 0 K, and converges to 0.83 MV/cm at 4 K. Paramagnetic (PM) phase is stable at temperatures greater than 5 K, while its boundary with AFM phase slightly shifts to the lower temperatures as the electric field increases. At a field of $E_{cr} \geq 0.83$ MV/cm, the AFM phase disappears at all considered temperatures and so the true FM phase becomes the only absolutely stable magnetic phase. The phase diagram proves that an electric field higher than $E_{cr}$ transforms the bulk EuTiO$_3$ into a true and relatively strong FM state at temperatures lower than 5 K. The result opens up the possibility to control bulk EuTiO$_3$ between different magnetic phases using external electric field. In particular, our calculations prove that it becomes possible to control the multiferroicity, including the content of FM and AFM phases, with the help of external electric fields.

**Figure 6.1b** illustrates *M* (solid curves) and anti-magnetization *L* (dashed curves) as a function of polarization induced by an external electric field and as a function of electric field itself at different temperatures from 1 – 5 K. One can see that at electric fields less than the critical value, only AFM magnetization exists. For the electric fields greater than the critical value, a FM magnetization occurs and increases as the strength of the electric field (or polarization) increases. An unusual cross-over from the first-order phase transition (corresponding to the FM magnetization appearance), to a second order transition appears with an increase in temperature. The decrease in AFM order for electric fields greater than the critical value follows the first order transition. The critical field value increases and the "gap" between the AFM and FM states shrinks as the temperature increases (compare the curves calculated for 1 K with the ones for 4 K).

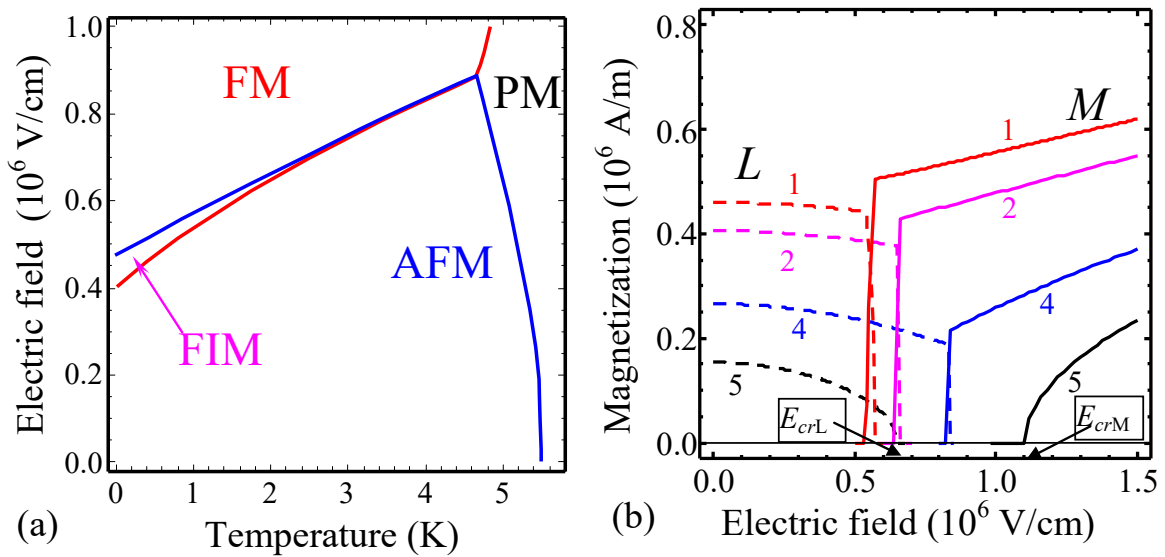

**Figure 6.1**. **Electric field control of bulk EuTiO$_3$ magnetic properties. (a)** Phase diagram of bulk EuTiO$_3$ in the coordinates of temperature versus external electric field. **(b)** Magnetization *M* (solid curves) and anti-magnetization *L* (dashed curves) as a function of external electric field at different temperatures of 1, 2, 4 and 5 K (numbers near the curves). (Reproduced from [M. D. Glinchuk et al, Phys.Rev. **B 89**, 014112 (2014)], with the permission of APS Publishing)

We can expect different trends for the value of critical field under the application of hydrostatic pressure, biaxial tensile or compressive strains to EuTiO$_3$. Our estimations show that the critical electric field should increase under the hydrostatic pressure, while it can change in an anisotropic manner for the biaxial tensile or compressive strains. Biaxial strains can lead to an increase in the critical field in some direction and a decrease in the other directions. The impact of shear strains can be even more complex.

Generally speaking, one can look for the fulfilment of expressions (6.6) in other paraelectric AFM oxides with ME coupling coefficients satisfying the conditions $\eta_{FM} < 0$ and $\eta_{AFM} > 0$, where the magnetization could be induced by an electric field $E > E_{cr}|_{M=0}$ [where $E_{cr}|_{M=0}$ is given by Eq.(6.6)], at some temperature range defined by the conditions $\alpha_M(T) > 0$ and $\alpha_L(T) < 0$. The search for such materials seems to be important both for understanding the mechanisms of ME coupling and for possible applications. The main current problem is the limited knowledge about ME coupling coefficients.

Let us discuss some more cases when one can expect *E*-field induced magnetization. In particular, such supposition can be made on the basis of data known for solid solutions $Sr_{1-x}Ba_xMnO_3$ [114] and $Sr_{1-x}Eu_xTiO_3$ [21]. Sakai et al [114] had shown the strong suppression of ferroelectricity observed at x≥0.4 and originated from $Mn^{+4}$ ions displacement upon the AFM order. This gives the direct evidence that $\eta_{AFM} > 0$ and with respect to the above written expression for the critical field, $\eta_{FM} < 0$. The assumption about different signs of $\eta_{FM}$ and $\eta_{AFM}$ also agrees with Smolenskii and Chupis [115], as well as Katsufuji [13] and Lee et al. Ryan et al [30] proposed to use the electric field to tune the magnetism in strained EuTiO$_3$ *thin film*. Critical electric field required for the origin of ferromagnetism varies from relatively high values 0.83MV/cm (for bulk EuTiO$_3$) to relatively small values 0.02 MV/cm (for bulk $Sr_{0.7}Ba_{0.3}MnO_3$). To resume the section, the phenomenological prediction can stimulate systematic *ab initio* calculations and experimental studies of the couplings in paraelectric antiferromagnet oxides.

## VII. MULTIFERROICS PROPERTIES OF BiFeO$_3$ AND ITS SOLID SOLUTIONS WITH RARE EARTH COMPOUNDS

### 7.1. Thermodynamic potential and phase diagram for multiferroic bismuth ferrite

Multiferroics, defined as materials with more than one ferroic long-range orders, are ideal systems for fundamental studies of couplings among the order parameters of different nature, e.g. FE polarization, structural AFD, FM and AFM order parameters [2, 9, 111, 116, 117, 118, 119, 120]. Bismuth ferrite BiFeO$_3$ (**BFO**) is the unique multiferroic [121, 122] with a strong ferroelectric polarization and antiferromagnetism at room temperature, as well as conduction and magnetotransport on domain walls [34, 123, 124]. The pronounced multiferroic properties and unusual domain structure evolution maintain in BFO thin films and heterostructures [35, 125, 126, 127, 128, 129]. Bulk BFO exhibits AFD order at temperatures below 1200 K; it is FE with a large spontaneous polarization below 1100 K and is AFM below Neel temperature $T_N \approx 650$ K [9, 130].

Recently we construct a comprehensive LGD thermodynamic potential and the phase diagram for pristine and slightly doped with La, a ferroelectric antiferromagnet at room temperature [131]. The role of the RM and RE couplings was established in Ref.[132]. The thermodynamic potential of LGD-type that describes AFM, FE and AFD properties of BFO, including the RM, RE and ME biquadratic couplings, and the AFD, FE, AFM contributions, as well as elastic energy has the form [132]:

$$\Delta G = \Delta G_{AFD} + \Delta G_{FE} + \Delta G_{AFM} + \Delta G_{BQC} + \Delta G_{ELS} \quad (7.1)$$

The AFD energy in the multiferroic AFM-FE-AFD $R3c$ phase is a six-order expansion on the oxygen tilt $\Phi_i$ and its gradients,

$$\Delta G_{AFD} = a_i^{(\Phi)}\Phi_i^2 + a_{ij}^{(\Phi)}\Phi_i^2\Phi_j^2 + a_{ijk}^{(\Phi)}\Phi_i^2\Phi_j^2\Phi_k^2 + g_{ijkl}^{(\Phi)}\frac{\partial \Phi_i}{\partial x_k}\frac{\partial \Phi_j}{\partial x_l} \quad (7.2)$$

Here $\Phi_i$ are components of pseudovectors, determining out-of-phase static rotations of oxygen octahedral groups.

The FE energy, $\Delta G_{FE}$, is a six-order expansion on the polarization vector $P_i$ and its gradients,

$$\Delta G_{FE} = a_i^{(P)}P_i^2 + a_{ij}^{(P)}P_i^2P_j^2 + a_{ijk}^{(P)}P_i^2P_j^2P_k^2 + g_{ijkl}^{(P)}\frac{\partial P_i}{\partial x_k}\frac{\partial P_j}{\partial x_l}. \quad (7.3)$$

The AFM energy, $\Delta G_{AFM}$, is a fourth-order expansion in terms of the AFM order parameter vector $L_i$, because this phase transition in BiFeO$_3$ is known to be the second order one, its gradient and gradient-related Lifshitz invariant [133, 134, 135].

$$\Delta G_{AFM} = a_i^{(L)}L_i^2 + a_{ij}^{(L)}L_i^2L_j^2 + g_{ijkl}^{(L)}\frac{\partial L_i}{\partial x_k}\frac{\partial L_j}{\partial x_l} + h_{ijkl}^{(L)}P_l\left(L_j\frac{\partial L_i}{\partial x_k} - L_i\frac{\partial L_j}{\partial x_k}\right). \quad (7.4)$$

In accordance with the classical LGD theory, we assume that the coefficients $a_i^{(\Phi)}$ and $a_k^{(P)}$ are temperature dependent according to Barrett law [87], $a_i^{(\Phi)} = \alpha_T^{(\Phi)}T_{q\Phi}\left(\coth(T_{q\Phi}/T) - \coth(T_{q\Phi}/T_\Phi)\right)$ and $a_k^{(P)} = \alpha_T^{(P)}\left(T_{qP}\coth(T_{qP}/T) - T_C\right)$, where $T_\Phi$ and $T_C$ are corresponding virtual Curie temperatures, $T_{q\Phi}$ and $T_{qP}$ are characteristic temperatures [100]. As it was shown recently [136] similar Barrett-type expressions can be used for the AFM coefficient $a_i^L(T)$ of pure bismuth ferrite $a_i^L(T) = \alpha_T^{(L)}T_L\left(\coth(T_L/T) - \coth(T_L/T_N)\right)$ with the Neel temperature $T_N = 645$ K and characteristic temperature $T_L = 550$ K. The expression $L \sim \sqrt{a_1^L(T)/a_{11}}$, being valid in the isotropic approximation, describes quantitatively both the temperature dependence of the AFM order parameter measured experimentally in BiFeO$_3$ by neutron scattering by Fischer et al. [130] and anomalous AFM contribution to the specific heat behaviour near the Neel temperature measured experimentally [137, 138, 139]. The gradient terms in the form of Lifshitz invariant,

$h_{ijkl}^{(L)}P_l\left(L_j\frac{\partial L_i}{\partial x_k} - L_i\frac{\partial L_j}{\partial x_l}\right)$, in Eq.(7.4) are of the so-called "flexo-type" (for classification see **Table I** in Ref.[140]) and are proportional to the third power of the order parameters $\sim PL\nabla L$. Their inclusion can induce an incommensurate spin modulation below the AFM transition, namely the cycloid spin order with a period about (62-64) nm [9, 116].

The AFD-FE-AFM coupling energy $\Delta G_{BQC}$ is a biquadratic form of the order parameters $L_i$, $P_i$ and $\Phi_i$:

$$\Delta G_{BQC} = \zeta_{ijkl}\Phi_i\Phi_j P_k P_l + \kappa_{ij}\Phi_i^2 L_j^2 + \lambda_{ij}P_i^2 L_j^2. \qquad (7.5)$$

For a given symmetry the coupling energy in Eq. (7.5) includes unknown tensorial coefficients $\zeta_{44}$, $\zeta_{11}$, $\zeta_{12}$ in Voight notations for the AFD-FE biquadratic couplings. Below, due to the lack of experimental data, the FE-AFM and AFD-AFM RM and biquadratic ME coupling constants are assumed to be isotropic, $\lambda_{ij} = \lambda\delta_{ij}$ and $\kappa_{ij} = \kappa\delta_{ij}$.

The elastic energy in the *R3c* phase is

$$\Delta G_{ELS} = -\left(s_{ijkl}\sigma_{ij}\sigma_{kl} + Q_{ijkl}\sigma_{ij}P_k P_l + R_{ijkl}\sigma_{ij}\Phi_k\Phi_l + Z_{ijkl}\sigma_{ij}L_k L_l\right). \qquad (7.6)$$

Here $s_{ijkl}$ are elastic compliance tensor components, $Q_{ijkl}$ are electrostriction tensor components, $R_{ijkl}$ are rotostriction tensor components, and $Z_{ijkl}$ are magnetostriction tensor components. All coefficients in the thermodynamic potential (7.1)-(7.6) were extracted from experimental data in Refs.[131, 132].

The calculated temperature dependence of the oxygen displacement $\Phi$ for the *R3c* phase is shown in **Fig. 7.1a.** Red and blue diamonds correspond to pure BiFeO$_3$ and BiFeO$_3$ doped with 5% of La respectively. The temperature dependence of the recalculated spontaneous polarization $P$ in *R3c* phase is shown in **Fig. 7.1b**. Red and blue diamonds correspond to experimental results for pure BiFeO$_3$ and doped with 5% of La respectively measured in this work.

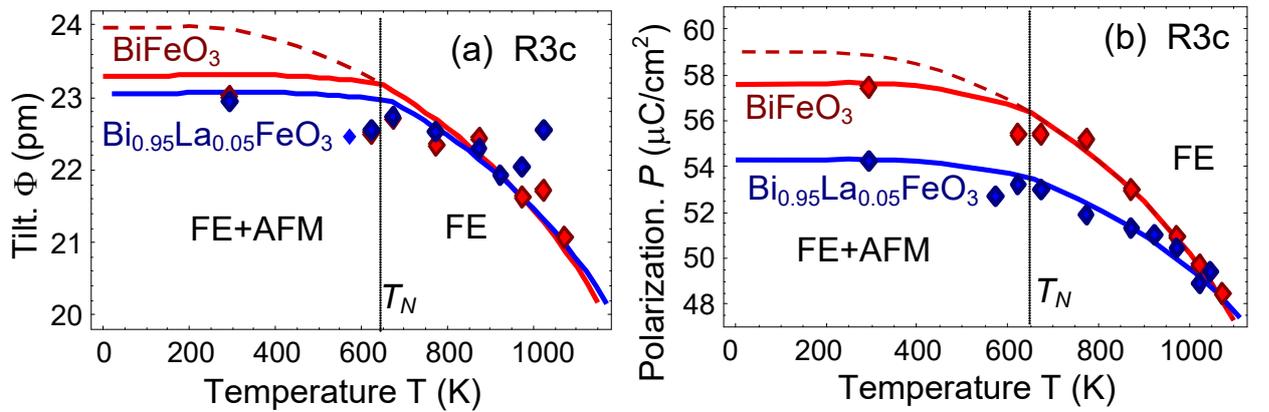

**Figure 7.1.** Temperature dependence of **(a)** AFD order parameter $\Phi$ (oxygen displacement from symmetric position) and **(b)** recalculated spontaneous polarization $P$ in *R3c* phase. Red and blue diamonds correspond to experimental results for pure BiFeO$_3$ and doped with 5% of La respectively measured in this work. Solid

curves represent the theoretical fitting with biquadratic AFD-AFM and FE-AFM couplings terms. Dashed curves for BiFeO$_3$ are calculated without biquadratic AFD-AFM and FE-AFM couplings terms. (Reproduced from [Dmitry V. Karpinsky et al, npj Computational Materials **3**:20 (2017)], with the permission of NPG Publishing).

We demonstrate that a strong biquadratic AFD-type coupling is the key to a quantitative description of BiFeO$_3$ multiferroic phase diagram including the temperature stability of the AFM, FE and AFD phases, as well as the for the prediction of novel intermediate structural phases (see **Fig. 7.2**). Furthermore, we show that RM coupling is very important to describe the FE polarization and AFD tilt behaviour in the *R3c* phase of BiFeO$_3$.

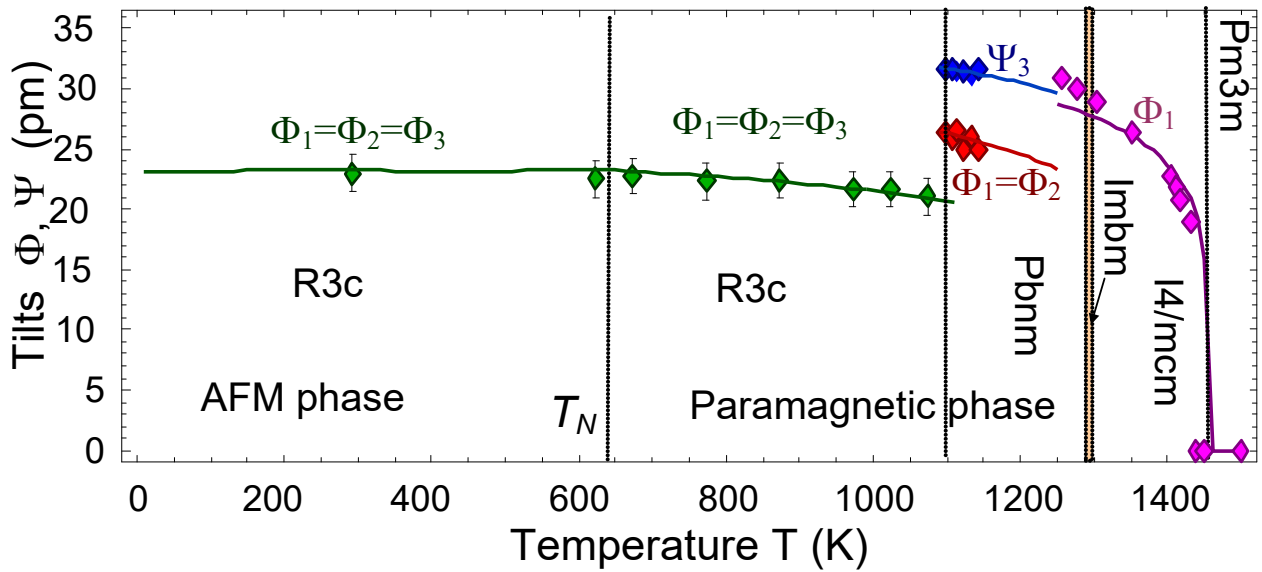

**Figure 7.2.** Temperature dependence of oxygen displacement components for different phases of BiFeO$_3$ along with the fitting (solid curves). Symbols' of different phases (*R3c*, *Pbnm*, *Imbm*, *I4/mcm* and *Pm3m*) are specified near the curves. Results for *Pbnm* phase are taken from [141], tilt for virtual *I4/mcm* phase is from [142]. (Reproduced from [Dmitry V. Karpinsky et al, npj Computational Materials **3**:20 (2017)], with the permission of NPG Publishing).

### 7.2. Rotomagnetic coupling in fine-grained multiferroic BiFeO$_3$: theory and experiment

Let us apply the thermodynamic approach based on the free energy (7.1)-(7.6) to a fine-grained BFO ceramics, for which the grain size $R$ varies from several tens nanometers to sub-microns, and the grains are separated by a stressed inter-grain shell of thickness $R_0 = (5 - 50)$ nm. The stresses can originate from different sources, such as surface tension itself, as well as from chemical pressure in the regions enriched by e.g. oxygen vacancies and/or other defects such as Fe clusters**.** Appeared [132] that the contributions of both these sources into the total stress are additive, and, therefore, hardly separable in many cases. Hence the significant part of the ceramics with densely packed identical spherical grains should be regarded affected by the surface, as well as by the chemical pressure created by the elastic defects accumulated in the grain shells and inter-grain spaces. In reality the grains are non-spherical, with distribution of sizes, and so packed much more densely reducing the part of the inter-grain space dramatically.

Therefore, for the case of the strained fine-grained ceramics Eq.(7.1) becomes affected by electrostriction, rotostriction and magnetostriction couplings according to Eqs.(7.5)-(7.6). A formal expression for the shift of AFM transition temperature related with the stresses near the grain boundaries can be obtained from the expression $a_j^{(L)}(T) = 0$. At that the approximate formulae, $a_j^{(L)} \approx a_T^{(L)}(T - T_N) - Z_{kljj}\sigma_{kl} + \kappa_{ij}\Phi_i^2 + \lambda_{ij}P_i^2$, is valid in the vicinity of Neel temperature. The expression is formal because the surface and gradient effects [72, 73] can contribute to the average values and their mean squire deviation in a complex and a priory nontrivial way. The concrete form of the expression for $a_j^{(L)}$ depends on the physical-chemical state of the grain core and surface.

Let us limit our consideration by the most common intrinsic surface stresses [72, 73, 79] coupled with Vegard strains (chemical pressure) [9, 143, 144] acting on both polarization **P**, tilt **Φ** and AFM order parameter **L** via the electrostriction, rotostriction and magnetostriction couplings, respectively. Also we regard that depolarizing field acting on ferroelectric polarization inside the grain is negligibly small due to the screening charges. Within these assumptions the radial component of the chemical pressure (denoted as $\sigma_{rr}^W(r)$) and intrinsic surface stress (denoted as $\sigma_{rr}^\mu(r)$) inside the core and shell regions acquires the form [132]:

$$\sigma_{rr}^\mu(r) \approx \begin{cases} -\dfrac{\mu}{R}, & r < R - R_0, \\ -\mu\left(\dfrac{1}{R - R_0} - \dfrac{1}{R}\right), & R - R_0 < r < R. \end{cases} \quad (7.7a)$$

$$\sigma_{rr}^W(r) \approx \begin{cases} 0, & r < R - R_0, \\ -G\dfrac{1+\nu}{1-\nu}W^m \delta N_m \dfrac{2R^3 + r^3}{3r^3}, & R - R_0 < r < R. \end{cases} \quad (7.7b)$$

Here $r$ is the distance from the grain centre to the observation point. The components of intrinsic surface stress tensor are regarded diagonal, i.e. $\mu_{kl} = \mu\delta_{kl}$ and $\mu$ is about $(1 - 10)$N/m. Poisson ratio is $\nu = -s_{12}/s_{11}$ for cubic m3m symmetry. $G = (c_{11} - c_{12})/2$ is the shear modulus.

The Vegard strain tensor $W_{kl}^m$ of $m$-th type defects is regarded diagonal, $W_{kl}^m = W^m \delta_{kl}$. For perovskites $ABO_3$ the Vegard strain tensor is often related with vacancies and its absolute value can be estimated as $|W| \propto (5 - 20)$ Å$^3$ for oxygen and cation vacancies [144]. Notably, "compositional" Vegard strains $\delta u_{kl} = W_{kl}^m \delta N_m$ can reach percents for vacancies concentration variation $\delta N_m \sim 10^{27} \, m^{-3}$ the near the surface. Despite the concentration is much higher than the defect concentration in a bulk [145], such values are typical for vacancies segregation near the surface due to the strong lowering of their formation energy at the surface [146, 147].

The total stress $\sigma_{rr}(r) = \sigma_{rr}^W(r) + \sigma_{rr}^\mu(r)$ averaged over the grain volume $V = \frac{4}{3}\pi R^3$ under the condition $R_0 \ll R$ has the form

$$\langle \sigma_{rr}(R) \rangle \approx \eta \frac{R_0}{R}, \qquad \eta = -\frac{\mu}{R_0} - 3G \frac{1+\nu}{1-\nu} W^m \delta N_m. \qquad (7.8)$$

As one can see from the explicit form of the "total stress" parameter $\eta$, its first term ($\sim \mu$) originates from the intrinsic surface stress, and the second term ($\sim W^m \delta N_m$) originates from the excess chemical pressure. Thus Eq.(7.8) proves that the of both chemical pressure and surface tension sources of the stresses contribute into the total stress additively, and, therefore hardly separable in many cases.

Eventually the renormalized AFM transition temperature for a quasi-spherical grain of radius $R$ covered by a thin shell of thickness $R_0$ acquires the form derived in Ref. [132]:

$$T_{AFM} \approx T_N - \frac{1}{\alpha_T^{(L)}} \left[ \frac{\kappa(2R_{12} + R_{11})}{|a_{11}^{(\Phi)}|} + \frac{\lambda(2Q_{12} + Q_{11})}{|a_{11}^{(P)}|} - (2Z_{12} + Z_{11}) \right] \eta \frac{R_0}{R}. \qquad (7.9)$$

The shift of $T_N$ in Eq.(7.9) contains three contributions, namely RM [proportional to $\kappa(2R_{12} + R_{11})$], RE [proportional to $\lambda(2Q_{12} + Q_{11})$] and MS [proportional to $(2Z_{12} + Z_{11})$] couplings with the total stress $\langle \sigma_{rr}(R) \rangle \sim \eta \frac{R_0}{R}$. According to the estimates in Ref. [132] the highest is the RM contribution, MS one is a bit smaller, and the RE contribution is about an order of magnitude smaller.

We show the dependence of the AFM transition temperature $T_{AFM}$ on the grain radius $R$ in **Fig. 7.3(a).** The RE, MS and RM contributions to $T_{AFM}$ are shown in **Fig. 7.3(b).** From **Fig. 7.3(b)** the size-induced increase of the AFM temperature is caused by the RM and MS couplings. The RE coupling leads to the decrease AFM transition, and the shift is several times smaller than the increase caused by RM coupling.

Temperature dependences of magnetization measured in the BFO ceramics under ZFC (lower curve) and FC condition (magnetic field of 1 kOe, upper curve) are shown in **Fig.7.4.** .One can see significantly shifted AFM transition temperature ($T_N \sim 690K$) as compared to the widely noted value of 640 K specific for BFO single crystals [148, 149].

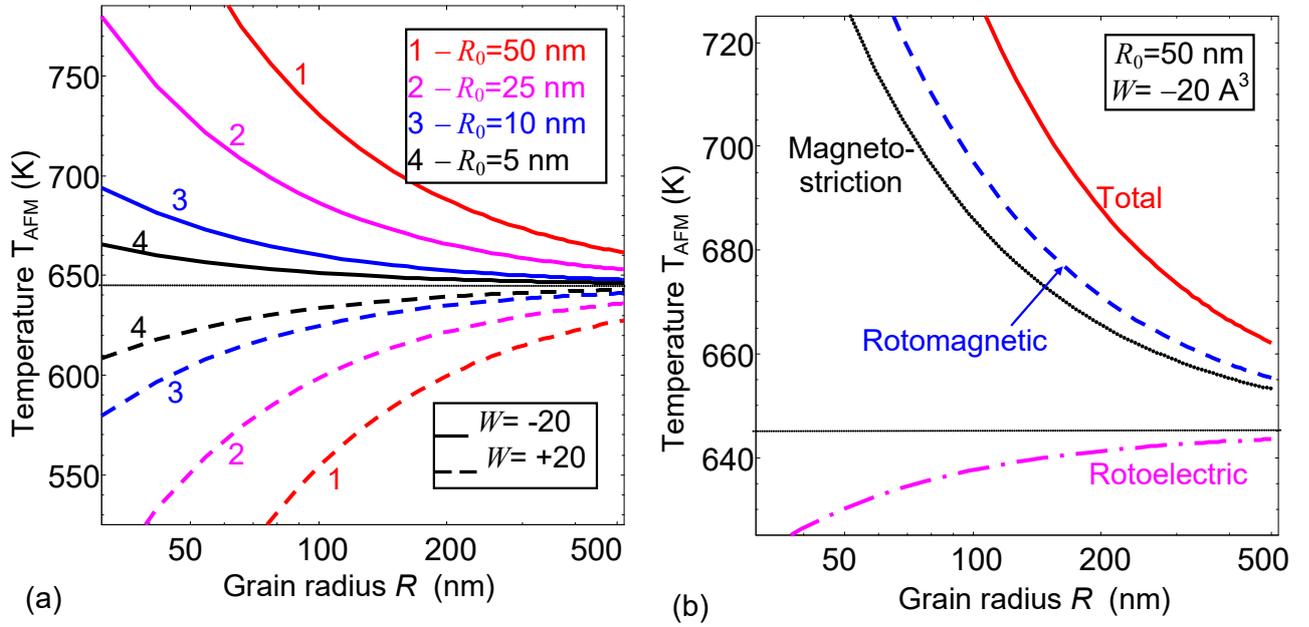

**Figure 7.3.** (a) Dependence of the AFM transition temperature $T_{AFM}$ vs. the grain radius R calculated from Eq.(7.9) for several shell thicknesses $R_0 = 50$ nm (curves 1), $R_0 = 25$ nm (curves 2), $R_0 = 10$ nm (curves 3), and $R_0 = 5$ nm (curves 4). Total Vegard coefficient $W = \sum_m W_m$ is equal to $-20$ Å$^3$ for solid curves and $+20$ Å$^3$ for dashed curves. (b) Separate contributions (RM, MS and RE) to the $T_{AFM}$. Surface tension coefficient $\mu = 5$ N/m and total defect concentration in the shell $\sum_m \delta N_m = 10^{27}$ m$^{-3}$. (Reproduced from [A. N. Morozovska et al, Phys.Rev. **B 97**,134115 (2018)], with the permission of APS Publishing)

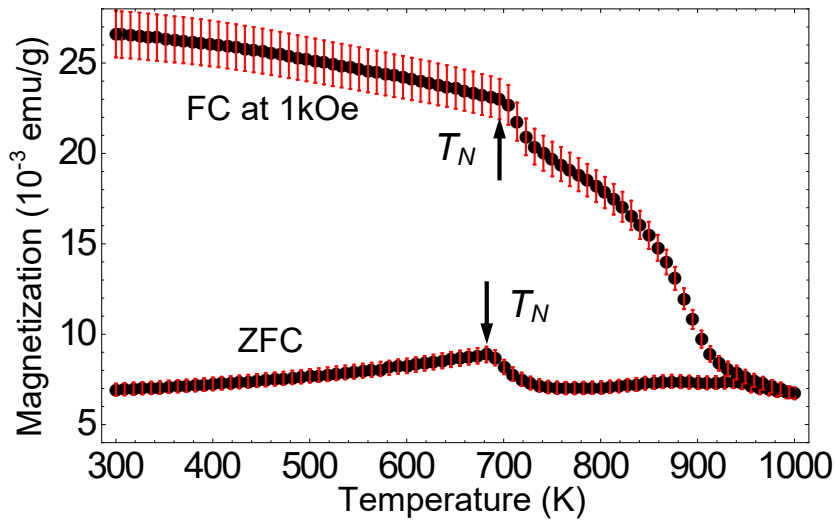

**Figure 7.4.** Temperature dependences of magnetization measured in the BFO ceramics under ZFC (lower curve) and FC condition (magnetic field of 1 kOe, upper curve). (Reproduced from [A. N. Morozovska et al, Phys.Rev. **B 97**,134115 (2018)], with the permission of APS Publishing)

To relate the above theoretical estimates with the experimental results shown in **Fig.7.4** we assume that several types of defects (oxygen vacancies and Fe clusters) are accumulated in the shells and their influence is synergetic. For the case the total defect concentration can reach relatively high values in the shell, $\sum_m \delta N_m = 10^{27}$ m$^{-3}$. In order to compare the above theory with the experimental results shown in **Fig.7.4** the observable physical quantities (e.g. magnetization **M**) should be averaged over the grain radius $R$ and shell thicknesses $R_0$ with a definite normalized distribution function $f(R, R_0)$. Since the $M^2 \sim (T - T_{AFM})$, the averaged AFM transition temperature is given by expression:

$$\langle T_{AFM} \rangle = \int_{R_{min}}^{R_{max}} dR \int_{R_0^{min}}^{R_0^{max}} dR_0 f(R, R_0) T_{AFM}(R, R_0). \quad (7.10)$$

According to **Fig. 7.5** the increase of $\langle T_{AFM} \rangle$ approximately on 45 K is possible for the ceramic with the average grain radius below 150 nm. However according the **Fig. 7.5** for the ceramics with the average grain size about 5 µm the Neel temperature should be about 650 K that is close to the single crystal value 645 K.

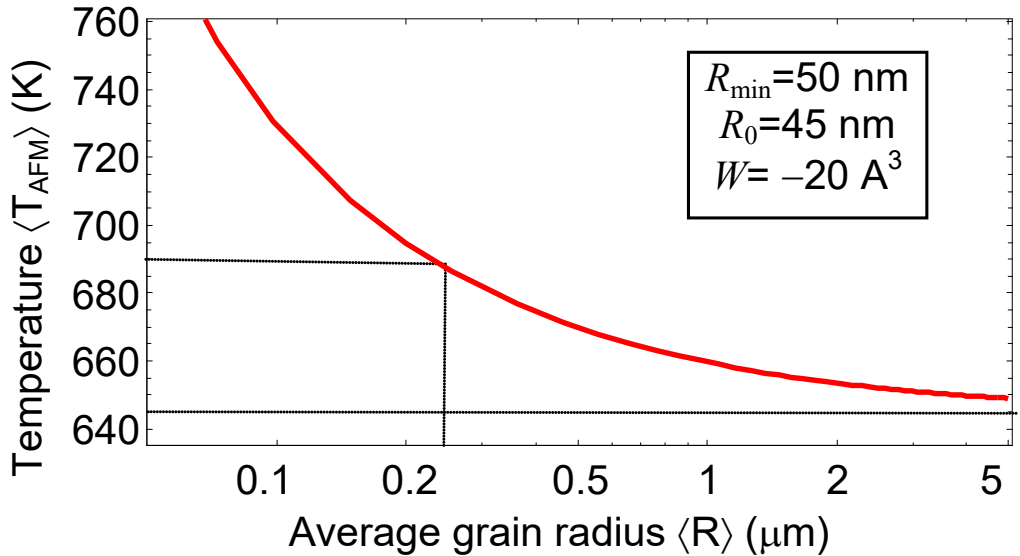

**Figure 7.5.** The dependence of the averaged transition temperature $\langle T_{AFM} \rangle$ on the average grain radius $\langle R \rangle$ calculated from Eq.(7.10) for minimal grain radius $R_{min} \approx 50$ nm, shell thickness $R_0 \approx 45$ nm, and Vegard coefficient $W = -20$ Å$^3$. (Reproduced from [A.N. Morozovska et al, Phys.Rev. **B 97**,134115 (2018)], with the permission of APS Publishing)

Hence the proposed theoretical model can explain experimental data shown in **Fig.7.4** only qualitatively, because it gives the increase of $\langle T_{AFM} \rangle$ approximately on 45 K for fine-grained ceramics with significant amount of grains with radius smaller than 250 nm. The one order of magnitude discrepancy between the average grain sizes required from the theoretical model (less than 500 nm) and experiment (about 5 μm) to reach the increase of $\langle T_{AFM} \rangle$ on 45 K evidently speaks in favor of strongly underestimated impact of the RM coupling by the chosen model parameters or unexpectedly high contribution of the small grains into the average magnetization (non-uniform distribution function of the grain sizes).

Using LGD theory for BiFeO$_3$ dense ceramics with quasi-spherical micron sized grain cores and nanosized inter-grain spaces we calculated a surprisingly strong size-induced increase of the AFM transition temperature caused by the joint action of RM and MS coupling with elastic stresses accumulated in the inter-grain spaces. The RE coupling leads to the decrease of AFM transition temperature, and the shift is several times smaller than the increase caused by RM coupling.

Also we performed experiments for dense BiFeO$_3$ ceramics, which revealed that the AFM transition was observed at $T_N$ ~690 K instead of $T_N$~645 K for a single crystal. To explain qualitatively the result we consider the possibility to control AFM properties of multiferroic BiFeO$_3$ via biquadratic AFD, RM, RE and MS couplings. To reach quantitative agreement between the theoretical model and experimental data one could also consider low symmetry phases [150, 151] with possibly higher impact of the RM coupling and other LGD parameters.

### 7.3. Determination of the AFD-AFM coupling constant for Bi$_{1-x}$R$_x$FeO$_3$ solid solutions

Using the theory of symmetry and the microscopic model Morozovska et al [136] predicted the possibility of a linear AFD-AFM effect in the perovskites with structural AFD and AFM long-range ordering and found the necessary conditions of its existence. The main physical manifestations of this effect are the smearing of the AFM transition and the jump of the specific heat near it. In the absence of external fields linear AFD-AFM coupling can induce a weak AFM ordering above the Neel temperature, but below the temperature of AFD transition. Therefore, there is the possibility of observing weak improper antiferromagnetism in multiferroics such as BiFeO$_3$ at temperatures $T > T_N$, for which the Neel temperature $T_N$ is about 645 K, and the AFD transition temperature is about 1200 K. By quantitative comparison with experiment we made estimations of the linear AFD-AFM effect in the solid solutions of multiferroic Bi$_{1-x}$R$_x$FeO$_3$ (R=La, Nd).

Available experimental results demonstrate noticeable features of the temperature dependencies of the specific heat in Bi$_{1-x}$R$_x$FeO$_3$ (R=La, Nd, x=0 – 0.2) solid solutions [152]. The features appears at the temperature of the AFM phase transition that is about (640-650) K. Corresponding experimental results are shown by symbols in the **Figs 7.6.** As one can see from the figure dashed curves calculated at zero AFD-AFM coupling ($\tilde{\chi} = 0$) and different composition x do not describe the specific hear smearing at temperatures $T > T_N$. Solid curves, calculated as $\tilde{\chi} = (2 -$ 2.5) SI units and $T_N = (645 - 655)$ K in dependence of x, describe the smearing effect adequately, proving the importance of the bilinear AFD-AFM effect for the understanding of the specific heat behaviour near the AFM phase transition. The inclusion of the bilinear AFD-AFM effect is necessary for the quantitative description of the experimental data.

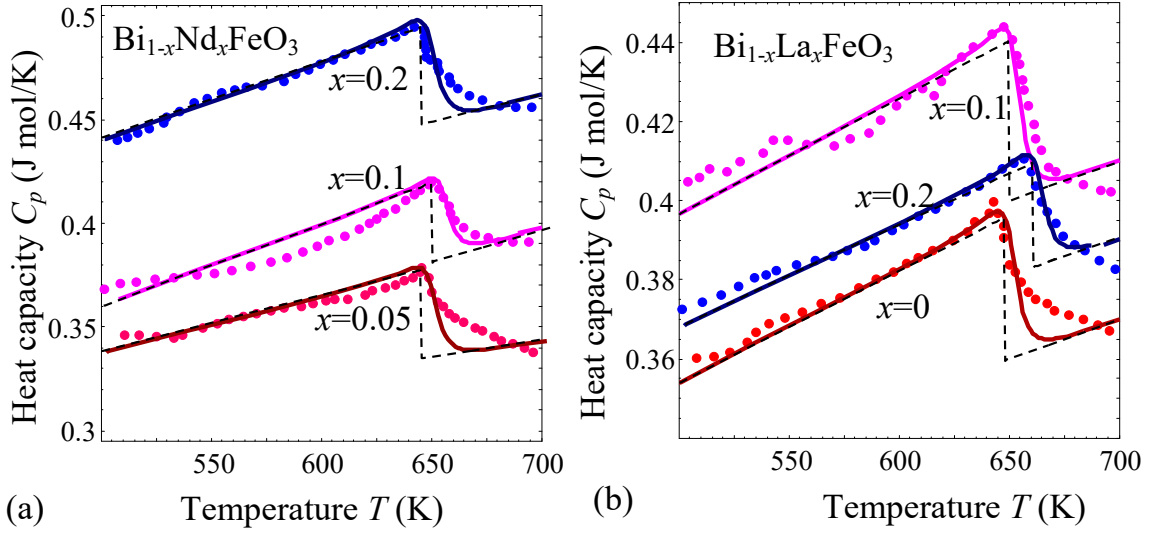

**Figure 7.6.** Temperature dependence of the specific heat near AFD phase transition of the solid solutions $Bi_{1-x}R_xFeO_3$ (R=La, Nd, x= 0 – 0.2). Symbols are experimental data for $Bi_{1-x}R_xFeO_3$ from Amirov et al [152] for heat capacity for several x values. Dashed curves are calculated by us for dimensionless coupling constant $\widetilde{\chi} = 0$. Solid curves correspond to different nonzero $\widetilde{\chi} = (2 - 2.5)$ SI units and $T_N = (645 - 655)$ K depending on the composition x, $T_L = 550$ K, $T_\Phi = 100$ K, $T_S = 1200$ K. (Reproduced from [A. N. Morozovska et al, Phys.Rev. **B 92**, 054421 (2015)], with the permission of APS Publishing)

To summarize the section, LGD thermodynamic potential is able to describe the sequence of serial and trigger-type phase transitions, the temperature-dependent behavior of the order parameters, and the corresponding susceptibilities of $Bi_{1-x}R_xFeO_3$. It can also be employed to predict the corresponding FE and AFD properties of $Bi_{1-x}R_xFeO_3$ thin films and nanoparticles by incorporating the gradient and surface energy terms that are strongly dependent on the shape, size and preparation method.

## VIII. CONCLUSIONS

The influence of RM, RE and ME coupling on phase diagram and properties of AFD perovskite oxides was considered in the framework of LGD theory. The main results we discuss and analyze in the review are the following.

(a) LGD approach predicts a surprisingly strong influence of the RM coupling practically on all the properties of EuTiO$_3$ in the temperature region of AFM and AFD phases coexistence, i.e. in a multiferroic state. In particular, the observed Neel temperature (T$_N$ = 5.5 K) was shown to be defined by RM coupling, while without the RM coupling T$_N$ appeared to be much higher (T$_N \approx$ 25.2 K). For weak RM coupling the AFM phase transition order appeared to be of the second order; while it becomes of the first order for the high enough RM coupling values. Therefore RM coupling strongly influences Neel temperature of AFD multiferroics and incipient ferroelectrics.

(b) LGD-based calculations show the presence of the triple AFD-FE-FM phase in Eu$_x$Sr$_{1-x}$TiO$_3$ nanosystems at low temperatures ($\leq$ 10 K). The polarization and magnetization values in the triple phase are calculated to be relatively high (~50 μC/cm$^2$ and ~0.5 MA/m). Therefore, the strong coupling between structural distortions, polarization and magnetization suggest the Eu$_x$Sr$_{1-x}$TiO$_3$ nanosystems as strong candidates for possible technical applications of this miltiferroic.

(c) LGD-based calculations had shown that the application of an electric field *E* in the absence of strain leads to the appearance of a FM phase due to the ME coupling. For *E*-fields higher than the critical one AFM phase disappears for all considered temperatures and FM becomes the only stable magnetic phase. The FM phase can also be induced by an *E*-field in other paraelectric AFM oxides with a positive AFM-type ME coupling coefficient and a negative FM-type ME coupling coefficient. The results show the possibility of controlling multiferroicity, including the FM and AFM phases, with help of an electric field application.

(d) It was shown on the example of non-ferroelectric SrTiO$_3$ that combined effect of flexoelectricity and rotostriction can lead to the appearance of improper spontaneous polarization and pyroelectricity in the vicinity of antiphase domain boundaries, structural twin walls, surfaces and interphases in the octahedrally tilted phase. The consideration of free charges and mobile oxygen vacancies contributions had shown essential increase of spontaneous polarization and pyroelectric coefficients values because they can effectively screen depolarization field.

(e) Low symmetry monoclinic phase with in-plane ferroelectric polarization is found to be stabilized by AFD oxygen octahedron tilts in Eu$_x$Sr$_{1-x}$TiO$_3$ thin films in particular due to flexoelectric coupling and rotostriction. The monoclinic phase is stable at moderate tensile strain over wide temperature range. The developed theory can be applied to typical AFD perovskites with octahedrally tilted phases such as incipient ferroelectrics SrTiO$_3$, CaTiO$_3$, antiferromagnet incipient ferroelectrics EuTiO$_3$, antiferromagnetic ferroelectric BiFeO$_3$ etc.

(f) The strong influence of RM, RE and ME phases on temperature stability of the AFM, FE, AFD and novel structural phases of multiferroic BiFeO$_3$ and Bi$_{1-x}$R$_x$FeO$_3$ (x = La, Nd) has been predicted and demonstrated within LGD approach.

(g) Recently the experimental and theoretical investigation of BiFeO$_3$ dense ceramics with quasi-spherical micron-sized grain core and nano-sized inter-grain spaces revealed a surprisingly strong size induced increase of AFM transition temperature ($T_N \sim 690$ K instead $T_N \sim 645$ K for single crystal) caused by the joint action of RM effect and magnetostriction coupled with elastic stress accumulated in the intergrain spaces. The RE coupling leads to decrease of AFM transition temperature, but the decrease is several times smaller than the increase caused by RM coupling.

Theoretical results obtained within LGD approach are in qualitative agreement with experimental results obtained earlier.

**Acknowledgements.** Authors are very grateful to Dr. E.A. Eliseev for technical support with manuscript preparation. A.N.M. work was partially supported by the National Academy of Sciences of Ukraine (project No. 0117U002612, 0118U003375 and 0118U003535)

# REFERENCES

[1] V. Gopalan and D.B. Litvin, Nature Materials **10**, 376–381 (2011).

[2] E.V. Balashova and A.K. Tagantsev. Phys. Rev. **B 48**, 9979 (1993).

[3] M. J. Haun, E. Furman, S.J. Jang and L. E. Cross, Ferroelectrics, **99**, 13 (1989).

[4] A.K. Tagantsev, E. Courtens and L. Arzel, Phys. Rev. B, **64**, 224107 (2001).

[5] . B. Houchmanzadeh, J. Lajzerowicz and E Salje, J. Phys.: Condens. Matter **3**, 5163 (1991)

[6] M. Daraktchiev, G. Catalan, J. F. Scott, Ferroelectrics, **375**, 122 (2008).

[7] Eugene A. Eliseev, Maya D. Glinchuk, Venkatraman Gopalan, Anna N. Morozovska. Rotomagnetic couplings influence on the magnetic properties of antiferrodistortive antiferromagnets. Journal of Applied Physics 118, 144101 (2015); doi: 10.1063/1.4932211

[8] T. Goto, T. Kimura, G. Lawes, A. P. Ramirez, and Y. Tokura, Phys. Rev. Lett. 92, 257201 – (2004)

[9] G. Catalan and James F. Scott. "Physics and applications of bismuth ferrite." Adv. Mater. **21**, 2463-2485. (2009).

[10] Zurab Guguchia, Hugo Keller, Jürgen KöHler, and Annette Bussmann-Holder. J. Phys.: Condens. Matter **24**, 492201 (2012).

[11] K. Caslin, R. K. Kremer, Z. Guguchia, H. Keller, J. Köhler, and A. Bussmann-Holder. J. Phys.: Condens. Matter **26**, 022202 (2014).

[12] Anna N. Morozovska,Eugene A. Eliseev, Maya D. Glinchuk, Olena M. Fesenko, Vladimir V. Shvartsman, Venkatraman Gopalan, Maxim V. Silibin, Dmitry V. Karpinsky. ROTOMAGNETIC COUPLING IN FINE GRAINED MULTIFERROIC $BiFeO_3$: THEORY AND EXPERIMENT (2018) (http://arxiv.org/abs/1803.02805)

[13] T. Katsufuji and H. Takagi, Phys. Rev. B **64**, 054415 (2001).

[14] V. V. Shvartsman, P. Borisov, W. Kleemann, S. Kamba, T. Katsufuji. Phys. Rev. B **81**, 064426 (2010).

[15] P. G. Reuvekamp, R. K. Kremer, J. Kohler, and A. Bussmann-Holder. Phys. Rev. B **90**, 094420 (2014)

[16] A. Bussmann-Holder, J. Kohler, R. K. Kremer, and J. M. Law, Phys. Rev. B **83**, 212102 (2011).

[17] M. Allieta, M. Scavini, L. J. Spalek, V. Scagnoli, H. C. Walker, C. Panagopoulos, S. S. Saxena, T. Katsufuji, and C. Mazzoli, Phys. Rev. B **85**, 184107 (2012).

[18] K. Z. Rushchanskii, N. A. Spaldin, andM. Lezaic, Phys. Rev. B **85**, 104109 (2012).

[19] V. Goian, S. Kamba, O. Pacherova, J. Drahokoupil, L. Palatinus, M. Dusek, J. Rohlıcek, M. Savinov, F. Laufek, W. Schranz, A. Fuith, M. Kachlık, K. Maca, A. Shkabko, L. Sagarna, A.Weidenkaff, and A. A. Belik, Phys. Rev. B **86**, 054112 (2012).

[20] A. P. Petrovic, Y. Kato, S. S. Sunku, T. Ito, P. Sengupta, L. Spalek, M. Shimuta, T. Katsufuji, C. D. Batista, S. S. Saxena, and C. Panagopoulos, Phys. Rev. B **87**, 064103 (2013).

[21] E. A. Eliseev, M. D. Glinchuk, V. V. Khist, C.-W. Lee, C. S. Deo, R. K. Behera, and A. N. Morozovska, J. Appl. Phys. **113**, 024107 (2013).

[22] Anna N. Morozovska, Maya D. Glinchuk, Rakesh K. Behera, Basyl Y. Zaylichniy, Chaitanya S. Deo, Eugene A. Eliseev. Phys.Rev. **B 84**, 205403 (2011)

[23] Anna N. Morozovska, Yijia Gu, Victoria V. Khist, Maya D. Glinchuk, Long-Qing Chen Venkatraman Gopalan, and Eugene A. Eliseev.¨ Low-symmetry monoclinic phase stabilized by oxygen octahedra rotations in thin strained $Eu_xSr_{1-x}TiO_3$ films." Phys.Rev. **B 87**, 134102 (2013).

[24] S.V. Vonsovskii, Magnetism (John Wiley and Sons, New York, 1974).

[25] J.H. Lee, Lei Fang, Eftihia Vlahos, Xianglin Ke, Young Woo Jung, L.F. Kourkoutis, Jong-Woo Kim, P.J. Ryan, Tassilo Heeg, M. Roeckerath, V. Goian, M. Bernhagen, R. Uecker, P.C. Hammel, K.M. Rabe, S. Kamba, J. Schubert, J.W. Freeland, D.A. Muller, C.J. Fennie, P. Schiffer, V. Gopalan, E. Johnston-Halperin, D. Schlom. Nature (London) **466**, 954 (2010).

[26] Lawes, G., A. B. Harris, T. Kimura, N. Rogado, R. J. Cava, A. Aharony, O. Entin-Wohlman, T. Yildirim, M. Kenzelmann, C. Broholm, and A. P. Ramirez. *Physical review letters* **95**, 087205 (2005).

[27] A S Moskvin, S-L Drechsler, Eur. Phys. J. **71**, 331 (2009)

[28] Andrew T. Mulder, Nicole A. Benedek, James M. Rondinelli, and Craig J. Fennie, Advanced Functional Materials, **23**, Issue 38, 4810, (2013).

[29] Maya D. Glinchuk, Eugene A. Eliseev, Yijia Gu, Long-Qing Chen, Venkatraman Gopalan and Anna N. Morozovska. Electric-field induced ferromagnetic phase in paraelectric antiferromagnets. *Physical Review B* 89, 014112 (2014).

[30] P. J. Ryan, J-W Kim, T. Birol, P. Thompson, J-H. Lee, X. Ke, P. S. Normile, E. Karapetrova, P. Schiffer, S. D. Brown, C. J. Fennie & D. G. Schlom. Nature Communications 4, 1334 (2013).

[31] A. Ohtomo, D. A. Muller, J. L. Grazul, H. Y. Hwang. Nature, **419**, 378, (2002).

[32] A. Ohtomo and H.Y. Hwang. Nature, **427**, 423 (2004).

33 J.W. Park, D.F. Bogorin, C. Cen, D.A. Felker, Y. Zhang, C.T. Nelson, C.W. Bark, C.M. Folkman, X.Q. Pan, M.S. Rzchowski, J. Levy and C.B. Eom. Nature Communications, **1**, 94 (2010).

[34] J. Seidel, L. W. Martin, Q. He, Q. Zhan, Y.-H. Chu, A. Rother, M. E. Hawkridge, P. Maksymovych, P. Yu, M. Gajek, N. Balke, S. V. Kalinin, S. Gemming, F. Wang, G. Catalan, J. F. Scott, N. A. Spaldin, J. Orenstein and R. Ramesh. "Conduction at domain walls in oxide multiferroics." Nature Materials, **8**, 229 (2009).

35 Ying-Hao Chu, Lane W. Martin, Mikel B. Holcomb, Martin Gajek, Shu-Jen Han, Qing He, Nina Balke, Chan-Ho Yang, Donkoun Lee, Wei Hu, Qian Zhan, Pei-Ling Yang, Arantxa Fraile-Rodríguez, Andreas Scholl, Shan X. Wang and R. Ramesh. "Electric-field control of local ferromagnetism using a magnetoelectric multiferroic." Nature Materials, **7**, no. 6: 478 (2008).

36 S. J. May P. J. Ryan, J. L. Robertson, J.-W. Kim, T. S. Santos, E. Karapetrova, J. L. Zarestky, X. Zhai, S. G. E. te Velthuis, J. N. Eckstein, S. D. Bader & A. Bhattacharya. Nature Materials, **8**, 892 (2009).

[37] M. Stengel, D. Vanderbilt, N. A. Spaldin, Nature Materials, **8**, 392 (2009).

[38] A. Vasudevarao, A. Kumar, L. Tian, J. H. Haeni, Y. L. Li, C.-J. Eklund, Q. X. Jia, R. Uecker, P. Reiche, K. M. Rabe, L. Q. Chen, D. G. Schlom, and Venkatraman Gopalan. Phys.Rev.Lett. **97**, 257602 (2006).

[39] E.A. Eliseev, A.N. Morozovska, M.D. Glinchuk, B.Y. Zaulychny, V.V. Skorokhod, R. Blinc. Phys. Rev. B. 82, 085408 (2010).

[40] D. A. Tenne, A. Bruchhausen, N. D. Lanzillotti-Kimura, A. Fainstein, R. S. Katiyar, A. Cantarero, A. Soukiassian, V. Vaithyanathan, J. H. Haeni, W. Tian, D. G. Schlom, K. J. Choi, D. M. Kim, C. B. Eom, H. P. Sun, X. Q. Pan, Y. L. Li, L. Q. Chen, Q. X. Jia, S. M. Nakhmanson, K. M. Rabe, X. X. Xi. Science, **313**, 1614 (2006).

[41] E.V. Bursian and O.I. Zaikovskii, Fiz. Tverd. Tela **10**, 1413 (1968) [Sov. Phys. Solid State **10**, 1121 (1968)]; E.V. Bursian, O.I. Zaikovsky, and K.V. Makarov, J. Phys. Soc. Japan, **28**, Suppl. 416 (1970).

[42] W. Ma and L E Cross. Appl. Phys. Lett. **79**, 4420 (2001).

[43] W. Ma and L.E. Cross. Appl. Phys. Lett. **88**, 232902 (2006).

[44] Craig J. Fennie and Karin M. Rabe, Phys. Rev. B, **72**, 100103(R) (2005).

[45] Nicole A. Benedek and Craig J. Fennie. Phys. Rev. Lett. **106**, 107204 (2011).

[46] L.Goncalves-Ferreira, Simon A. T. Redfern, Emilio Artacho, and Ekhard K. H. Salje. Phys. Rev. Lett. **101**, 097602 (2008).

47 Eric Bousquet, Matthew Dawber, Nicolas Stucki, Celine Lichtensteiger, Patrick Hermet, Stefano Gariglio, Jean-Marc Triscone and Philippe Ghosez, Nature, **452**, 732 (2008).

[48]. M.S. Majdoub, P. Sharma, and T. Cagin, Phys. Rev. **B 77**, 125424 (2008).

[49]. G. Catalan, B. Noheda, J. McAneney, L. J. Sinnamon, and J. M. Gregg, Phys. Rev B **72**, 020102 (2005).

[50] E.A. Eliseev, A.N. Morozovska, M.D. Glinchuk, and R. Blinc. Phys. Rev. B. **79**, 165433, (2009).

[51]. D. Lee, A. Yoon, S. Y. Jang, J.-G. Yoon, J.-S. Chung, M. Kim, J. F. Scott, and T. W. Noh. Phys. Rev. Lett. **107**, 057602 (2011)

[52] A.N. Morozovska, E.A. Eliseev, M.D. Glinchuk, Long-Qing Chen, Venkatraman Gopalan // Interfacial Polarization and Pyroelectricity in Antiferrodistortive Structures Induced by a Flexoelectric Effect and Rotostriction/ Phys.Rev.B. **85**, 094107 (2012)

[53] A.N. Morozovska, E.A. Eliseev, S.V. Kalinin, Long-Qing Chen and Venkatraman Gopalan. Surface polar states and pyroelectricity in ferroelastics induced by flexo-roto field. Appl. Phys. Lett. **100**, 142902 (2012).

[54] A.N. Morozovska, E.A. Eliseev, Maya D. Glinchuk, S.V. Kalinin, Long-Qing Chen, and Venkatraman Gopalan. Impact of Free Charges on Polarization and Pyroelectricity in Antiferrodistortive Structures and Surfaces Induced by a Flexoelectric Effect. Ferroelectrics, **438:1**, 32-44 (2012)

[55] P. Zubko, G. Catalan, A. Buckley, P.R. L. Welche, J. F. Scott. Phys. Rev. Lett. **99**, 167601 (2007).

[56] Sandra Van Aert, Stuart Turner, Rémi Delville, Dominique Schryvers, Gustaaf Van Tendeloo, Ekhard K. H. Salje. "Direct observation of ferrielectricity at ferroelastic domain boundaries in CaTiO3 by electron microscopy." *Advanced Materials* 24, no. 4: 523-527 (2012). DOI: 10.1002/adma.201103717

[57]. P.A. Fleury and J. M. Worlock, Phys. Rev. **174**, 613 (1968).

58. J.F. Nye. Physical properties of crystals: their representation by tensors and matrices (Oxford: Clarendon Press, 1985).

[59] A.K. Tagantsev, and G. Gerra, J. Appl. Phys. **100**, 051607 (2006).

[60] C.H. Woo and Yue Zheng, Appl. Phys. A **91**, 59 (2007)

[61] A.M. Bratkovsky, and A.P. Levanyuk, Journal of Computational and Theoretical Nanoscience, 6, 465 (2009).

[62] G. Rupprecht and R.O. Bell, Phys. Rev. **135**, A748 (1964).

[63] A.N. Morozovska, E.A. Eliseev, M.D. Glinchuk, Long-Qing Chen, Venkatraman Gopalan. Interfacial Polarization and Pyroelectricity in Antiferrodistortive Structures Induced by a Flexoelectric Effect and Rotostriction (v2) / http://arxiv.org/abs/1108.0019

[64] A.S. Yurkov, JETP Letters, Vol. **94**, No. 6, pp. 455 (2011).

[65] E. Heifets, R.I. Eglitis, E.A. Kotomin, J. Maier, G. Borstel. Surface Science **513**, 211 (2002)

[66] R. Herger, P. R. Willmott, O. Bunk, C. M. Schleputz, B. D. Patterson, B. Delley. Phys. Rev. Lett. **98,** 076102 (2007)

[67] A. Kholkin, I. Bdikin, T.Ostapchuk, and J.Petzelt, Appl. Phys. Lett. **93**, 222905 (2008).

[68] S. Dai, M. Gharbi, P. Sharma, H. S. Park. J. Appl.Phys. **110,** 104305 (2011)

[69] H.W. Jang, A. Kumar, S. Denev, M. D. Biegalski, P. Maksymovych, C.W. Bark, C. T. Nelson, C. M. Folkman, S.H. Baek, N. Balke, C. M. Brooks, D.A. Tenne, D. G. Schlom, L. Q. Chen, X. Q. Pan, S.V. Kalinin, V. Gopalan, and C. B. Eom. Phys. Rev. Lett. 104, 197601 (2010).

[70] Eugene A. Eliseev, Anna N. Morozovska, Yijia Gu, Albina Y. Borisevich, Long-Qing Chen and Venkatraman Gopalan, and Sergei V. Kalinin. Phys.Rev. B **86**, 085416 (2012)

[71] J. F. Scott, E. K. H. Salje, M. A. Carpenter. Phys. Rev. Lett. 109, 187601 (2012).

[72]. A. N. Morozovska, E. A. Eliseev, and M.D. Glinchuk, *Phys. Rev.* **B 73**, 214106 (2006).

[73]. A. N. Morozovska, M. D. Glinchuk, and E.A. Eliseev., *Phys. Rev.* **B 76**, 014102 (2007).

[74]. M.D. Glinchuk, E.A. Eliseev, A.N. Morozovska, and R. Blinc, *Phys. Rev.* **B 77**, 024106 (2008).

[75]. S. P. Lin, Yue Zheng, M. Q. Cai, and Biao Wang. Appl. Phys. Lett. **96**, 232904 (2010)

[76]. Yue Zheng and C. H. Woo, J. Appl. Phys. **107**, 104120 (2010).

[77] A.N. Morozovska, E.A. Eliseev, R. Blinc, M.D. Glinchuk. Phys. Rev. **B 81**, 092101 (2010).

78. Feng Liu and M. G. Lagally, Phys. Rev. Lett. **76**(17), 3156 (1996).

[79]. V.A. Shchukin, D. Bimberg. Rev. Mod. Phys. **71**(4), 1125-1171 (1999).

[80]. Ji Zang, Minghuang Huang, and Feng Liu, Phys. Rev. Lett. **98**, 146102 (2007)

[81]. Ji Zang, and Feng Liu, Nanotechnology **18**, 405501(2007)

[82] A.N. Morozovska, E.A. Eliseev, S.L. Bravina, A.Y. Borisevich, and S.V. Kalinin. J. Appl. Phys. **112**, 064111 (2012)

[83] Zurab Guguchia, Alexander Shengelaya, Hugo Keller, Jurgen Kohler, and Annette Bussmann-Holder. Phys. Rev. **B 85**, 134113 (2012).

[84] B.I. Shklovskii and A.L. Efros. Electronic properties of doped semiconductors. (Springer-Verlag, Berlin 1984). 388 *Pages.*

[85] Harry Fried and M. Schick. Phys. Rev. **B 38**, 954–956 (1988)

[86] M. J. Haun, E. Furman, T. R. Halemane and L. E. Cross, Ferroelectrics, **99**, 55 (1989),

[87] J. H. Barrett, Phys. Rev. **86**, 118 (1952)

[88] N.A. Pertsev, A.G. Zembilgotov, A. K. Tagantsev. Phys. Rev. Lett. **80**, 1988 (1998).

[89] G.A. Korn, and T.M. Korn. Mathematical handbook for scientists and engineers (McGraw-Hill, New-York, 1961).

[90]. W. Ma, M. Zhang, Z. Lu. Phys. Stat. Sol. (a). **166**, № 2, 811-815 (1998).

[91] K.Uchino, Eiji Sadanaga, Terukiyo Hirose. J.Am.Ceram.Soc. **72** [8], 1555-58 (1989).

[92]. Wh. Ma, Appl. Phys. A, **96**, 915-920 (2009).

93. M. A. McLachlan, D.W. McComb, M.P. Ryan, E.A. Eliseev, A.N. Morozovska, E. Andrew Payzant, Stephen Jesse, Katyayani Seal, Arthur P. Baddorf, Sergei V. Kalinin. Adv. Func. Mater. **21**, 941–947 (2011).

[94] I. E. Dzyaloshinskii, Sov. Phys. JETP **5**, 1259–1272 (1957);

[95] T. Moriya, Phys. Rev. **120**, 91–98 (1960).

[96] The boundary of ferroelectric phase with radial polarization $P_\rho$ can be estimated from the condition

$$0 = \alpha_P(T,x) + \frac{\mu}{R_e}(Q_{11}(x)+Q_{12}(x)) + \frac{2Q_{12}(x)}{s_{11}(x)+s_{12}(x)}\frac{R_i^2}{R_e^2}u_c - \eta_{12}(x)\frac{\alpha_\Phi(T,x)}{\beta_\Phi} + \frac{1}{\varepsilon_0\varepsilon_b}$$

[97] Darrell G. Schlom, Long-Qing Chen, Chang-Beom Eom, Karin M. Rabe, Stephen K. Streiffer, and Jean-Marc Triscone. Annu. Rev. Mater. Res. 37, 589–626 (2007).

[98] D.D. Fong, G.B. Stephenson, S.K. Streiffer, J.A. Eastman, O. Auciello, P.H. Fuoss, and C. Thompson, Science **304**, 1650 (2004).

[99] M. J. Highland, T. T. Fister, D. D. Fong, P. H. Fuoss, Carol Thompson, J. A. Eastman, S. K. Streiffer, and G. B. Stephenson. Phys. Rev. Lett. **107**, 187602 (2011)

[100] Yijia Gu, Karin Rabe, Eric Bousquet, Venkatraman Gopalan, and Long-Qing Chen. Phys. Rev. B **85**, 064117 (2012).

[101] Yurong Yang, Wei Ren, Massimiliano Stengel, X. H. Yan, and L. Bellaiche. Phys. Rev. Lett. 109, 057602 (2012).

[102] N. A. Pertsev, A. K. Tagantsev, and N. Setter, Phys. Rev. B **61**, R825 (2000).

[103] Oswaldo Diéguez, Silvia Tinte, A. Antons, Claudia Bungaro, J. B. Neaton, Karin M. Rabe, and David Vanderbilt. Phys. Rev. B 69, 212101 (2004).

[104] Y.L. Li, L. Q. Chen. Appl. Phys. Lett. 88, 072905 (2006).

[105] J.S. Speck, W. Pompe. J.Appl.Phys. **76**(1), 466 (1994).

[106] T. Yamada, T. Kiguchi, A. K. Tagantsev, H. Morioka, T. Iijima, H. Ohsumi, S. Kimura, M. Osada, N. Setter, and H. Funakubo. Integrated Ferroelectrics 115, no. 1: 57-62 (2010).

[107] Makoto Iwata and Yoshihiro Ishibashi. Japanese Journal of Applied Physics 44, No. 5A, pp. 3095–3098 (2005).

[108] B. Noheda, D. E. Cox, G. Shirane, J. A. Gonzalo, L. E. Cross and S-E. Park. Appl. Phys. Lett. 74, 14, 2059 (1999).

[109] F. Cordero, F. Trequattrini, F. Craciun and C. Galassi, J. Phys.: Condens. Matter 23, 415901 (2011).

[110] Pingping Wu, Xingqiao Ma, Yulan Li, Venkatraman Gopalan, and Long-Qing Chen. Appl. Phys. Lett. 100, 092905 (2012).

[111] M. Fiebig. "Revival of the magnetoelectric effect." *Journal of Physics D: Applied Physics* 38, no. 8, R (2005).

[112] S.M. Wu, Shane A. Cybart, P. Yu, M. D. Rossell, J. X. Zhang, R. Ramesh, and R. C. Dynes. *Nature materials* 9, no. 9: 756-761 (2010)

[113] Y. Tokunaga, Y. Taguchi, T. Arima, and Y. Tokura, Nature Physics **8**, 838-844 (2012).

[114] H. Sakai, J. Fujioka, T. Fukuda, D. Okuyama, D. Hashizume, F. Kagawa, H. Nakao, Y. Murakami, T. Arima, A. Q. R. Baron, Y. Taguchi, and Y. Tokura. Physical Review Letters 107, no. 13 (2011): 137601.

[115] G. A., Smolenskiĭ, and I. E. Chupis. "Ferroelectromagnets." *Soviet Physics Uspekhi* 25, no. 7: 475 (1982).

[116] N. A. Spaldin and M. Fiebig, "The renaissance of magnetoelectric multiferroics." *Science* 309, no. 5733, 391-392(2005).

[117] J.M. Rondinelli, N.A. Spaldin. "Structure and properties of functional oxide thin films: insights from electronic-structure calculations." *Advanced materials* 23, no. 30, 3363-3381 (2011).

[118] A. P. Pyatakov, A. K.Zvezdin, "Magnetoelectric and multiferroic media." *Physics-Uspekhi* 55, no. 6, 557-581 (2012).

[119] J. F. Scott. "Data storage: Multiferroic memories." *Nature materials* 6, no. 4, 256 (2007).

[120] Manfred Fiebig, Thomas Lottermoser, Dennis Meier, Morgan Trassin. The evolution of multiferroics. Nature Reviews Materials volume 1, Article number: 16046 (2016), doi:10.1038/natrevmats.2016.46

[121] Donna C. Arnold, Kevin S. Knight, Gustau Catalan, Simon AT Redfern, James F. Scott, Philip Lightfoot, and Finlay D. Morrison. "The β-to-γ Transition in BiFeO3: A Powder Neutron Diffraction Study." *Advanced Functional Materials* 20, no. 13 (2010): 2116-2123.

[122] R. Palai, R. S. Katiyar, Hans Schmid, Paul Tissot, S. J. Clark, Jv Robertson, S. A. T. Redfern, G. A. Catalan, and J. F. Scott. "β phase and γ− β metal-insulator transition in multiferroic Bi Fe O 3." *Physical Review B* 77, no. 1: 014110 (2008).

[123] Q. He, C.-H. Yeh, J.-C. Yang, G. Singh-Bhalla, C.-W. Liang, P.-W. Chiu, G. Catalan, L.W. Martin, Y.-H. Chu, J. F. Scott, and R. Ramesh. "Magnetotransport at domain walls in BiFeO 3." *Physical review letters* 108, no. 6: 067203 (2012).

[124] G. Catalan, J. Seidel, R. Ramesh, and J. F. Scott. "Domain wall nanoelectronics." *Reviews of Modern Physics* 84, no. 1: 119 (2012).

[125] J. Wang, J. B. Neaton, H. Zheng, V. Nagarajan, S. B. Ogale, B. Liu, D. Viehland, V. Vaithyanathan, D. G. Schlom, U. V. Waghmare, N. A. Spaldin, K. M. Rabe, M. Wuttig, R. Ramesh "Epitaxial BiFeO3 multiferroic thin film heterostructures." *Science* 299, no. 5613: 1719-1722 (2003).

[126] Y.-H. Chu, Qian Zhan, Lane W. Martin, Maria P. Cruz, Pei-Ling Yang, Gary W. Pabst, Florin Zavaliche, Seung-Yeul Yang, Jing-Xian Zhang, Long-Qing Chen, Darrell G. Schlom, I.-Nan Lin, Tai-Bor Wu, and Ramamoorthy Ramesh, "Nanoscale domain control in multiferroic BiFeO3 thin films." *Advanced Materials* 18, no. 17: 2307-2311 (2006).

[127] Peter Maksymovych, Mark Huijben, Minghu Pan, Stephen Jesse, Nina Balke, Ying-Hao Chu, Hye Jung Chang, Albina Y. Borisevich, Arthur P. Baddorf, Guus Rijnders, Dave H. A. Blank, Ramamoorthy Ramesh, and Sergei V. Kalinin. "Ultrathin limit and dead-layer effects in local polarization switching of BiFeO3" Phys. Rev. B, 85, 014119 (2012).

[128] C. Beekman, W. Siemons, M. Chi, N. Balke, J. Y. Howe, T. Z. Ward, P. Maksymovych, J. D. Budai, J. Z. Tischler, R. Xu, W. Liu, and H. M. Christen. "Ferroelectric Self-Poling, Switching, and Monoclinic Domain Configuration in BiFeO3 Thin Films." *Advanced Functional Materials* 26, no. 28: 5166-5173 (2016).

[129] Morgan Trassin, Gabriele De Luca, Sebastian Manz, and Manfred Fiebig. "Probing ferroelectric domain engineering in BiFeO3 thin films by second harmonic generation." Advanced Materials 27, no. 33: 4871-4876 (2015).

[130] P. Fischer, M. Polomska, I. Sosnowska, M. Szymanski. "Temperature dependence of the crystal and magnetic structures of BiFeO3." *Journal of Physics C: Solid State Physics* 13, no. 10, 1931 (1980).

[131] Dmitry V. Karpinsky, Eugene A. Eliseev, Fei Xue, Maxim V. Silibin, A. Franz, Maya D. Glinchuk, Igor O. Troyanchuk, Sergey A. Gavrilov, Venkatraman Gopalan, Long-Qing Chen, and Anna N. Morozovska. "Thermodynamic potential and phase diagram for multiferroic bismuth ferrite ($BiFeO_3$)". npj Computational Materials **3**:20 (2017); doi:10.1038/s41524-017-0021-3

[132] Anna N. Morozovska, Eugene A. Eliseev, Maya D. Glinchuk, Olena M. Fesenko, Vladimir V. Shvartsman, Venkatraman Gopalan, Maxim V. Silibin, Dmitry V. Karpinsky. Rotomagnetic coupling in fine-grained multiferroic BiFeO3: theory and experiment. Phys.Rev. **B 97**, 134115 (2018)

[133] Pierre Tolédano, "Pseudo-proper ferroelectricity and magnetoelectric effects in TbMnO3". Phys. Rev. **B 79**, 094416 (2009).

[134] A. Brooks Harris, "Ferroelectricity induced by incommensurate magnetism." *Journal of Applied Physics* **99**, no. 8: 08E303 (2006)]

[135] Benjamin Ruette, S. Zvyagin, Alexander P. Pyatakov, A. Bush, J. F. Li, V. I. Belotelov, A. K. Zvezdin, and D. Viehland. "Magnetic-field-induced phase transition in $BiFeO_3$ observed by high-field electron spin resonance: Cycloidal to homogeneous spin order." *Physical Review B 69*, no. 6: 064114 (2004)

[136] Anna N. Morozovska, Victoria V. Khist, Maya D. Glinchuk, Venkatraman Gopalan, and Eugene A. Eliseev. "Linear antiferrodistortive-antiferromagnetic effect in multiferroics: Physical manifestations." *Physical Review B* 92, no. 5: 054421 (2015).

[137] J.R. Chen, W.L. Wang, J.-B. Li, G.H. Rao. "X-ray diffraction analysis and specific heat capacity of (Bi$_{1-x}$La$_x$)FeO3 perovskites". Journal of Alloys and Compounds 459, 66–70 (2008).

[138] S. N. Kallaev, Z. M. Omarov, R. G. Mitarov, A. R. Bilalov, G. G. Gadzhiev, L. A. Reznichenko, R. M. Ferzilaev, S. A. Sadykov, Physics of the Solid State 56, 1412-1415 (2014).

[139] A. A. Amirov, A. B. Batdalov, S. N. Kallaev, Z. M. Omarov, I. A. Verbenko, O. N. Razumovskaya, L. A. Reznichenko, and L. A. Shilkina. "Specific features of the thermal, magnetic, and dielectric properties of multiferroics BiFeO 3 and Bi 0.95 La 0.05 FeO 3." Physics of the Solid State 51, no. 6: 1189-1192 (2009).

[140] E.A. Eliseev, S.V. Kalinin, Yijia Gu, M. D. Glinchuk, V.V. Khist, A. Borisevich, Venkatraman Gopalan, Long-Qing Chen, and A. N. Morozovska. "Universal emergence of spatially modulated structures induced by flexoantiferrodistortive coupling in multiferroics." *Physical Review B* 88, 224105 (2013).

[141] D.C.Arnold, K.S.Knight, F.D.Morrison and P.Lightfoot, Phys.Rev.Lett., **2009,** 102, 027602

[142] Igor A. Kornev and L. Bellaiche. Phys. Rev. B **2009,** 79, 100105(R)

[143] X. Zhang, A. M. Sastry, W. Shyy, "Intercalation-induced stress and heat generation within single lithium-ion battery cathode particles." J. Electrochem. Soc. **155**, A542 (2008).

[144] Daniel A. Freedman, D. Roundy, and T. A. Arias, "Elastic effects of vacancies in strontium titanate: Short-and long-range strain fields, elastic dipole tensors, and chemical strain." Phys. Rev. B **80**, 064108 (2009).

[145] O. Volnianska and P. Boguslawski. Magnetism of solids resulting from spin polarization of p orbitals. J. Phys.: Condens. Matter **22,** 073202 (2010).

[146] H. Jin, Y. Dai, BaiBiao Huang, and M.-H. Whangbo, "Ferromagnetism of undoped GaN mediated by through-bond spin polarization between nitrogen dangling bonds." Appl. Phys. Lett. **94**, 162505(2009)

[147] Fenggong Wang, Zhiyong Pang, Liang Lin, Shaojie Fang, Ying Dai, and Shenghao Han. "Magnetism in undoped MgO studied by density functional theory." Phys. Rev. **B 80**, 144424 (2009).

[148] M. Ramazanoglu, W. Ratcliff, II, Y. J. Choi, Seongsu Lee, S.-W. Cheong, and V. Kiryukhin Temperature-dependent properties of the magnetic order in single-crystal BiFeO3. Phys. Rev. **B 83**, 174434 (2011) https://doi.org/10.1103/PhysRevB.83.174434

[149] Manoj K. Singh, W. Prellier, M. P. Singh, Ram S. Katiyar, and J. F. Scott. Spin-glass transition in single-crystal BiFeO3 Phys. Rev. **B 77**, 144403 (2008). https://doi.org/10.1103/PhysRevB.77.144403

[150] Tom T.A. Lummen, Yijia Gu, Jianjun Wang, Shiming Lei, Fei Xue, Amit Kumar, Andrew T. Barnes, Eftihia Barnes, Sava Denev, Alex Belianinov, Martin Holt, Anna N. Morozovska, Sergei V. Kalinin, Long-Qing Chen, Venkatraman Gopalan. Thermotropic Phase Boundaries in Classic Ferroelectrics. Nature Communications, **5**, Article number: 3172, (2014) DOI: 10.1038/ncomms4172

[151] T. T. A. Lummen, J. Leung, A. Kumar, X. Wu, Y. Ren, B. K. VanLeeuwen, R. C. Haislmaier, M. Holt, K. Lai, S. V. Kalinin, V. Gopalan, "Emergent Low Symmetry Phases and Large Property Enhancements in Ferroelectric KNbO3 Bulk Crystals" Adv. Mater., 29, 1700530 (2017). https://doi.org/10.1002/adma.201700530

[152] A. A. Amirov, A. B. Batdalov, Z. M. Omarov, S. N. Kallaev,I. A. Verbenko, L. A. Reznichenko, The heat capacity of multiferroic Bi1-xRexFeO3 (Re = La, Nd; x = 0-0,2)

Phase transitions, ordered states and new materials Number 03 (2010). URL: http://ptosnm.ru/ru/issue/2010/3/48/publication/514